\documentclass[letter]{article}
\usepackage[utf8]{inputenc}
\usepackage{mathtools}
\usepackage[square,sort,comma,numbers]{natbib}
\usepackage[margin=25mm]{geometry}
\usepackage{graphics}
\usepackage{color}
\usepackage{mathrsfs}
\usepackage{floatrow}   
\usepackage{graphicx,xcolor} 
\usepackage{subfigure}
\usepackage[framemethod=tikz]{mdframed}
\usepackage{amssymb,amsmath,amsfonts,amstext}
\usepackage{stmaryrd}
\usepackage{verbatim}
\usepackage{amsthm}
\usepackage{esint}
\usepackage{caption}
\usepackage{multirow}
\usepackage{bm}
\usepackage{authblk}

\graphicspath{ {./Figures/} }

\usepackage{float}
\usepackage[linesnumbered,ruled,vlined]{algorithm2e}

\makeatletter

\providecommand{\keywords}[1]
{
  \small	
  \textbf{\textit{Keywords---}} #1
}

\floatstyle{ruled}
\newfloat{algorithm}{tbp}{loa}
\floatname{algorithm}{Algorithm}
\makeatother

\title{\textbf{Predicting the unobserved: a statistical mechanics framework for non-equilibrium material response with quantified uncertainty}}

\author[1]{Shenglin Huang}
\author[2]{Ian R.~Graham}
\author[3]{Robert A.~Riggleman}
\author[1]{Paulo Arratia}
\author[4]{Steve Fitzgerald} 
\author[1]{Celia Reina \thanks{creina@seas.upenn.edu}}
\affil[1]{Department of Mechanical Engineering and Applied Mechanics, University of Pennsylvania, Philadelphia, PA 19104, USA}
\affil[2]{Department Physics and Astronomy, University of Pennsylvania, Philadelphia, PA 19104, USA}
\affil[3]{ Department of Chemical and Biomolecular Engineering, University of Pennsylvania, Philadelphia, Pennsylvania 19104, USA}
\affil[4]{Department of Applied Mathematics, University of Leeds, Leeds LS2 9JT, United Kingdom}

\begin{document}

\maketitle


\begin{abstract}

Can far-from-equilibrium material response under arbitrary loading be inferred from equilibrium data and vice versa? Can the effect of element transmutation on mechanical behavior be predicted? Remarkably, such extrapolations are possible in principle for systems governed by stochastic differential equations, thanks to a set of exact relations between probability densities for trajectories derived from the path integral formalism \citep{chen2007exact,nummela2007exact,kieninger2021path}. In this article, we systematically investigate inferences (in the form of ensemble-averages) drawn on system/process $S$ based on stochastic trajectory data of system/process $\tilde{S}$, with quantified uncertainty, to directly address the aforementioned questions. Interestingly, such inferences and their associated uncertainty do not require any simulations or experiments of $S$. The results are exemplified over two illustrative examples by means of numerical simulations: a one-dimensional system as a prototype for polymers and biological macromolecules, and a two-dimensional glassy system. In principle, the approach can be pushed to the extreme case where $\tilde{S}$ is simply comprised of Brownian trajectories, i.e., equilibrium non-interacting particles, and $S$ is a complex interacting system driven far from equilibrium. In practice, however, the ``further’’ $\tilde{S}$ is from $S$, the greater the uncertainty in the predictions, for a fixed number of realizations of system $\tilde{S}$. 

\end{abstract}



\section{Introduction}
\noindent

Predicting the far-from-equilibrium response of materials is of great scientific and practical interest. Any dynamically-loaded system, from an RNA strand being stretched by optical tweezers to steel being cold-rolled, is driven away from equilibrium. 
Moreover, one may be interested in equilibrium and non-equilibrium quantities like free energies or rheological/flow properties, and wish to infer them from experiments or simulations of potentially different systems or loading conditions. On the simulation side, molecular dynamic simulations have a computational cost and associated time scale limitations that currently precludes the direct exploration of material behavior at low strain rates \citep{yan2016time}. Similarly, it is difficult in many dissipative experimental systems to probe the free energy landscape by evaluating the work at infinitesimally slow loading rates \citep{collin2005verification}. 
 Ultimately, extrapolations of material behavior over loading conditions or material systems is crucial for the inverse problem of material design.

Attempting to make predictions about one system based on knowledge of another has a long history: approximating the system of interest as a perturbation on a more easily-soluble one is a standard technique across mechanics and physics (\cite{hashin1963variational,zwanzig1954}), and the idea finds applications in equilibrium free energy calculations with techniques such as thermodynamic integration \citep{rickman2002free}, umbrella sampling \citep{torrie1977nonphysical} or metadynamics \citep{laio2002escaping,laio2008metadynamics}. Other approaches used to infer equilibrium information from non-equilibrium data, coming from either simulations or experiments, include fluctuation theorems, such as the Jarzynski relation \citep{jarzynski1997equilibrium} or Crooks' fluctuation theorem \citep{crooks1999entropy}. More recently, in the context of simulating non-equilibrium behavior, hyperdynamics methods \citep{voter,kim2013local} have emerged as a means to accelerate the simulation of rare events and alleviate the time-scale bottleneck. These may be abstractly viewed as non-equilibrium analogues of the previous approaches, 
where the potential is modified to have a faster-evolving system, e.g., by converting a large energy barrier into a smaller one,  and the results are then corrected to emulate those of the original system.
These approaches are distinct from methods such as transition path sampling \citep{bolhuis2002transition}, forward flux sampling \citep{allen2009forward}, milestoning \citep{faradjian2004computing}, parallel replica dynamics \citep{voter1998parallel}, and parallel trajectory splicing \citep{perez2016long}, which leave the potential untouched and extend the temporal reach of the simulation via sophisticated algorithms that preserve the statistics of the original system. For a review on accelerated molecular dynamics methods we refer the reader to \cite{perez2009accelerated} and \cite{voter2002extending}.

In 2007, an exact correspondence between trajectory probabilities in two different systems was derived, and exploited (\cite{chen2007exact}; \cite{nummela2007exact}-- see also earlier works with similar ideas, e.g.,  \cite{zuckerman1999dynamic}) to accelerate the sampling of trajectories that overcome both energetic barriers (e.g., surface diffusion in a crystal) and entropic barriers (e.g., nanopore traversal by a polymer; see also \cite{shin2010polymer}). The fundamental idea in each case is this: add a {\it bias} to the system potential $V\to \tilde V \coloneqq V+V_{bias}$, perform simulations of the system governed by $\tilde V$, and use the exact relations (discussed in detail below) to reweight the probability of the simulated trajectories so that they are statistically equivalent to those of the original system governed by $V$. The potential $V_{bias}$ can be highly general, and could for example be selected to ``fill in'' a deep potential well in which the system is trapped, or to point the system toward a particular location of interest in space. This is distinct from other accelerated molecular dynamics formulations  as the reweighting is exact for each trajectory, it does not require an \emph{a priori} knowledge of the reaction coordinates, it could potentially be applied to experimental data, and no time rescaling is performed. There is also no requirement that the system's evolution can be characterized by long sojourns in deep potential wells, punctuated by occasional transitions between them, meaning that path integral hyperdynamics can be applied even to systems with no well-defined transition state, such as the energetic barriers mentioned above. Furthermore, we are interested not only in accelerating rare event sampling (though this provides an intuitive example), but more generally in the statistical connections between distinct systems. Rather than accelerating the simulation of a system's evolution, we focus on the computation of ensemble-average quantities, arbitrarily far from equilibrium. The method's principal shortcoming lies in the reweighting of the trajectories: adding too extreme a bias will destroy the statistics of the quantity one is trying to compute for a fixed computational cost \citep{ikonen2011diffusion}; and this issue limits in practice the approach to systems of relatively small size and short time simulations. In other words, choosing a large bias will greatly accelerate the sampling of rare trajectories, at the cost of introducing great uncertainty in the statistical conclusions. Here, our interest lies not only in accelerated sampling strategies, but also in understanding non-equilibrium material behavior, and its connection to equilibrium properties. We extend the path integral hyperdynamics approach to include time-dependent, out-of-equilibrium boundary conditions, of interest in mechanics, and fully quantify, for the first time, the statistical uncertainty that the biasing procedure introduces. 

The paper is organized as follows. In section \ref{Sec:Reweighting} below, we give a detailed derivation of the procedure to obtain expectation values of a given statistical observable $\mathcal O$ in system $S$ from data generated by stochastic simulations of system $\tilde{S}$. Next, in Section \ref{Sec:Uncertainty}, we quantify the uncertainty introduced by the biasing procedure in terms of the difference of interparticle potentials and loading conditions between the target ($S$) and simulated ($\tilde{S}$) systems/processes. These results are then exemplified in Section \ref{Sec:Ex1} over a 1D mass-spring chain subject to a non-equilibrium, time-dependent boundary condition. While this example is simple, it has proved useful as a prototype for a polymeric chain and biological macromolecules, and it is here used to showcase the various possibilities offered by the formalism. In particular, we firstly investigate element transmutation, in which systems $S$ and $\tilde{S}$ have different interatomic potentials. Secondly, we consider going from equilibrium to non-equilibrium conditions for the same material, and finally, we take the method to its extreme by letting system $\tilde{S}$ be a simple Brownian motion, and $S$ a nonlinear interacting chain with time-dependent boundary conditions. A second physical system is explored in Section \ref{Sec:Ex2}, where the caging in two-dimensional glassy systems is predicted from the liquid state. Finally, Section \ref{Sec:Conclusions} contains our conclusions, and several appendices give further technical details of the calculations for completeness.  

\section{Predicting the non-equilibrium behavior of a material/process from that of
a reference one} \label{Sec:Reweighting}

The goal of the present manuscript is to predict the evolution of an observable $\mathcal{O}(t)$ (e.g., instantaneous stress, work done on the system,
diffusion coefficient) for a material and loading conditions $S$, from that of $\tilde{S}$. This general objective thus includes, as particular cases, virtual element transmutation for given loading conditions (e.g., response for different interparticle potentials), as well as different loading conditions for a given material (e.g., predicting the response to an excitation from equilibrium behavior). In all cases, the material systems are here considered to be composed of particles, whose motion obey overdamped Langevin dynamics. That is, the position of each particle $\mathbf{r}_i$ evolves
as
\begin{equation} \label{Eq:Langevin}
    \eta \dot{\mathbf{r}}_i = -\frac{\partial V(\mathbf{r},\boldsymbol \lambda (t))}{\partial \mathbf{r}_i} +\sqrt{2k_B T \eta}\, \dot{\boldsymbol \xi}_i, \quad i=1, ..., N
\end{equation}
where $\eta$ is the effective viscosity, $k_B$ the Bolztmann constant, $T$ the temperature and $\dot{\boldsymbol\xi}_i$ is a vector of independent white noise (i.e., each component is assumed Gaussian with zero mean and variance given by $\langle \dot{\boldsymbol \xi}_{i}(t) \dot{\boldsymbol \xi}_{j}(t') \rangle= \delta_{ij} \delta(t-t')  \mathbf{I}_d$, with $\mathbf{I}_d$ being the $d\times d$ identity matrix and $d$ being the dimension of the problem). In addition, the total potential energy of the system $V$ is considered to depend on all particles positions $\mathbf{r}$ as well as, potentially, time-dependent boundary conditions $\boldsymbol \lambda(t)$. These equations are commonly used to model systems in aqueous solution, such as biological macromolecules \citep{raj2011phase} and colloids \citep{markutsya2014characterization}, or as the basis of stochastic thermostats in molecular dynamics simulations \citep{hijazi2018fast}, when the system is coupled to a heat bath. Although the proposed framework may be easily generalized to the case where external forces are present \citep{nummela2007exact}, or to the underdamped case where inertia is not negligible \citep{chen2007exact, kieninger2021path}, in what follows we consider the simpler case of Eq.~\eqref{Eq:Langevin}.

We here aim at predicting observables of material/process $S$, (characterized by potential $V$ and loading protocol $\boldsymbol \lambda(t)$), from data on material/process $\tilde{S}$ (with potential $\tilde{V}$ and protocol $\tilde{\boldsymbol \lambda}(t)$), both evolving according to the overdamped Langevin dynamics. To achieve this goal, we resort to the path integral formalism, which provides an equivalent representation to the stochastic dynamics given by Eq.~\eqref{Eq:Langevin} \citep{chaichian2018path}. More precisely, for system $S$, the probability density of trajectories $\mathbf{r}(t)$, $0<t<\tau$ for given initial conditions $\mathbf{r}(0) = \mathbf{r}^0$, is given by
\begin{equation} \label{Eq:PathV}
    \mathcal{P}(\mathbf{r}(t)|\mathbf{r}^0) = \frac{1}{\mathcal{Z}}e^{-\beta \mathcal{I}}, \quad \text{where} \quad  \mathcal{I} = \frac{1}{4\eta}\int_0^\tau \sum_{i=1}^N \left|\eta \dot{\mathbf{r}}_i + \frac{\partial V}{\partial \mathbf{r}_i} \right|^2\, dt,
\end{equation}
where $\mathcal{Z}$ is a normalization factor that ensures that the path probability distribution is normalized to one, and $\beta=(k_B T)^{-1}$ is the inverse temperature. Similarly, for system $\tilde{S}$ with potential $\tilde{V}$, and same time range as system $S$, the path probability density reads
\begin{equation} \label{Eq:PathVtilde}
    \tilde{\mathcal{P}}(\mathbf{r}(t)|\mathbf{r}^0) = \frac{1}{\mathcal{Z}}e^{-\beta\tilde{\mathcal{I}}}, \quad \text{where} \quad  \tilde{\mathcal{I}} = \frac{1}{4\eta}\int_0^\tau \sum_{i=1}^N \left|\eta \dot{\mathbf{r}}_i + \frac{\partial \tilde{V}}{\partial \mathbf{r}_i} \right|^2\, dt.
\end{equation}
The derivations and precise meaning of Eqs.~\eqref{Eq:PathV} and \eqref{Eq:PathVtilde} is provided in \ref{Sec:ProofPath}, where it is observed that the normalization factor $\mathcal{Z}$ is identical for both expressions (when using the It\^o interpretation for the integrals inside the exponentials).

Then, the expected value of given observable $\mathcal{O}$, in each ensemble ($S$ or $\tilde{S}$) with the same initial conditions, is given by $ \langle \mathcal{O} (\tau) \rangle_S = \int \mathcal{D}\mathbf{r} \ \mathcal{O}(\tau)\, \mathcal{P}(\mathbf{r}(t)|\mathbf{r}_0)$ and $ \langle \mathcal{O} (\tau) \rangle_{\tilde{S}} = \int \mathcal{D}\mathbf{r} \ \mathcal{O}(\tau)\, \tilde{\mathcal{P}}(\mathbf{r}(t)|\mathbf{r}_0)$, respectively, where $\int \mathcal{D}\mathbf{r}$ denotes integration over all paths $\mathbf{r}$ with $\mathbf{r}(0) = \mathbf{r}^0$. These two observables may then be related in a straight-forward manner as (see \ref{Sec:ChenHoringProof} for detailed proof)
\begin{equation} \label{Eq:Bias}
\begin{split}
& \langle \mathcal{O} \rangle_S = \left\langle \mathcal{O} \frac{\mathcal{P}}{\tilde{\mathcal{P}}}\right\rangle_{\tilde{S}} = \left\langle \mathcal{O} e^{-\beta\mathcal{I}_{bias}}\right\rangle_{\tilde{S}}, \quad \text{with} \\
&\mathcal{I}_{bias}= \mathcal{I}-\mathcal{\tilde{I}}=\frac{1}{4\eta}\int_0^\tau \sum_{i=1}^N \frac{\partial V_{bias}}{\partial \mathbf{r}_i}\cdot\left( \frac{\partial V_{bias}}{\partial \mathbf{r}_i} - 2 \sqrt{2k_BT\eta}\,\, \dot{\boldsymbol\xi}_i\right) \, dt, 
 \end{split}
\end{equation}
and $V_{bias} = \tilde{V}(\mathbf{r},\tilde{\boldsymbol\lambda})-V(\mathbf{r},\boldsymbol\lambda)$. In other words, the evolution of a material/process can be reweighted to obtain the ensemble average of any observable for a different system or process. Equation \eqref{Eq:Bias} has actually been previously derived by \cite{chen2007exact} with the purpose of accelerating dynamic simulations, though the possible change in boundary conditions $\boldsymbol\lambda$ was there not discussed (albeit trivial). A pseudo-code describing in detail the steps to realize Eq.~\eqref{Eq:Bias} using an Euler-Maruyama time discretization scheme \citep{kloeden1992stochastic} is shown in Algorithm \ref{Alg:LangevinPrediction}. There, observables $\mathcal{O}$ of very different nature are considered for completeness, though their treatment is identical. These could be the expected evolution of a single particle, or an average over the whole system, and they may also be instantaneous in nature (e.g., force at a given time), or dependent on the full evolution (e.g., work done over a given time interval, mean squared displacement or particle overlap). Examples of all of the above will be provided in Sections \ref{Sec:Ex1} and \ref{Sec:Ex2}. We remark that although the theoretical description above and pseudo-code solely consider changes in the potential energy, both the temperature and the viscosity may be varied as well; in this case, the normalization factors $\mathcal{Z}$ of the path probability distributions will be distinct for the two systems, and hence their ratio, now distinct from one, will have to be included in $\mathcal{P}/\mathcal{\tilde{P}}$. Similarly, the interparticle forces may not necessarily be potential in nature, though that will be the case for the examples considered in Sections \ref{Sec:Ex1} and \ref{Sec:Ex2}.

\begin{algorithm}
\caption{Pseudo-code for predicting material/process $S$ from material/process $\tilde{S}$ with overdamped Langevin dynamics}
\label{Alg:LangevinPrediction}
\LinesNumbered
\SetKwBlock{Begin}{}{}
\SetAlgoLined
\Begin{

    \raggedright
    \everypar={\nl}
    \SetAlgoVlined

    \tcp{Overdamped Langevin dynamics for material/process $\tilde{S}$ and prediction for material/process $S$}
    $\sigma = 2 k_B T \eta $
    
    \For{all $N_R$ realizations}
    {
      Set initial condition $\mathbf{r}^0$ and $t^0=0$;
      
      \For{$n$ from $0$ to $n_T = T/ \Delta t$}
          {
            $\tilde{\boldsymbol\lambda}^{n+1} = \tilde{\boldsymbol\lambda}^n + \dot{\tilde{\boldsymbol\lambda}}(t^n) \Delta t$;
            
            $\boldsymbol\lambda^{n+1} = \boldsymbol\lambda^n + \dot{\boldsymbol\lambda}(t^n) \Delta t$;
            
            \For{$i$ from 1 to $N$}
            {
                Generate noise $\Delta \boldsymbol\xi_i^n$ with standard normal distribution;
                
                Compute $\nabla_i \tilde{V}^n = \left. \frac{\partial \tilde{V}}{\partial \mathbf{r}_i} \right|_{\tilde{\mathbf{r}}^n, \tilde{\boldsymbol\lambda}^n}$,
                $\nabla_i V^n = \left. \frac{\partial V}{\partial \mathbf{r}_i} \right|_{\tilde{\mathbf{r}}^n, \boldsymbol\lambda^n}$
                and $\nabla_i V_{bias}^n = \nabla_i \tilde{V}^n - \nabla_i V^n$;
                
                $\tilde{\mathbf{r}}_i^{n+1} = \tilde{\mathbf{r}}_i^{n} + \frac{1}{\eta} \left[ - \nabla_i \tilde{V}^n \Delta t 
                + \sqrt{\sigma} \Delta \boldsymbol\xi_i^n \right]$
                \tcp*{Update new displacement field}
            }
            
            \tcp{Compute the observables of interest, such as}
            
            $\widetilde{\mathbf{F}}_{ex}^n = \frac{\partial \tilde{V}^n}{\partial \tilde{\boldsymbol\lambda}^n}$, 
            \quad
            $\mathbf{F}_{ex}^n = \frac{\partial V^n}{\partial \boldsymbol\lambda^n}$
            \tcp*{external force}
            
            $\widetilde{W}^{n+1} = \widetilde{W}^n + \widetilde{\mathbf{F}}_{ex}^n \cdot \dot{\tilde{\boldsymbol\lambda}}^n \Delta t$, 
            \quad $W^{n+1} = W^n + \mathbf{F}_{ex}^n \cdot \dot{\boldsymbol\lambda}^n \Delta t$
            \tcp*{total work}
            
            $\widetilde{\mathcal{MSD}}^n = \frac{1}{N} \sum_{i=1}^N \left\| \tilde{\mathbf{r}}^n_i - \tilde{\mathbf{r}}_i^0 \right\|^2$
            \tcp*{mean squared displacement}
            
            $\widetilde{Q}^n(a) =  \frac{1}{N} \sum_{i=1}^N H( a - \left\| \tilde{\mathbf{r}}^n_i - \tilde{\mathbf{r}}_i^0 \right\|)$
            \tcp*{particle overlap}
            
            \BlankLine
            \tcp{Compute the bias factors}
            $\mathcal{I}_{bias}(t^{n+1}) =
            \mathcal{I}_{bias}(t^{n}) + 
            \frac{1}{4\eta} \sum_{i=1}^{N} \nabla_i V_{bias}^n \cdot\left( \nabla_i V_{bias}^n \Delta t
            - 2 \sqrt{\sigma}\,\, \Delta  \boldsymbol\xi^n_i \right)$;
            
            $\mathcal{P}_{bias}(t^{n+1}) = e^{-\beta \mathcal{I}_{bias} (t^{n+1})}$
            
            \BlankLine
            \tcp{Summing the observables of interest over all realizations, with $\tilde{\mathcal{O}}^n$ denoting, for instance, $\tilde{\mathbf{r}}^n$, $\widetilde{\mathbf{F}}_{ex}^n$, $\widetilde{W}^n$, $\widetilde{\mathcal{MSD}}^n$ or $\widetilde{Q}^n(a)$, and $\mathcal{O}^n$ denoting $\tilde{\mathbf{r}}^n$, $\mathbf{F}_{ex}^n$, $W^n$, $\widetilde{\mathcal{MSD}}^n$ or $\widetilde{Q}^n(a)$}
            
            $\left< \tilde{\mathcal{O}}^n \right>_{\tilde{S}}
            = \left< \tilde{\mathcal{O}}^n \right>_{\tilde{S}} +  \tilde{\mathcal{O}}^n $,
            
            $\left< \mathcal{O}^n \mathcal{P}_{bias}(t^n)\right>_{\tilde{S}}
            = \left< \mathcal{O}^n \mathcal{P}_{bias}(t^n)\right>_{\tilde{S}} +  \mathcal{O}^n \mathcal{P}_{bias}(t^n)$,

            $\mathcal{N}(t^n)
            = \mathcal{N}(t^n) +  \mathcal{P}_{bias}(t^n)$,
            
            \BlankLine
            \tcp{Update time}
            $t^{n+1} = t^n + \Delta t$;
          }
    }
    \tcp{Compute the ensemble average for all times, with $\tilde{\mathcal{O}}$ denoting, for instance, $\tilde{\mathbf{r}}$, $\widetilde{\mathbf{F}}_{ex}$, $\widetilde{W}$, $\widetilde{\mathcal{MSD}}$ or $\widetilde{Q}(a)$, and $\mathcal{O}$ denoting $\tilde{\mathbf{r}}$, $\mathbf{F}_{ex}$, $W$, $\widetilde{\mathcal{MSD}}$ or $\widetilde{Q}(a)$}
    
    $\left< \tilde{\mathcal{O}} \right>_{\tilde{S}} = \frac{1}{N_R} \left< \tilde{\mathcal{O}} \right>_{\tilde{S}}$
    \tcp*{for process $\tilde{S}$}
    
    $\mathcal{N} = \frac{1}{N_R}\mathcal{N}$
    
    $\left< \mathcal{O} \mathcal{P}_{bias} \right>_{\tilde{S}} = \frac{1}{N_R \mathcal{N}} \left< \mathcal{O} \mathcal{P}_{bias} \right>_{\tilde{S}}$
    \tcp*{prediction for process $S$}
    
}
\end{algorithm}
\begin{algorithm}
\LinesNumbered
\SetKwBlock{Begin}{}{}
\Begin{
    \raggedright
    \everypar={\nl}
    \SetAlgoVlined
    
    \tcp{Overdamped Langevin dynamic simulations for material/process $S$ (for comparison, optional)}
    \For{all $N_R$ realizations}
    {
      Set initial conditions $\mathbf{r}^0$ and $t^0=0$;
      
      \For{$n$ from $0$ to $n_T = T/ \Delta t$}
          {
            $\boldsymbol\lambda^{n+1} = \boldsymbol\lambda^n + \dot{\boldsymbol\lambda}(t^n) \Delta t$;
            
            \For{$i$ from 1 to $N$}
            {
                Generate noise $\Delta \boldsymbol\xi_i^n$ according to a standard normal distribution;
                
                Compute $\nabla_i V^n = \left. \frac{\partial V}{\partial \mathbf{r}_i} \right|_{\mathbf{r}^n, \boldsymbol\lambda^n}$;
                
                $\mathbf{r}_i^{n+1} = \mathbf{r}_i^{n} + \frac{1}{\eta} \left[ - \frac{\partial V}{\partial \mathbf{r}_i} \Delta t 
                + \sqrt{\sigma} \Delta \boldsymbol\xi_i^n \right]$
                \tcp*{Update new displacement field}
            }
            
            \tcp{Compute the observables of interest}
            
            $\mathbf{F}_{ex}^n = \frac{\partial V^n}{\partial \boldsymbol\lambda^n}$
            \tcp*{external force}
            
            $W^{n+1} = W^n + \mathbf{F}_{ex}^n \cdot \dot{\boldsymbol\lambda}^n \Delta t$
            \tcp*{total work}
            
            $\mathcal{MSD}^n = \frac{1}{N} \sum_{i=1}^N \left\| \mathbf{r}^n_i - \mathbf{r}_i^0 \right\|^2$
            \tcp*{mean squared displacement}
            
            $Q^n(a) =  \frac{1}{N} \sum_{i=1}^N H( a - \left\| \mathbf{r}^n_i - \mathbf{r}_i^0 \right\|)$
            \tcp*{particle overlap}
            
            \BlankLine
            \tcp{Summing the observables of interest over all realizations, with $\mathcal{O}^n$ denoting, for instance, $\mathbf{r}^n$, $\mathbf{F}_{ex}^n$, $W^n$, $\mathcal{MSD}^n$ or $Q^n(a)$}
            
            $\left< \mathcal{O}^n \right>_S
            = \left< \mathcal{O}^n \right>_S +  \mathcal{O}^n $,
            
            \BlankLine
            \tcp{Update time}
            $t^{n+1} = t^n + \Delta t$;
          }
    }
    \tcp{Compute the ensemble average for all times, with $\mathcal{O}$ denoting, for instance, $\mathbf{r}$, $\mathbf{F}_{ex}$, $W$, $\mathcal{MSD}$ or $Q(a)$}
    
    $\left< \mathcal{O} \right>_S = \frac{1}{N_R} \left< \mathcal{O} \right>_S$
    \tcp*{for process $S$}
    
}
\end{algorithm}

An attentive reader may discover that the calculation of $\mathcal{I}_{bias}$ effectively requires the calculation of the forces for the potential $V$ associated to the system one is ultimately interested in, hence wondering about the computational efficiency of such an approach. While this is certainly the case, the benefits could be manifold. First, an appropriate choice for $\tilde{V}$ could greatly accelerate the dynamics for a system that is trapped in an energetic or entropic barrier, as noted in the introduction. This has been exploited by several authors \citep{chen2007exact,nummela2007exact}. Second, since the forces are computed from the trajectories of system $\tilde{S}$, these calculations could be done \emph{a posteriori} and be trivially parallelized in time. This second point is quite interesting on its own, as perfect time parallelization has always been thought of being impossible due to the sequential nature of time. Yet, temporal parallelization of molecular dynamic simulations is very active area of research aimed at enabling long-time simulations \citep{perez2016long}. Finally, an entire family of interparticle potentials characterized by a proportionality constant, may be predicted with minimal added computational cost compared to that of a single potential. An example of the latter will be provided in Section \ref{Sec:Ex2}, where this point will explained in further detail.

\begin{figure}[H]
    \includegraphics[width=0.9\textwidth]{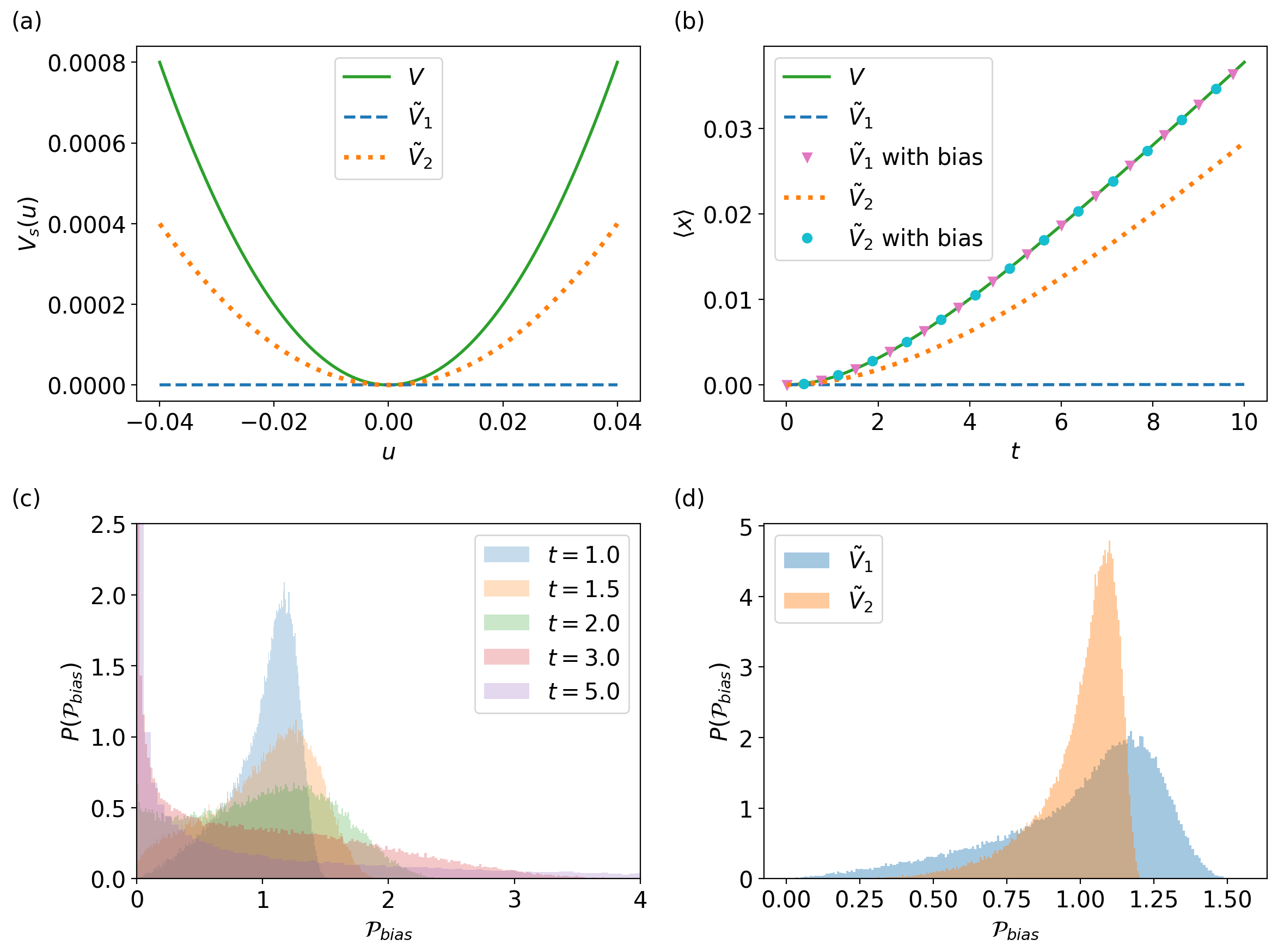}
    \caption{(a) Interparticle potentials for the simulated ($\tilde{V}_1$ and $\tilde{V}_2$) and targeted ($V$) systems. (b) Ensemble average of the displacement for the simulated ($\tilde{V}_1$ and $\tilde{V}_2$) and targeted systems ($V$), as well as the predictions from the first to the latter ($\tilde{V}_1$ with bias and $\tilde{V}_2$ with bias). (c) Time evolution of the probability density for $\mathcal{P}_{bias}$ from $\tilde{V}_1$ to $V$. (d) Comparison between the probability distribution of $\mathcal{P}_{bias}$ at $t=1$ from $\tilde{V}_1$ to $V$ and from $\tilde{V}_2$ to $V$. Parameters are chosen as $N=1$, $\eta = 5$, $\beta = 10^4$, $\tilde{V}_{s1}(u)=0$, $\tilde{V}_{s2}(u)=\frac{1}{4} u^2$, $V_s(u) = \frac{1}{2} u^2$, $N_R=10^5$.}
    \label{fig:sampling}
\end{figure}

While the appeal of the path reweighting strategy is clear from the above discussion, and its applications are, in appearance, limitless, it is worth noting the sampling issues that arise as the two interparticle potentials and/or boundary conditions diverge from each other. To illustrate this point, consider a single degree of freedom system ($N=1$) consisting of a mass connected to two linear springs, with spring constant $k=1$. The end of one of the springs is held fixed, while the end of the second is being pulled at a constant velocity $v_p$, so that the resulting equation of motion is  
\begin{equation}
\eta \dot{x} = - 2k x +k v_p t+ \sqrt{2 k_B T \eta}\, \dot{\xi} .
\end{equation}
In this example, the aim is to predict the behavior of this system from a particle undergoing Brownian motion, i.e., $\tilde{V}_{1}=0$ and an intermediate harmonic potential $\tilde{V}_{2}$ with spring constant $\tilde{k}=1/2$. Figures \ref{fig:sampling}(a)-(b) shows the three interparticle potentials considered as well as the results of the Langevin simulations for each system, together with the predictions based on Eq.~\eqref{Eq:Bias}. While the predictions and the validations are indistinguishable, the normalized histogram of the path probability ratio $\mathcal{P}_{bias}\coloneqq \mathcal{P}/\tilde{\mathcal{P}}$ in Figures \ref{fig:sampling}(c)-(d), clearly show that these become more heavy-tailed with time for a given potential $\tilde{V}$, or for a fixed time, as the difference between the potentials $V$ and $\tilde{V}$ become larger. On a practical level, heavier tails imply larger errors for the empirical averages of Eq.~\eqref{Eq:Bias} for a fixed number of realizations of system $\tilde{S}$. Quantifying \emph{a priori} these estimates is therefore crucial to assess the accuracy of the predictions without the need of validations. This is precisely the goal of the next section.

\section{Uncertainty Quantification}
\label{Sec:Uncertainty}

From Eq.~\eqref{Eq:Bias} it is immediate that $\left\langle e^{-\beta\mathcal{I}_{bias}}\right\rangle_{\tilde{S}}=1$. Although this identity is exactly satisfied for the exact probability density, an empirical average, i.e., $\mathcal{N}\coloneqq\sum_{r=1}^{N_R} \frac{1}{N_R} e^{-\beta\mathcal{I}_{bias,r}}$, where $N_R$ is the number of realizations, may deviate from 1 if insufficient samples are used. In general, more realizations will be needed the further apart systems/processes $\tilde{S}$ and $S$ are, as noted in the previous section. Such a distance can be measured by means of the Kullback-Leibler divergence, or relative entropy, between two probability distributions $\mathcal{\tilde{P}}$, and $\mathcal{P}$, which, in this case, is equal to
\begin{equation}
    D_{KL}\left(\tilde{\mathcal{P}} \| \mathcal{P} \right) = \left \langle \log \frac{\tilde{\mathcal{P}}}{\mathcal{P}}\right \rangle_{\tilde{S}} = \beta \langle{I}_{bias} \rangle_{\tilde{S}}.
\end{equation}

In practice, we will quantify the uncertainty of our predictions (for general observables $\mathcal{O}$) using the standard error of the mean of $\mathcal{P}_{bias}\coloneqq \mathcal{P}/\tilde{\mathcal{P}}=e^{-\beta \mathcal{I}_{bias}}$

\begin{equation}
\sigma_{\mathcal{N}}   = \frac{\sigma_{\mathcal{P}_{bias}}}{\sqrt{N_R}}, 
\end{equation}
where $\sigma^2_{\mathcal{P}_{bias}}$ is the variance of $\mathcal{P}_{bias}$, and $\mathcal{N} = \sum_{r=1}^{N_R} \frac{1}{N_R} \mathcal{P}_{bias,r} $. We use this standard error $\sigma_{\mathcal{N}}$ as an estimate for $\left| \mathcal{N} - 1 \right|$.

\subsection{Linear uncertainty propagation estimate} \label{Sec:LinearUQ}
The simplest approach to compute $\sigma^2_{\mathcal{P}_{bias}}$ is to consider the time-discretized evaluation of $\mathcal{P}_{bias}$ as a nonlinear function of $\Delta \xi^n = \xi^{n+1}_i-\xi^n_i$, where the indices $i$ and $n$ here refer to the degrees of freedom $i=1, ..., N d$ ($d$ is the dimension of the problem), and discrete time $t^n$, respectively, and to use the classical formulas for propagation of uncertainty \citep[Chapter 3]{taylor1997introduction}, i.e.,
\begin{equation}
    \sigma^2_{\mathcal{P}_{bias}} = \sum_{i=1}^{Nd} \sum_{n=0}^{n_T-1} \left(\frac{\partial \mathcal{P}_{bias}}{\partial \Delta \xi^n_i} \Big |_{\Delta \boldsymbol \xi =\mathbf{0}}\right)^2 \sigma^2_{\Delta \xi^n_i} = \sum_{i=1}^{Nd} \sum_{n=0}^{n_T-1} \left(\frac{\partial \mathcal{P}_{bias}}{\partial \Delta \xi^n_i} \Big |_{\Delta \boldsymbol \xi =\mathbf{0}}\right)^2 \Delta t. 
\end{equation}
Here, we have made use of the fact that $\Delta \xi^n_i$ are uncorrelated, and $\sigma^2_{\Delta \xi^n_i}=\Delta t$. The notation $\Big |_{\Delta \boldsymbol \xi = \mathbf{0}}$ is used to denote that the partial derivatives are evaluated at the mean value of $\Delta \boldsymbol \xi$, which is zero.

As discussed in detail in \ref{Sec:ProofPath} and \ref{Sec:ChenHoringProof},  $\mathcal{P}_{bias}$ may be expressed using the It{\^o} interpretation as
\begin{equation}
\begin{split}
    &\mathcal{P}_{bias} = \exp \left[-\frac{\beta}{4\eta} \sum_{j=1}^{Nd} \sum_{m=0}^{n_T-1}  \frac{\partial V_{bias}(\mathbf{x}^m)}{\partial x_j^m} \left[ \frac{\partial V_{bias}(\mathbf{x}^m)}{\partial x_j^m}\Delta t - 2\sqrt{\sigma} \Delta\xi^m_j  \right]\right], \quad \text{with} \\
    & x^m_j = x^0_j + \sum_{p=0}^{m-1} \Delta x^p_j, \quad \Delta x^p_j = x^{p+1}_j-x^p_j \quad \text{and} \quad \Delta x^p_j = -\frac{\partial \tilde{V}(\mathbf{x}^p)}{\partial x_p^j} \frac{\Delta t}{\eta}+ \frac{\sqrt{\sigma}}{\eta} \Delta \xi^p_j,
\end{split}
\end{equation}
where $\sigma=2 k_B T \eta$.

Noting that $\mathcal{P}_{bias}$ depends on $\Delta \xi^n_i$ directly and through $x^m_j$ for $m>n$, one obtains for the simplest case of $\tilde{V}=0$ that
\begin{equation}
    \frac{\partial \mathcal{P}_{bias}}{\partial \Delta \xi^n_i} = \mathcal{P}_{bias} \left[ - \frac{\beta}{4 \eta} \left( \sum_{j=1}^{Nd} \sum_{m=n+1}^{N_T-1}  2 \frac{\partial V_{bias}}{\partial x^m_j} \frac{\partial^2 V_{bias}}{\partial x_j^m \partial x_i^m} \frac{\sqrt{\sigma}}{\eta} \Delta t - \frac{\partial V_{bias}}{\partial x_i^n} 2 \sqrt{\sigma} \right) \right]. 
\end{equation}
For such a case, the variance of $\mathcal{P}_{bias}$ may then be easily computed as
\begin{equation} \label{Eq:UP}
    \sigma^2_{\mathcal{P}_{bias}} = \frac{\beta}{2\eta}
    \mathcal{P}_{bias}^2 \sum_{i=1}^{Nd}
    \sum_{n=0}^{n_T-1}\left[\frac{\partial V_{bias}}{\partial x_i^n}
    -\sum_{j=1}^N\sum_{m=n+1}^{n_T-1}\frac{\partial V_{bias}}{\partial x_j^m}
    \frac{\partial^2 V_{bias}}{\partial x_j^m \partial x_i^m } \frac{\Delta t}{\eta}  
    \right]^2  \Delta t \Bigg |_{\Delta \boldsymbol \xi =\mathbf{0}}.
\end{equation}

As will be seen later on in Fig.~\ref{fig:Var_Pbias_prop} by means of an example, this uncertainty propagation method, albeit simple for $\tilde{V}=0$, does not capture well the variance of $\mathcal{P}_{bias}$ even for very simple physical systems. This is likely due to the importance of nonlinear effects induced by the exponential of $\mathcal{P}_{bias}$. Furthermore, its generalization to simulated systems with $\tilde{V}\neq0$ is rather intricate, due to the many nested sums that this would require.

\subsection{Nonlinear uncertainty quantification estimate}
\label{Sec:NonlinearUQ}

To resolve the above two issues, we resort instead to perform a linear approximation to the bias potential gradient in the exponent of $\mathcal{P}_{bias}$ to then estimate $\sigma^2_{\mathcal{P}_{bias}}$ directly as $\langle \mathcal{P}_{bias}^2\rangle_{\tilde{S}}-\langle \mathcal{P}_{bias} \rangle_{\tilde{S}}^2$. Specifically, we expand the bias potential gradient linearly from a reference trajectory $\mathbf{x}_r(t)$ as
\begin{equation}
\label{Eq:dVExpansion}
    \nabla V_{bias}(\mathbf{x},t) 
    = \left. \nabla V_{bias} \right|_{\mathbf{x}_r(t)} 
    + \left. \nabla \nabla V_{bias}\right|_{\mathbf{x}_r(t)} \delta \mathbf{x} + O(\left\|\delta \mathbf{x}\right\|^2) ,
\end{equation}
where $\delta \mathbf{x} = \mathbf{x} - \mathbf{x}_r$. Similarly, the noise term in the Langevin equations associated to system $\tilde{S}$ may be approximated as
\begin{equation} \label{Eq:noise_approx}
\begin{split}
    \sqrt{\sigma} \Delta \boldsymbol\xi^n 
    & = \eta \left( \mathbf{x}^{n+1} - \mathbf{x}^{n}\right)
    + \nabla \tilde{V}(\mathbf{x}^{n}, t^n) \Delta t \\
    & = \eta \left( \Delta \mathbf{x}^{n}_r +  \delta \mathbf{x}^{n+1} 
    - \delta \mathbf{x}^{n} \right)
    + \left. \nabla \tilde{V} \right|_{\mathbf{x}_r^n} \Delta t
    +  \left. \nabla \nabla \tilde{V} \right|_{\mathbf{x}_r^n}   \delta \mathbf{x}^{n} 
    \Delta t
    + O(\left\| \delta \mathbf{x} \right\|^2 \Delta t),
\end{split}
\end{equation}
where we have used an It{\^o} representation for the discretized equations and defined $\Delta \mathbf{x}^n_r = \mathbf{x}^{n+1}_r-\mathbf{x}^n_r$. Using the above two approximations, and the change of variables $\delta \mathbf{y}^n = \left( \eta / \sqrt{\sigma \Delta t} \right) \delta \mathbf{x}^n$, introduced for convenience,  the exponents in $\mathcal{P}_{bias}$ and $\tilde{\mathcal{P}}$ have a quadratic form in $\delta \mathbf{y}^n$. As a result, both $\langle \mathcal{P}_{bias}\rangle_{\tilde{S}}$ and $\langle \mathcal{P}_{bias}^2 \rangle_{\tilde{S}}$ may be computed analytically. These derivations are quite involved and are therefore relayed to \ref{Seq:UQ}. The resulting expression for $\langle \mathcal{P}_{bias}\rangle_{\tilde{S}}$ is
\begin{equation}
\langle \mathcal{P}_{bias}\rangle_{\tilde{S}}  =   \frac{1}{\sqrt{\det\mathbf{A}}} 
    e^{ \frac{1}{2} \mathbf{b}^\text{T} \mathbf{A}^{-1} \mathbf{b} + c  }.
\end{equation}
Here, $c$ is a constant defined as,
\begin{equation} \label{Eq:c}
     c = -\frac{1}{2\sigma} \sum_{n=0}^{n_T - 1}
    \left\| \eta \frac{\Delta \mathbf{x}^{n}_r}{ \Delta t} 
    + \left. \nabla V \right|_{\mathbf{x}_r^n}
    \right\|^2 \Delta t,
\end{equation}
the vector $\mathbf{b}$ consists of $n_T$ small $Nd$-dimensional vectors,
\begin{equation} \label{Eq:b}
\begin{split}
    \mathbf{b}  =  
    \begin{pmatrix}
    \mathbf{b}^{1} \\
    \mathbf{b}^{2} \\
    \vdots  \\
    \mathbf{b}^{n_T}
    \end{pmatrix},
\end{split}
\end{equation}
with
\begin{equation}
\begin{split}
    \mathbf{b}^n = 
     \left\{ \begin{aligned}
    & - \sqrt{\frac{\Delta t}{\sigma}}  \left[ 
    \frac{\Delta t}{\eta} \left. \nabla \nabla V \right|_{\mathbf{x}_r^n} \left( \eta \frac{\Delta \mathbf{x}^{n}_r}{ \Delta t} 
    + \left. \nabla V \right|_{\mathbf{x}_r^n} \right)
    - \Delta \left( \eta \frac{\Delta \mathbf{x}^{n-1}_r}{ \Delta t} 
    + \left. \nabla V \right|_{\mathbf{x}_r^{n-1}} \right) \right] ,
    \quad && n < n_T \\
    & - \sqrt{\frac{\Delta t}{\sigma}}
    \left( \eta \frac{\Delta \mathbf{x}^{n_T-1}_r}{ \Delta t} 
    + \left. \nabla V \right|_{\mathbf{x}_r^{n_T-1}} 
    \right) , 
    \quad && n = n_T
    \end{aligned} \right.
\end{split}
\end{equation}
and the matrix $\mathbf{A}$ may be written as, 
\begin{equation} \label{Eq:A}
\begin{split}
    \mathbf{A} &  =  
    \begin{pmatrix}
    \mathbf{A}^{11} & \mathbf{A}^{12} & \cdots & \mathbf{A}^{1 n_T} \\
    \mathbf{A}^{21} & \mathbf{A}^{22} & \cdots & \mathbf{A}^{2 n_T} \\
    \vdots & \vdots & \ddots & \vdots \\
    \mathbf{A}^{n_T 1} & \mathbf{A}^{n_T 2} & \cdots & \mathbf{A}^{n_T n_T}
    \end{pmatrix}
\end{split}
\end{equation}
with each $Nd \times Nd$ matrix block defined as,
\begin{equation}
\label{Eq:Apq}
\begin{split}
    \mathbf{A}^{pq} & =  
    \left\{ \begin{aligned}
    & \mathbf{I} 
    + \mathbf{\Gamma}^n \mathbf{\Gamma}^n, 
    \quad \quad 
    && (p,q) = (n, n) \text{ with } 
    n = 1, \cdots, n_T-1\\
    & - \mathbf{\Gamma}^n,
    && (p,q) = (n+1, n) \text{ or } (n, n+1) \text{ with } 
    n = 1, \cdots, n_T-1 \\
    & \mathbf{I},
    && (p,q) = (n_T, n_T) \\
    & \mathbf{0},
    && \text{otherwise} \\
    \end{aligned} \right. 
\end{split}
\end{equation}
with
\begin{equation}
\label{Eq:Gamma}
    \mathbf{\Gamma}^n
    = \mathbf{I} 
    - \frac{1}{\eta} 
    \left. \nabla \nabla V \right|_{\mathbf{x}_r^n} \Delta t.
\end{equation}
The matrix $\mathbf{A}$ is thus symmetric. Remarkably, as detailed in \ref{Seq:UQ}, it may be shown that the equations $\det (\mathbf{A}) = 1$ and $\mathbf{b}^\text{T} \mathbf{A}^{-1}\mathbf{b} = - 2c$ are identically satisfied in this discretized setting leading to $\langle \mathcal{P}_{bias}\rangle_{\tilde{S}}=1$. We remark that in these derivations, the ensemble average for $\langle \mathcal{P}_{bias}\rangle_{\tilde{S}}$ is analytically computed, i.e., it does not result from an empirical average.

As for $\langle \mathcal{P}_{bias}^2\rangle_{\tilde{S}}$, this reads
\begin{equation}
    \langle \mathcal{P}_{bias}^2\rangle_{\tilde{S}}  =  \frac{1}{\sqrt{\det (\mathbf{A}_{sq})}} 
    e^{ \frac{1}{2} \mathbf{b}^\text{T}_{sq} \mathbf{ A}_{sq}^{-1} \mathbf{b}_{sq} + c_{sq} },
\end{equation}
where the coefficients $\mathbf{A}_{sq}$, $\mathbf{b}_{sq}$ and $c_{sq}$ can be expressed as,
\begin{align}\label{Eq:Asq}
    &\mathbf{A}_{sq} = 2\mathbf{A} - \tilde{\mathbf{A}},  \\ \label{Eq:bsq}
    &\mathbf{b}_{sq} = 2\mathbf{b} -\tilde{\mathbf{b}},\\  \label{Eq:csq}
    &c_{sq} = 2c - \tilde{c}.
\end{align}
Here, $\tilde{\mathbf{A}}$, $\tilde{\mathbf{b}}$ and $\tilde{c}$ are the defined as the analogues of $\mathbf{A}$, $\mathbf{b}$, and $c$, respectively, with the potential $V$ replaced by $\tilde{V}$.

Therefore, the resulting expression for the variance of $\mathcal{P}_{bias}$ is,
\begin{equation}
\label{Eq:VarPb_}
\begin{split}
    \sigma^2_{\mathcal{P}_{bias}}
    & = \left< \mathcal{P}_{bias}^2 \right>_{\tilde{S}} 
    -  \left< \mathcal{P}_{bias} \right>_{\tilde{S}}^2 \\
    &  = \frac{1}{\sqrt{\det(\mathbf{A}_{sq})}} 
    e^{ \frac{1}{2} \mathbf{b}^\text{T}_{sq} \mathbf{A}_{sq}^{-1} \mathbf{b}_{sq} + c_{sq} }
    - 1.
\end{split}
\end{equation}
The details of its implementation are detailed in  Algorithm~\ref{Alg:UQ}.

\begin{figure}[H]
    \centering
    \includegraphics[width=0.45\textwidth]{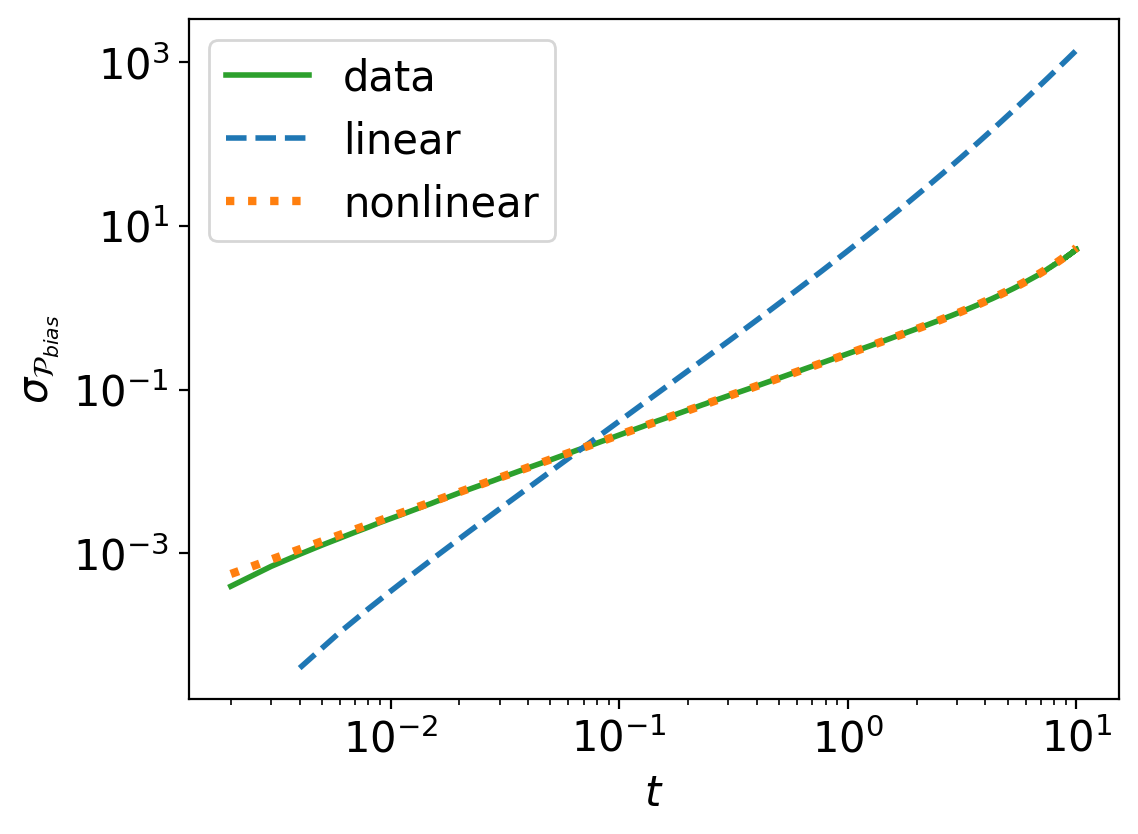}
    \caption{Prediction of the standard deviation of $\mathcal{P}_{bias}$ with the linear (blue dashed line) and nonlinear (orange dotted line) uncertainty estimates, together with $\sigma_{\mathcal{P}_{bias}}$, as directly computed from simulation data (green solid line). The simulation parameters are $N=1$, $N_R=10^5$, $k_B T = 10^{-4}$, $\eta = 5$, $v_p = 0.01$, $\Delta t = 10^{-3}$, from Brownian motion to quadratic potential with $k=1$. For the nonlinear uncertainty quantification estimates, the time discretization used is $n_T=100$ and the reference trajectory is chosen as $\mathbf{x}_r=\mathbf{0}$.}
    \label{fig:Var_Pbias_prop}
\end{figure}

To illustrate the improved performance of this estimate compared to the classical uncertainty propagation approach, consider the single one degree of freedom system discussed at the end of Section \ref{Sec:Reweighting} with $\tilde{V}=0$. 
Figure~\ref{fig:Var_Pbias_prop} shows the value of $\sigma_{\mathcal{P}_{bias}}$ as a function of time,  directly obtained from numerical simulations, together with the estimates given by Eq.~\eqref{Eq:UP} and Eq.~\eqref{Eq:VarPb_}. As it may there be observed, the difference in accuracy is striking, even for this rather simple example. Consequently, the later examples will make use of the nonlinear propagation of uncertainty strategy described in this section. 

We remark that the accuracy of the uncertainty quantification (UQ) estimates here derived will strongly depend on the form of the bias potential that the trajectories explore. Equation \eqref{Eq:VarPb_} is nominally exact for a quadratic bias (such as the one used in Fig.~\ref{fig:Var_Pbias_prop}), though it is expected to only give an estimate for general potentials. In the example of Section \ref{Sec:Ex1}, quartic potentials will be considered, while Section \ref{Sec:Ex2} will examine systems with Hertzian and Lennard-Jones potentials. These last two potentials strongly deviate from quadratic, with the Hertzian not even displaying a unique minimum, and will thus allow us to explore the degree of accuracy of these estimates for a wide range of systems.

\begin{algorithm}
\LinesNumbered
\setcounter{AlgoLine}{0}
\SetKwBlock{Begin}{}{}
\Begin{

    \raggedright
    \everypar={\nl}
    \SetAlgoVlined

    
    Choose a reference trajectory $\mathbf{x}_r(t)$ 
    
    \For{all selected $t < T$ at which to estimate the uncertainty}
    {
      Set timestep as $\Delta t = t / n_T$;
      
      \For{all $n$ from $0$ to $n_T$}
      {
        $\tau^n = n \Delta t$;
        
        Compute $\mathbf{x}_r^n$, $\dot{\mathbf{x}}_r^n$, $\left. \nabla V \right|_{\mathbf{x}_r^n}$, $\left. \nabla \tilde{V} \right|_{\mathbf{x}_r^n}$, $\left. \nabla \nabla V \right|_{\mathbf{x}_r^n}$, $\left. \nabla \nabla \tilde{V} \right|_{\mathbf{x}_r^n}$ at time $\tau^n$;
      }
      Compute $c$ and $c_{sq}$ as in Eqs.~\eqref{Eq:c} and \eqref{Eq:csq};
      
      Initialize $\mathbf{b}$ and $\mathbf{b}_{sq}$ as zero vectors with size $N d n_T$, $\mathbf{A}$ and $\mathbf{A}_{sq}$ as zero matrices with size $N d n_T \times N d n_T$
      
      \For{all $n$ from $1$ to $n_T-1$}
      {
        Compute vector components $\mathbf{b}^n$ and $\mathbf{b}^n_{sq}$ as in Eqs.~\eqref{Eq:b} and \eqref{Eq:bsq};
        
        Compute $\mathbf{\Gamma}^n$ and $ \mathbf{\tilde \Gamma}^n $ as in Eq.~\eqref{Eq:Gamma};
        
        Compute matrix blocks $\mathbf{A}^{pq}$ and $\mathbf{A}_{sq}^{pq}$ for $(p, q) = (n,n), (n+1, n)$ and $(n, n+1)$ as in Eqs.~\eqref{Eq:Apq} and \eqref{Eq:Asq};
      }
      Compute $\mathbf{b}^{n_T}$, $\mathbf{b}^{n_T}_{sq}$ as in Eqs.~\eqref{Eq:b}, \eqref{Eq:bsq} and $\mathbf{A}^{n_T n_T}$, $\mathbf{A}^{n_T n_T}_{sq}$ as in Eqs.~\eqref{Eq:Apq}, \eqref{Eq:Asq};
      
      Solve $\mathbf{u}_{sq}$ from $\mathbf{A_{sq} u_{sq}} = \mathbf{b_{sq}}$;
      
      Compute $\mathbf{b}^\text{T}_{sq}\mathbf{A}_{sq}^{-1} \mathbf{b}_{sq} = \mathbf{b}_{sq} \cdot \mathbf{u}_{sq}$;
      
      Compute $\det \left( \mathbf{A}_{sq} \right)$;
      
      $\sigma^2_{\mathcal{P}_{bias}} (t)
      = \frac{1}{\sqrt{\det\left(\mathbf{A}_{sq}\right)}} 
      \exp \left( \frac{1}{2} \mathbf{b}^\text{T}_{sq} \mathbf{A}_{sq}^{-1} \mathbf{b}_{sq} + c_{sq} \right)
    - 1.$
    }

}
\caption{Pseudo-code for nonlinear uncertainty quantification}
\label{Alg:UQ}
\end{algorithm}

\subsection{Connection between the two strategies} \label{Sec:Nonlinar2Linear}

While the two UQ approaches previously discussed in Sections \ref{Sec:LinearUQ} and \ref{Sec:NonlinearUQ} are quite distinct in their starting point and final conclusions, Eq.~\eqref{Eq:UP} may actually be recovered following the same strategy as for the nonlinear estimate under various simplifying assumptions. In particular, assuming $\mathbf{x}_r(t)=\mathbf{0}$ (hence, $\Delta \mathbf{x}_r^n=\mathbf{0}$, and $\delta\mathbf{x}^n=\mathbf{x}^n$), $\tilde{V}=0$ (hence, $V=-V_{bias}$), and only keeping linear terms in the integrand of $\mathcal{I}_{bias}$, such integrand can be approximated as
\begin{equation} \label{Eq:Ibas}
\begin{split}
    & \nabla V_{bias}(\mathbf{x}^n,t^n) \cdot
    \left(\nabla V_{bias}(\mathbf{x}^n,t^n) \Delta t 
    - 2\sqrt{\sigma} \Delta \boldsymbol\xi^n \right) \\
    & \simeq \left \| \left. \nabla V_{bias} \right|_{\mathbf{x}_r^n=\mathbf{0}} 
    \right\| ^2 \Delta t 
    - 2 \left. \nabla V_{bias} \right|_{\mathbf{x}_r^n=\mathbf{0}} \cdot
    \left[ \eta \left( \mathbf{x}^{n+1} 
    - \mathbf{x}^{n} \right)
    - \left. \nabla \nabla V_{bias} \right|_{\mathbf{x}_r^n=\mathbf{0}}
     \mathbf{x}^{n} \Delta t
    \right].
\end{split}
\end{equation}
Next, following the discrete Langevin equation for system $\tilde{S}$, we replace in the above expression $\eta(\mathbf{x}^{n+1}-\mathbf{x}^n) = \sqrt{\sigma} \Delta \boldsymbol \xi^n$, and $\mathbf{x}^n=\sum_{m=0}^{n-1} \Delta \mathbf{x}^m=\sum_{m=0}^{n-1} \frac{\sqrt{\sigma}}{\eta} \Delta \boldsymbol \xi^m$. Furthermore, we note that expanding around $\mathbf{x}_r=\mathbf{0}$ is identical to expanding around $\Delta \boldsymbol \xi = \mathbf{0}$. Then, $\mathcal{P}_{bias}$ can be written as a function of $\Delta \boldsymbol \xi$ and $\left< \mathcal{P}_{bias} \right>^2_{\tilde{S}}$ and $\left< \mathcal{P}_{bias}^2 \right>_{\tilde{S}} $ may be computed analytically, giving (for details see \ref{Sec:Connection}) 
\begin{equation} \label{Eq:<Pbias>2}
\begin{split}
    \left< \mathcal{P}_{bias} \right>^2_{\tilde{S}} 
    & \simeq \left. \exp \left[ - \frac{1}{\sigma}  \sum_{n=0}^{n_T-1} \left \|  \nabla V_{bias}(\mathbf{x}^n, t^n) 
    \right\| ^2 \Delta t \right]
    \right|_{\Delta \boldsymbol\xi = \mathbf{0}} \\
    & \quad 
    \exp \left[ 
     \frac{\Delta t}{ \sigma} \sum_{n=0}^{n_T-1} 
    \left. \left( \nabla V_{bias}(\mathbf{x}^n, t^n)
    - \frac{1}{\eta} \sum_{m=n+1}^{n_T-1} 
    \nabla \nabla V_{bias}(\mathbf{x}^m, t^m) \nabla V_{bias}(\mathbf{x}^m, t^m) \Delta t \right)^2  \right|_{\Delta \boldsymbol\xi = \mathbf{0}}\right]
\end{split}
\end{equation}

\begin{equation} \label{Eq:<Pbias2>}
\begin{split}
    \left< \mathcal{P}_{bias}^2 \right>_{\tilde{S}}
    &  \simeq \left. \exp \left[ - \frac{1}{\sigma}  \sum_{n=0}^{n_T-1} \left \|  \nabla V_{bias}(\mathbf{x}^n, t^n) 
    \right\| ^2 \Delta t \right]
    \right|_{\Delta \boldsymbol\xi = \mathbf{0}} \\
    & \quad 
    \exp \left[ 
     \frac{2\Delta t}{\sigma} \sum_{n=0}^{n_T-1} 
    \left. \left( \nabla V_{bias}(\mathbf{x}^n, t^n)
    - \frac{1}{\eta} \sum_{m=n+1}^{n_T-1} 
    \nabla \nabla V_{bias}(\mathbf{x}^m, t^m) \nabla V_{bias}(\mathbf{x}^m, t^m) \Delta t \right)^2
    \right|_{\Delta \boldsymbol\xi = \mathbf{0}}  \right]
\end{split}
\end{equation}

Finally, assuming that the second exponential terms in both $\left< \mathcal{P}_{bias} \right>^2_{\tilde{S}}$ and $\left< \mathcal{P}_{bias}^2 \right>_{\tilde{S}}$ are small, we approximate $\exp \left(x\right)\simeq 1 +x$, obtaining the following expression for the variance of $\mathcal{P}_{bias}$
\begin{equation}
\begin{split}
    \sigma_{\mathcal{P}_{bias}}^2
    & = \left< \mathcal{P}_{bias}^2 \right>_{\tilde{S}} 
    - \left< \mathcal{P}_{bias} \right>_{\tilde{S}}^2  \\
    & = \left. \exp \left[ - \frac{1}{\sigma}  \sum_{n=0}^{n_T-1} \left \|  \nabla V_{bias}(\mathbf{x}^n, t^n) 
    \right\| ^2 \Delta t \right]
    \right|_{\Delta \boldsymbol\xi = \mathbf{0}} \\
    & \quad 
    \left. \left[ 
     \frac{\Delta t}{\sigma} \sum_{n=0}^{n_T-1} 
    \left( \nabla V_{bias}(\mathbf{x}^n, t^n)
    - \frac{1}{\eta} \sum_{m=n+1}^{n_T-1} 
    \nabla V_{bias}(\mathbf{x}^m, t^m) \cdot
    \nabla \nabla V_{bias}(\mathbf{x}^m, t^m) \Delta t \right)^2  \right]\right|_{\Delta \boldsymbol\xi = \mathbf{0}}.
\end{split}
\end{equation}
This exactly corresponds to the variance given by the linear variance propagation method in Eq.~\eqref{Eq:UP}.

\noindent

\section{Example 1: Pulling experiments on one-dimensional mass-spring chains} \label{Sec:Ex1}

\subsection{Model description and overview of the cases to be examined}

\begin{figure}[H]
    \centering
    \includegraphics[width=0.45\textwidth]{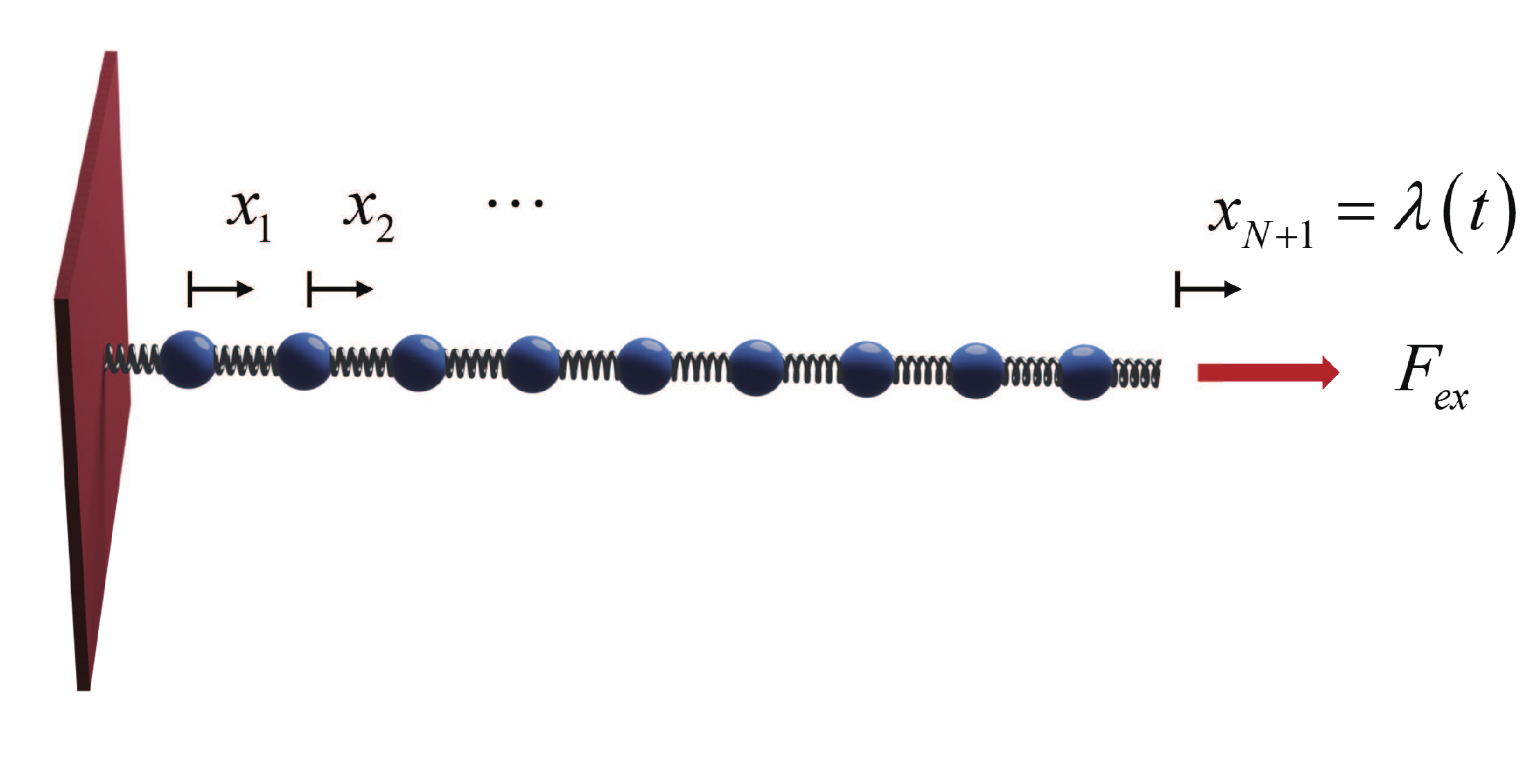}
    \caption{Schematics of the 1D mass-spring system used in Example 1. Reprinted from \citep{huang2021harnessing} Harnessing fluctuation theorems to discover free energy and dissipation potentials from non-equilibrium data, Vol 145, Shenglin Huang, Chuanpeng Sun, Prashant K.~Purohit, Celia Reina, Page 3, Copyright (2021), with permission from Elsevier.}
    \label{fig:MassSpringSystem}
\end{figure}

The first example that we examine is a prototype for polymers \citep{doi1988theory}, and biological macromolecules such as DNA and coiled-coil proteins \citep{raj2011phase}. In particular, the model consists of $N$ particles following Langevin dynamics, connected through $N+1$ identical springs, as shown in  Fig.~\ref{fig:MassSpringSystem}. The first spring is fixed to a wall, while the last one has a prescribed displacement boundary condition of the form $x_{N+1}=\lambda(t)=v_p t$, where $v_p$ is a constant pulling velocity. Denoting by $\tilde{V}_s$ the potential energy of an individual spring for the system being simulated (system $\tilde{S}$), the governing equations read
\begin{equation}
    \left\{ \begin{aligned}
    & \eta \dot{x}_i 
    = \tilde{V}'_s(x_{i+1} - x_i)
    - \tilde{V}'_s(x_{i} - x_{i-1})
    + \sqrt{\sigma} \dot{\xi}_i ,
    && i = 1, 2, \cdots, N-1 \\
    & \eta \dot{x}_N
    = \tilde{V}'_s(\lambda(t) - x_N)
    - \tilde{V}'_s(x_{N} - x_{N-1})
    + \sqrt{\sigma} \dot{\xi}_N .
    \end{aligned} \right. ,
\end{equation}
where $\sigma=2k_bT\eta$. 

In the following, we will consider three specific combinations of systems $S$ and $\tilde{S}$ to demonstrate the versatility of the approach and the quality of the UQ estimates. These include
\begin{enumerate}
    \item [$\bullet$] Element transmutation. In this first example, we will aim at predicting the behavior of material $S$ from $\tilde{S}$ (different potential energies), while subjected to the same pulling protocol. This is similar to the implementation of hyperdynamics \citep{chen2007exact} or path reweighting method \citep{kieninger2021path}, where the potential is the only varying parameter between the simulated and the predicted system.
    \item[$\bullet$] Predicting the far-from-equilibrium response of a material from its equilibrium behavior. Here, the potential energy will remain the same, while the pulling velocity will change from 0 (equilibrium) to a finite value (away from equilibrium).
    \item[$\bullet$] For the last example, we will change both the potential energy and the pulling protocol. In particular, we will consider the extreme case of predicting the non-equilibrium response of a nonlinear interacting mass-spring chain from independent Brownian particles, i.e., $\tilde{V}=0$.
\end{enumerate}

In each of these cases, we will consider observables that widely range in nature: from the expected evolution of each 
particle, to the instantaneous force exerted on the system at a given time, to the work done over a given time interval, which of course depends on the full evolution.  For all cases, we choose $N=10$ particles, temperature $k_B T = 10^{-4}$, viscosity $\eta = 5$, interparticle potentials $V_s$ and $\tilde{V}_s$ of quartic form, i.e., $\frac{1}{2}k_2 x^2 +\frac{1}{4} k_4 x^4$, and we will perform $N_R=10^5$ realizations of system $\tilde{S}$. The Langevin equations will be numerically simulated using an Euler-Maruyama scheme \citep{kloeden1992stochastic} with a time step $\Delta t = 10^{-3}$. For comparison purposes, material/process $S$ will also be simulated, and the same number of realizations will be used to directly estimate the ensemble averages. 

For the uncertainty quantification estimates, we choose the reference trajectory $\mathbf{x}_r(t)=\mathbf{0}$ and use 100 time steps in the discretization for each evaluated time point. That is, for each time at which the variance of $\mathcal{P}_{bias}$ is computed, a value of $n_T=100$ is chosen in the calculation of the coefficients $\mathbf{A}_{sq}$, $\mathbf{b}_{sq}$ and $c_{sq}$, according to Eqs.~\eqref{Eq:Asq}, \eqref{Eq:bsq} and \eqref{Eq:csq}, respectively. We remark that although $\mathbf{x}_r(t)=\mathbf{0}$ does not correspond in all cases to the expected trajectory of system/process $\tilde{S}$, this choice greatly simplifies the calculations and leads to remarkably accurate estimates for $\sigma^2_{\mathcal{P}_{bias}}$ and $\left| \mathcal{N}-1 \right|$.

\subsection{Case 1: Element transmutation: from material $\tilde{S}$ to material $S$ subjected to the same pulling protocol} \label{Sec:Case1}

\begin{figure}[H]
    \centering
    \includegraphics[width=\textwidth]{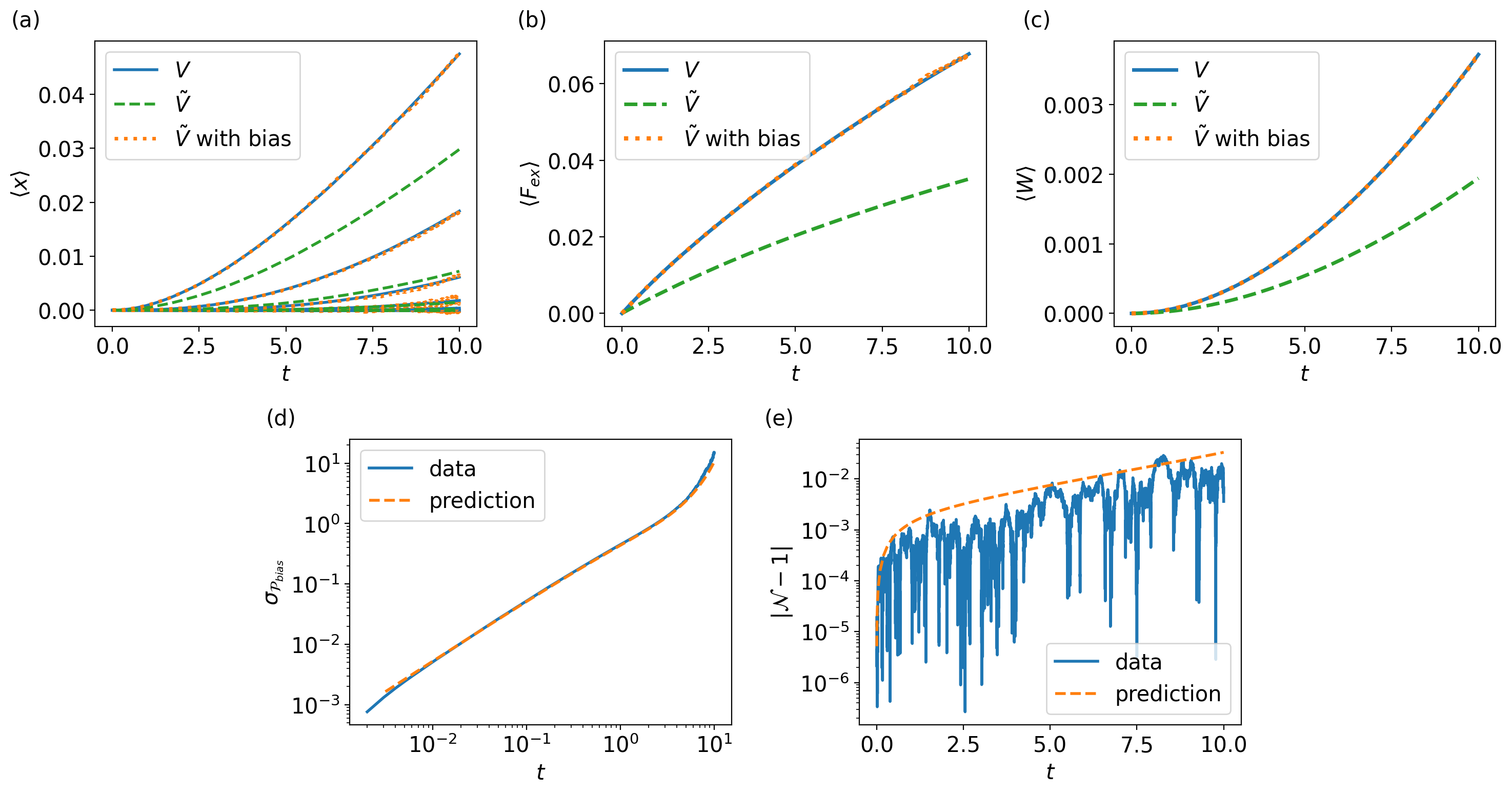}
    \caption{Prediction of the non-equilibrium behavior of material $S$ with quartic interatomic potential $V_s(u) = \frac{1}{2} k_{2} u^2 + \frac{1}{4} k_{4} u^4$ from material $\tilde{S}$ with quadratic potential $\tilde{V}_s(u) = \frac{1}{2} \tilde{k}_{2} u^2$ under the same pulling velocity $\dot{\lambda}(t) = \dot{\tilde{\lambda}}(t) = v_p(t)$, where $\tilde{k}_{2} = 0.5$, $k_{2} = 1$, $k_{4}=100$ and $v_p(t) = 0.01$. (a-c) Results for the ensemble averages of three different observables: (a) the displacement of each particle $\left< x_i \right>$, with $i=1, \dots, 10$ from bottom to top, (b) the external force $\left< F_{ex} \right>$ and (c) the external work $\left< W \right>$. The blue solid lines and green dashed lines are the results from Langevin simulation for material $S$ and  $\tilde{S}$, respectively. The orange dotted lines are the predictions for material $S$ from material $\tilde{S}$. (d, e) Validation of the uncertainty quantification estimates by evaluating the time evolution of (d) the standard deviation of $\mathcal{P}_{bias}$ and (e) the deviation of the empirical factor $\mathcal{N}$ from one. The solid blue lines are the exact value from the data and the orange dotted lines are the predictions using the nonlinear uncertainty quantification method.}
    \label{fig:1D_V_to_V}
\end{figure}

This first case demonstrates the capability of the method for predicting the non-equilibrium response of a system with an anharmonic potential, namely $V_s(u) = \frac{1}{2} k_{2} u^2 + \frac{1}{4} k_{4} u^4$ (material $S$) from a harmonic one, $\tilde{V} = \frac{1}{2} \tilde{k}_{2} u^2$, (material $\tilde{S}$), under the same pulling protocol $\dot{\lambda}(t) = \dot{\tilde{\lambda}}(t) = v_p$. Here, we choose $\tilde{k}_{2} = 0.5$, $k_{2} = 1$, $k_{4}=100$ and $v_p = 0.01$. Figures ~\ref{fig:1D_V_to_V} (a-c) show the results for the ensemble averages of the displacement of each particle $\left< x_i \right>$, the external force applied on the system $\left< F_{ex} \right>$ and the external work $\left< W \right>$. The blue solid lines, green dashed lines and orange dotted lines represent, respectively, the results from Langevin simulation for system $S$ with potential $V$, those for system $\tilde{S}$ with potential $\tilde{V}$, and the prediction of system $S$ from $\tilde{S}$ using Eq.~\eqref{Eq:Bias}. In particular, the lines from bottom to the top in Fig.~\ref{fig:1D_V_to_V}(a) represent the particles from 1 to $N=10$. As it is there observed, all predictions are in excellent agreement with the validation data. Only very minor differences for the average displacement and external force are observed after around $t\sim8$, where the predictions become slightly more stochastic in nature. These increase in the errors are to be expected as noted at the end of Section \ref{Sec:Reweighting} and can be predicted using the nonlinear uncertainty quantification estimates discussed in Sec.~\ref{Sec:NonlinearUQ}. We recall that these estimates aim at predicting the deviation of the empirical average $\mathcal{N} = \sum_{r=1}^{N_R} \frac{1}{N_R} \mathcal{P}_{bias,r}$ from one through $\sigma_{\mathcal{N}}=\sigma_{\mathcal{P}_{bias}}/\sqrt{N_R}$, which itself is used as a measure of how well the path probability distribution of the system $S$ (the one not simulated) is captured. Figures~\ref{fig:1D_V_to_V}(d, e) show the estimated growing standard deviation of factor $\mathcal{P}_{bias}$ in a log-log scale and the deviation of $\mathcal{N}$ from one in a log-linear scale, respectively, compared to the corresponding values directly calculated from the Langevin simulation. Remarkably, without simulation data for neither system $S$ nor system $\tilde{S}$, the derived estimates perfectly predict the standard deviation of $\mathcal{P}_{bias}$, which spans four decades, and the deviation of $\mathcal{N}$. As time increase, the deviation of $\mathcal{N}$ finally reaches $3\%$, which is tolerable and agrees well with the small errors in the observables. We note that the accuracy of these results is striking, as the bias potential is not purely quadratic (for which the estimates would be exact), but instead include a non-negligible quartic contribution.

\subsection{Case 2: From equilibrium to non-equilibrium for the same material}

\begin{figure}[H]
    \centering
    \includegraphics[width=\textwidth]{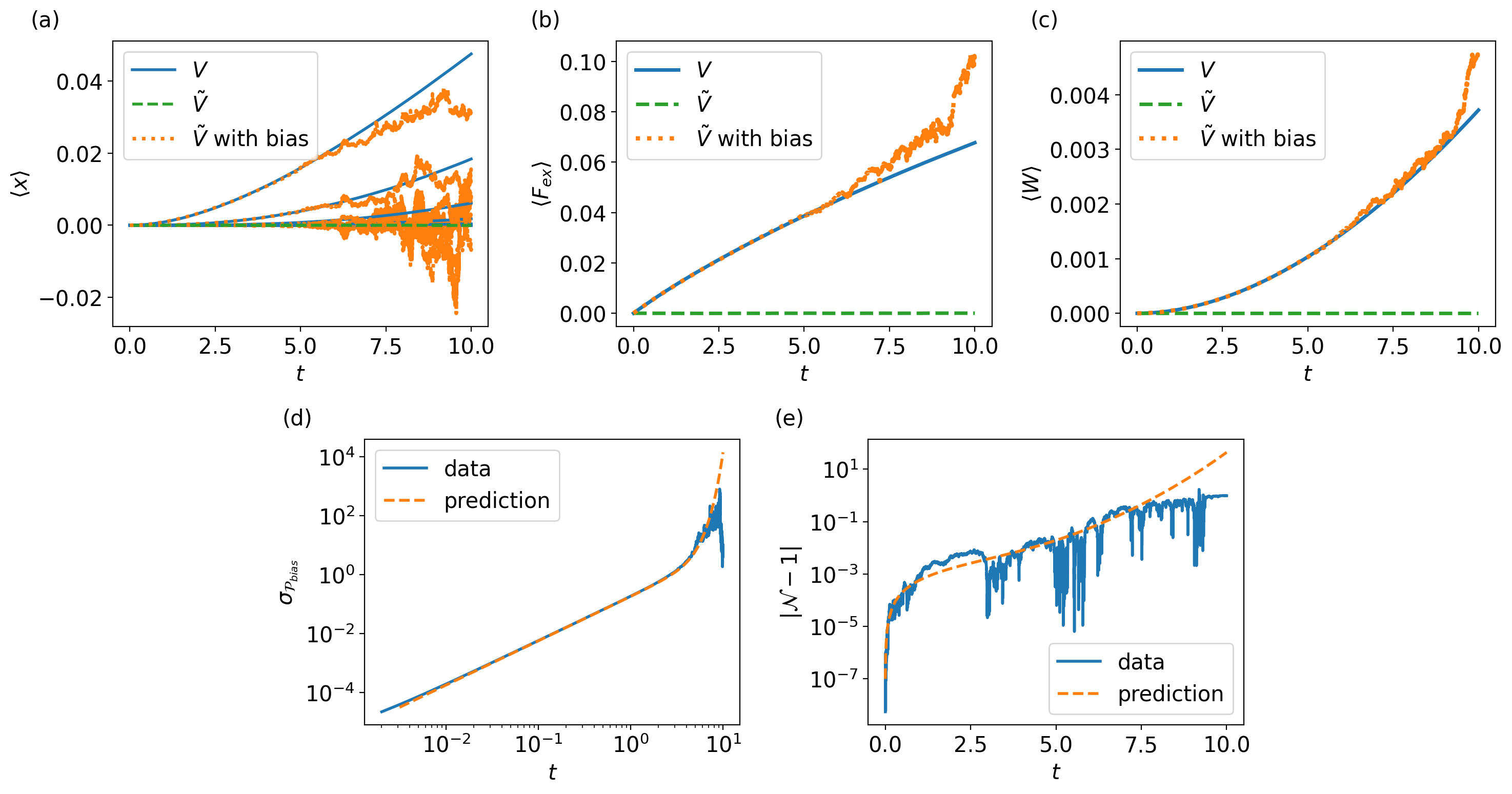}
    \caption{Predicton of the non-equilibrium behavior under pulling velocity $\dot{\lambda}(t) = v_p(t) >0 $ (process $S$) from the equilibrium response, i.e., $\dot{\tilde{\lambda}}(t) = 0$, (process $\tilde{S}$) for the same material with quartic interatomic potential $V_s(u) = \tilde{V}_s(u) = \frac{1}{2} k_{2} u^2 + \frac{1}{4} k_{4} u^4$, where $k_{2} = 1$, $k_{4}=100$ and $v_p(t) = 0.01$. (a-c) Results for the ensemble averages of three different observables: (a) the displacement of each particle $\left< x_i \right>$, (b) the external force $\left< F_{ex} \right>$ and (c) the external work $\left< W \right>$. The blue solid lines and green dashed lines are the results from Langevin simulation with process $S$ and process $\tilde{S}$, respectively. The orange dotted lines are the prediction for process $S$ from process $\tilde{S}$. (d, e) Validation of the uncertainty quantification estimates by evaluating the time evolution of (d) the standard deviation of $P_{bias}$ and (e) the deviation of the empirical factor $\mathcal{N}$ from one. The solid blue lines are the exact value from the data and the orange dotted lines are the predictions using the nonlinear uncertainty quantification method.}
    \label{fig:1D_eq_to_noneq}
\end{figure}

In contrast to the first case, this second case illustrates the prediction of the non-equilibrium response for a given system (finite pulling velocity) given its equilibrium behavior (zero pulling velocity). The interatomic potential considered is $V_s(u) = \tilde{V}_s(u) = \frac{1}{2} k_{2} u^2 + \frac{1}{4} k_{4} u^4$, with $k_{2}=1$ and $k_{4}=100$, and the pulling velocity for the aimed non-equilibrium process is $\dot{\lambda}(t) = v_p=0.01$. Figures~\ref{fig:1D_eq_to_noneq}(a-c) depict the prediction for the average displacement of each particle, average external force and average work. While all the predictions have good agreements with the true results at the beginning, the errors start to become more significant from about $t \sim 6$, at which point the ensemble averages appear to be more stochastic in nature. This increased stochasticity results from a decrease in the number of trajectories that contribute in practice to the ensemble average and is directly related to the heavier tails of $\mathcal{P}_{bias}$ discussed in Section \ref{Sec:Reweighting}. The larger errors, as compared to Case 1 studied in Section \ref{Sec:Case1}, are also understandable from a sampling perspective. In the process to be predicted, all the particles (especially the last few ones) are moving rightward in the most probable trajectories induced by the positive pulling velocity. However, these trajectories are highly unlikely to be observed at equilibrium, where the right end is fixed. Hence, the predicted evolution of the observables (especially the average displacement for the last particle and, consequently, the average external force and work) are biased. Figures~\ref{fig:1D_eq_to_noneq}(d, e) show the growing standard deviation of factor $\mathcal{P}_{bias}$ and the deviation of the normalization factor $\mathcal{N}$ from one. The latter reaches an error of 0.1 at around $t\sim8$, at which time the prediction for the observables becomes very poor and are no longer reliable. Here, again, the uncertainty quantification estimates provide an excellent prediction of the sampling errors.

\subsection{Case 3: From Brownian particles to the non-equilibrium response of an interacting particle system}

\begin{figure}[H]
    \centering
    \includegraphics[width=\textwidth]{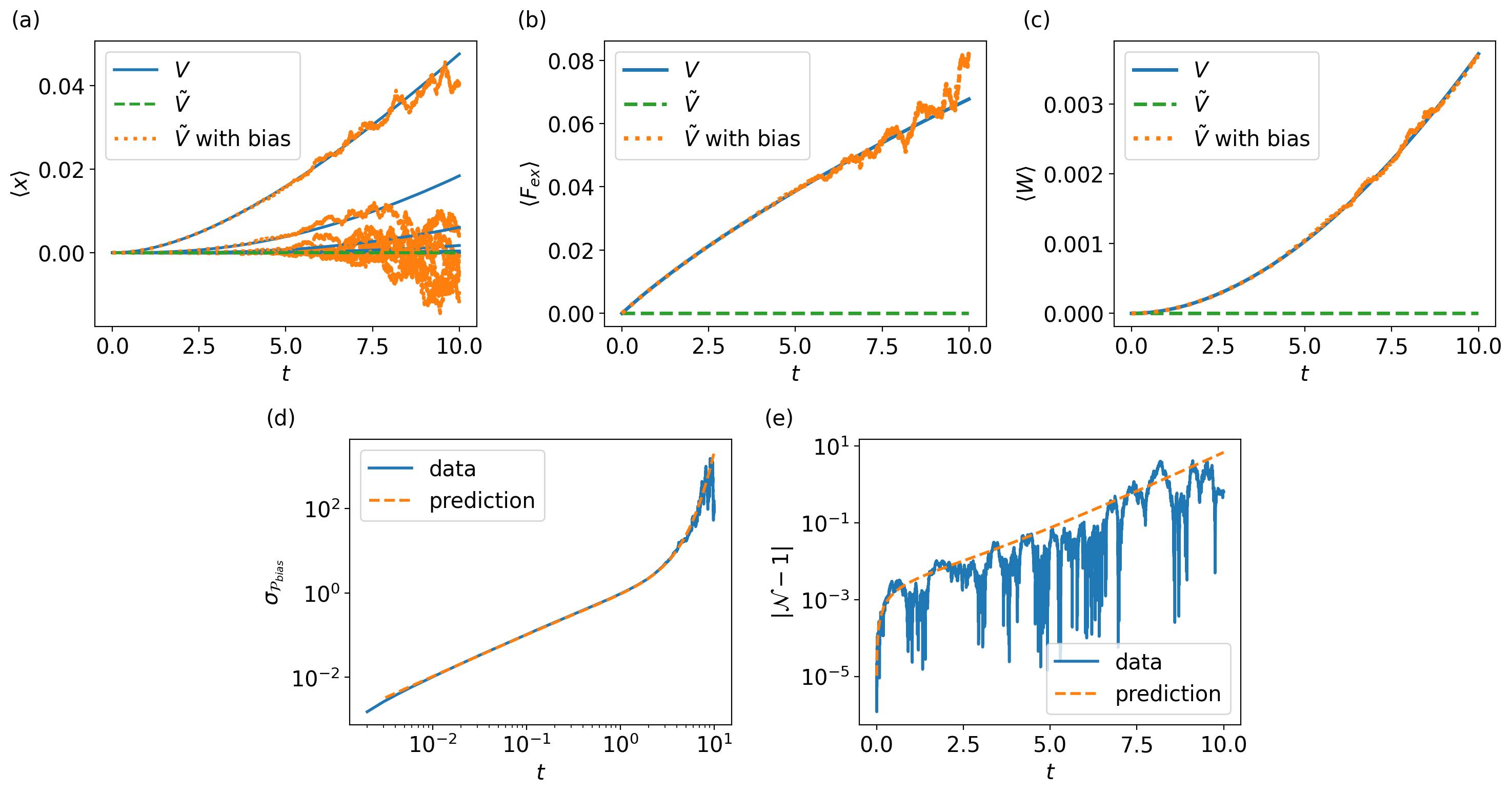}
    \caption{Prediction of the non-equilibrium behavior for material $S$ with quartic interatomic potential $V_s(u) = \frac{1}{2} k_{2} u^2 + \frac{1}{4} k_{4} u^4$, with $k_{2} = 1$, $k_{4}=100$, under pulling velocity $\dot{\lambda}(t) = v_p(t)= 0.01$ from Brownian trajectories. (a-c) Results for the ensemble averages of three different observables: (a) the displacement of each particle $\left< x_i \right>$, (b) the external force $\left< F_{ex} \right>$ and (c) the external work $\left< W \right>$. The blue solid lines and green dashed lines are the results from Langevin simulations of system/process $S$ and $\tilde{S}$, respectively. The orange dotted lines are the prediction for material/process $S$ from $\tilde{S}$. (d, e) Validation of the uncertainty quantification estimates by evaluating the time evolution of (d) the standard deviation of $P_{bias}$ and (e) the deviation of the empirical factor $\mathcal{N}$ from one. The solid blue lines are the exact value from the data and the orange dotted lines are the predictions using the nonlinear uncertainty quantification method.}
    \label{fig:1D_Brownian_to_noneq}
\end{figure}

The third and final case considered is aimed at demonstrating an extreme example for the path reweighting strategy. Specifically, we choose to predict the non-equilibrium behavior of an anharmonic chain from independent Brownian particles, i.e., $\tilde{V}=0$.  Here, material/process $S$ is also set as a quartic interatomic potential $V_s(u) = \frac{1}{2} k_{2} u^2 + \frac{1}{4} k_{4} u^4$ with $k_{2}=1$ and $k_{4}=100$ with pulling velocity $\dot{\lambda}(t) = v_p=0.01$. Figures~\ref{fig:1D_Brownian_to_noneq}(a-c) depict the predictions for the average displacement, average external force and average work. Despite the extreme nature of the example, the predictions are still reasonably good up to $t=10$, and actually better than that of Case 2 above. Here, the Brownian particles can freely move, while those of Case 2 are constrained due to the boundary conditions and interatomic potential for system $\tilde{S}$. This significantly reduces the sampling errors that govern the accuracy of the predictions. Finally, Figures \ref{fig:1D_Brownian_to_noneq}(d, e) show the growing standard deviation of $\mathcal{P}_{bias}$ with time and the deviation of $\mathcal{N}$ with one. Again, the UQ estimates perfectly predict both quantities throughout six decades. Moreover, the error of $|\mathcal{N}-1|$ reaches $10\%$ around $t\sim6$, after which the number of realizations of system $\tilde{S}$ is insufficient to accurately make predictions of system $S$.

\section{Example 2: Caging in two-dimensional glassy systems} \label{Sec:Ex2}

One of the most ubiquitous examples of out-of-equilibrium behavior, and one that we are regularly familiar with from everyday experience, is that of glasses \citep{Stillinger2013, Char2017}. Glassy dynamics is observed  in a wide range of length scales (from nanoparticles to grains)  spanning a variety of industries from building materials (concrete) to paints (colloidal suspensions), and household goods (foams \& gels)\citep{Bonn2017, Nic2018}. Despite the ubiquity of glassy systems, the study of equilibrium statistical mechanics leaves us ill-equipped to answer many questions surrounding the theory of glasses, though there has been significant recent progress \citep{berthier2011theoretical}. In these final examples, we aim to predict features of the glassy dynamics from the motions of particles in the equilibrium, fluid phase of 2D glass-formers. Supercooled liquids above their glass transition temperature exhibit a characteristic onset of ``caged'' dynamics, whereby particles vibrate in the cage formed by their neighbors before thermal fluctuations enable hopping past the cage. This phenomenon of particle localization is usually called caging, and can be observed in many dynamical quantities. The mean squared displacement ($\mathcal{MSD}$), familiar to many from the study of Brownian motion, is sufficient to explain such behavior. The emergence of sub-diffusive (slope $<$ 1) trends in the log-log fits to $\mathcal{MSD}$ is a hallmark of the onset of particle caging, where as the observance of a plateau is indicative of strongly caged dynamics, and the duration of the cage grows at the materials cooled towards the glass transition.

\subsection{Model description and overview of the cases to be examined}

\begin{figure}[H]
    \centering
    \includegraphics[width=0.7\textwidth]{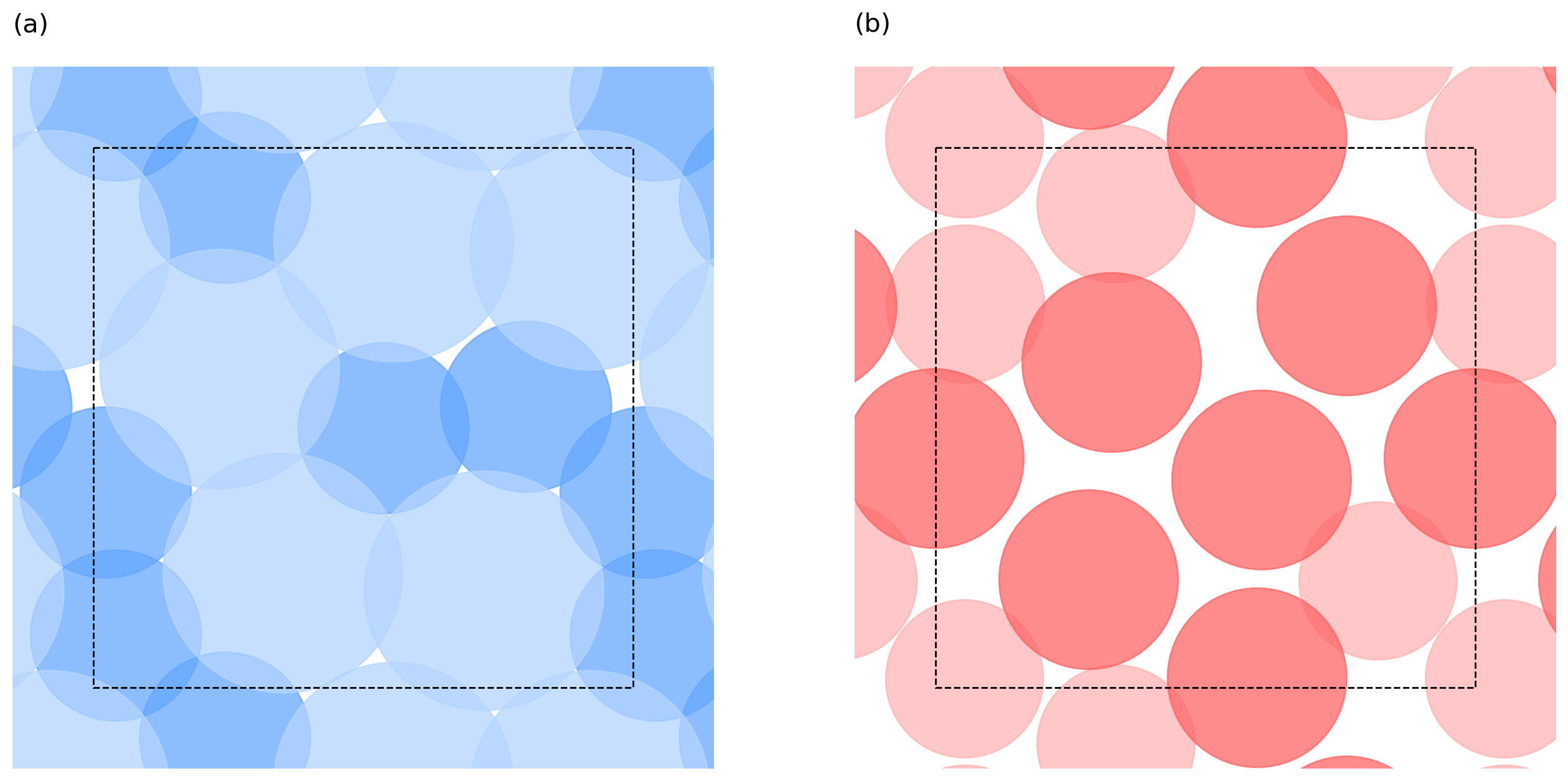}
    \caption{Initial configurations used in Example 2. (a) Hertzian system in a 50:50 mixture. Dark blue particles are A type and light blue are B type. (b) Lennard-Jones system in a 60:40 mixture. Dark red are A type and light red are B type. The radii of the particles reflect a size of 0.5$\sigma_{ii}$ for particle species $i$. The black dashed square is the boundary of the simulation box.}
    \label{fig:Particles}
\end{figure}

Here, we explore two model systems, a bidisperse mixture of Hertzian disks and the standard Kob-Anderson type Lennard-Jones glass, shown in Fig.~\ref{fig:Particles}. The Hertzian system is defined by an interparticle potential
\begin{equation} \label{Eq:HertzPot}
    V_{ij}(r) = \begin{cases} 
        0.4\epsilon\left(1-\frac{r}{\sigma_{ij}}\right)^{2.5} & r\leq \sigma_{ij} \\
        0 & r > \sigma_{ij}
   \end{cases}
\end{equation}
with a 50:50 A:B mixture, where $\sigma_{AB} = 1.2\sigma_{AA}$ and $\sigma_{BB} = 1.4\sigma_{AA}$. A constant packing fraction of 1.0$\sigma^{-2}_{AA}$ is used. The Lennard-Jones system uses the usual interparticle potential with a numerical cut-off distance
\begin{equation} \label{Eq:LJPot}
    V_{ij}(r) = \begin{cases} 
        4\epsilon_{ij}\left[\left(\frac{\sigma_{ij}}{r}\right)^{12} - \left(\frac{\sigma_{ij}}{r}\right)^6\right] & r\leq 2.5\sigma_{ij} \\
        0 & r > 2.5\sigma_{ij}
   \end{cases}
\end{equation}
 in a 60:40 A:B mixture, where the interaction lengths are set to $\sigma_{AB} = 0.8\sigma_{AA}$ and $\sigma_{BB} = 0.88\sigma_{AA}$, the energies are set to $\epsilon_{AB} = 1.5\epsilon_{AA}$ and $\epsilon_{BB} = 0.5\epsilon_{AA}$, and a packing fraction of 1.1$\sigma^{-2}_{AA}$ is selected. We choose this species ratio and packing fraction to reduce the chance of crystallization within the system, as the usual 80:20 mixture used in 3D is more prone to do so in 2D \citep{Bruening2008, Flenner2015}. Both systems are composed of 10 particles and $\sigma_{AA}$ is taken to be 1 universally. Additionally, the simulations are performed in a square, periodic box, and no external forcing is applied to the configurations. As above, the Euler-Mayurama method \citep{kloeden1992stochastic} is used to integrate the overdamped Langevin dynamics of the system with parameters $\eta=5$, and $\Delta t = 10^{-3}$. We compare our results among two different temperatures $k_B T \in \{ 10^{-1}, 10^{-2} \}$. In the unbiased simulations, we choose pairings of $\tilde{\epsilon}$ and $k_B T$ such that we observe phenomenology consistent with dynamical arrest if we were to increase $\tilde{\epsilon}$ or decrease $k_B T$. Thus, we use $\tilde{\epsilon}=10.0$ when $k_B T=10^{-1}$ and $\tilde{\epsilon}=1.0$ when $k_B T=10^{-2}$ in the Hertzian case, and $\tilde{\epsilon}_{AA}=0.1$ when $k_B T=10^{-1}$ and $\tilde{\epsilon}_{AA}=0.01$ when $k_B T=10^{-2}$ in the Lennard-Jones model. A total of $N_R = 10^7$ realizations of the liquid phase in each system were simulated to perform predictions, and another $N_R = 10^5$ are produced to validate the accuracy of these results.

To analyze the results of our simulations we compute two dynamical quantities, the mean squared displacement $\mathcal{MSD}$ and the particle overlap $Q(a)$, defined as
\begin{align} \label{Eq:MSD}
&\mathcal{MSD}= \left\langle \frac{1}{N} \sum_{i=1}^N \|\Delta \mathbf{r}_i \|^2 \right\rangle, \\ \label{Eq:Q}
&Q(a) = \left\langle \frac{1}{N} \sum_{i=1}^N H( a - \| \Delta \mathbf{r}_i \|)\right\rangle, 
\end{align}
where $H$ is the Heaviside step function. The overlap $Q(a)$ quantifies the fraction of particles that have moved some distance $a$ within the simulation box. The overlap function in this context conveys essentially the same information as the self-intermediate scattering function, also commonly used in the glassy literature. We choose $a=0.1$ in all simulations to display the onset of caged dynamics at accessible timescales. 

While the glass transition can normally be observed by lowering the system temperature until plateaus are viewed in quantities like the $\mathcal{MSD}$ and $Q(a)$, we here 
vary instead the energy scale $\epsilon$ relative to $k_B T$ to increase the height of energetic barriers and induce caging. We prepare the initial state of each system to be completely force balanced, thus making the initial dynamics nearly degenerate regarding the energy scale $\epsilon$. Initially, states are sampled from a spatially uniform distribution and the configurations are subsequently quenched to their inherent structure using gradient descent \citep{Tsalikis2008}. In each system, we only employ one initial configuration, where we observe phenomenology consistent with caged dynamics to generate our realizations.

Since we only modify the potential through a scaling coefficient between our reference and target systems, we can pull the difference in $\tilde{V}$ and $V$ through the integration. This leaves us with two terms in the computation of $\mathcal{I}_{bias}$, denoted below as $\mathcal{I}_1$ and $\mathcal{I}_2$
\begin{equation} \label{Eq:SquareSplitPre}
\frac{\partial V_{bias}}{\partial \mathbf{r}_i} = \left(1-\frac{\epsilon}{\tilde{\epsilon}}\right)\frac{\partial \tilde{V}}{\partial \mathbf{r}_i} = -\left(\chi - 1\right)\frac{\partial \tilde{V}}{\partial \mathbf{r}_i},
\end{equation}
\begin{equation} \label{Eq:SquareSplit}
\begin{split}
&\mathcal{I}_{bias}=\frac{(\chi - 1)^2}{4\eta}\int_0^\tau \sum_i^N \left\| \frac{\partial \tilde{V}}{\partial \mathbf{r}_i} \right\|^2 \, dt + (\chi - 1) \sqrt{\frac{k_BT}{2\eta}} \int_0^\tau \sum_i^N \frac{\partial \tilde{V}}{\partial \mathbf{r}_i} \cdot \dot{\boldsymbol\xi}_i \, dt
= \mathcal{I}_2 (\chi - 1)^2 - \mathcal{I}_1 (\chi - 1).
 \end{split}
\end{equation}
With this equation in hand, we can compute any member of this family of potentials at virtually no added computational cost; we simply compute the forces for $\tilde{V}$ once, and then apply Eq.~(\ref{Eq:SquareSplit}) for each desired $V$. In our results, we employ this method to generate predictions at 10 different target potentials with ratios $V/\tilde{V}=\chi$ sampled logarithmically from 1.259 to 10.0.

\begin{figure}[H]
    \centering
    \includegraphics[width=\textwidth]{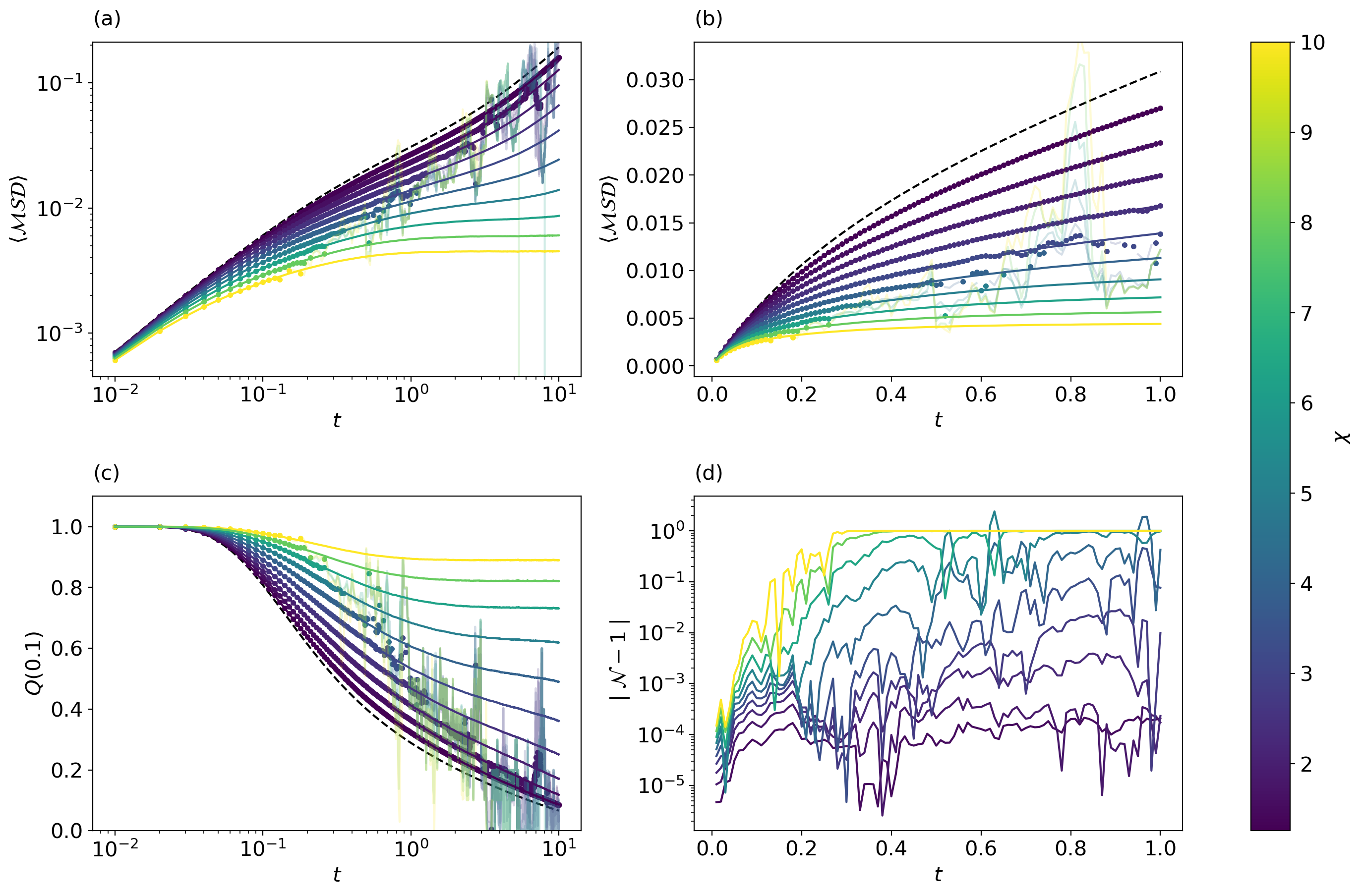}
    \caption{Observables for the Hertzian case with $k_B T = 10^{-1}$, highlighting the transition to caging. Mean squared displacement (a,b) and particle overlap (c) calculated from $\tilde{V}$ (black dashed line) and validation datasets (bold lines). Prediction points in which $|\mathcal{N}-1|<0.1$ are plotted with solid dots, otherwise they are plotted with faint lines to highlight the prediction's deviation from validation. (d) Deviation in $ \mathcal{N}$ from unity, indicating when we can be confident in the accuracy of the predicted observables. }
    \label{fig:HertzianResults}
\end{figure}

\subsection{Results}

First, we look at both systems with $k_B T = 10^{-1}$. In the Hertzian case, shown in Fig.~\ref{fig:HertzianResults}, we find good agreement between our prediction and the validation data over a wide range of times. In biases as high as $\chi=10.0$, the curves track well, and we can observe dramatically slower dynamics as indicated by the slopes of $\mathcal{MSD}$ in Figure \ref{fig:HertzianResults}(a,b). From Figure \ref{fig:HertzianResults}(a) we can see in our simulation of $\tilde{V}$ (black dashed line) initially diffusive behavior until $t\sim0.2$, followed by a period of slight sub-diffusivity, and the reemergence of diffusive behavior after $t\sim3.0$. As we increase $\chi$, the dynamics become more sluggish until we observe a prolonged plateau for $\chi>5.0$. We find the same signatures of caged dynamics in $Q(0.1)$ shown in Figure \ref{fig:HertzianResults}(c). The length of time before the accuracy of the method breaks down is strongly dependent upon the value of $\chi$ used, where a $\chi$ of 10.0 breaks down at $t\sim0.2$, while $\chi=1.259$ is accurate well beyond $t=10.0$. While the caging plateau seen beyond $\chi\sim5.0$ appears to be just out of reach, the sub-diffusive behavior that we can predict is still strong evidence for predicting the onset of caged dynamics from diffusive dynamics.

\begin{figure}[H]
    \centering
    \includegraphics[width=\textwidth]{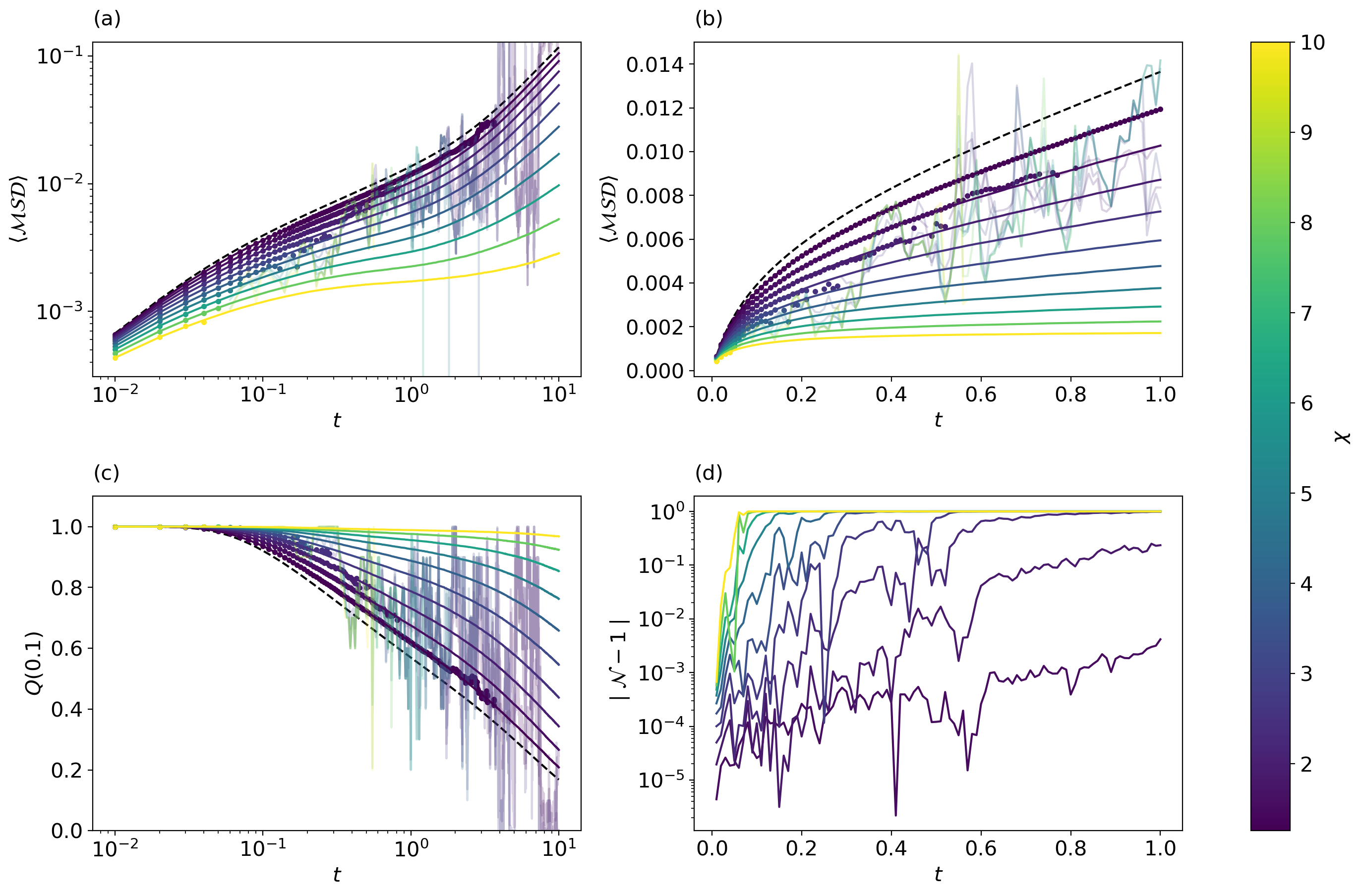}
    \caption{Observables for the Lennard-Jones case with $k_B T = 10^{-1}$, highlighting the transition to caging. Mean squared displacement (a,b) and particle overlap (c) calculated from $\tilde{V}$ (black dashed line) and validation datasets (bold lines). Prediction points in which $|\mathcal{N}-1|<0.1$ are plotted with solid dots, otherwise they are plotted with faint lines to highlight the prediction's deviation from validation. (d) Deviation in $ \mathcal{N}$ from unity, indicating when we can be confident in the accuracy of the predicted observables. }
    \label{fig:LJResults}
\end{figure}

\begin{figure}[H]
    \centering
    \includegraphics[width=\textwidth]{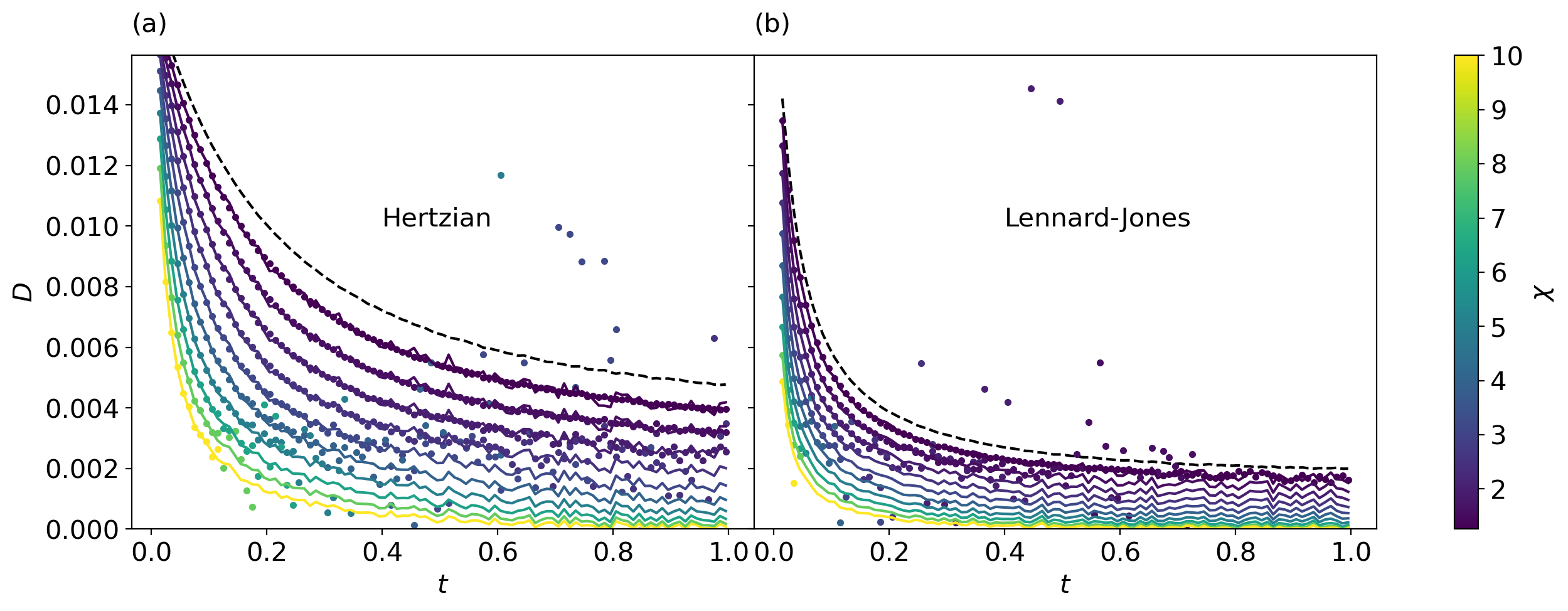}
    \caption{Instantaneous diffusion coefficient calculated for both the Hertzian (a) and Lennard-Jones (b) models, at $k_B T = 10^{-1}$. Prediction points are plotted as dots, and validation as solid lines. Y-scales in both plots are shared. The predictions appear to breakdown at a fixed diffusion rate of about $D\sim 0.003$ that is independent of the potential employed.}
    \label{fig:Diffusion}
\end{figure}

Similar tends are seen in the data from the Lennard-Jones model, shown in Figure \ref{fig:LJResults}. We find that the predictions break down at shorter times than seen in the Hertzian case. This is likely due to the more rapid emergence of large forces in the Lennard-Jones model compared to the Hertzian system when deviating from the minima of the potential. To further explore this relationship between the potentials, dynamics, and confidence in our predictions, we compute the instantaneous diffusion coefficients $D$ from our $\mathcal{MSD}$ curves using a simple forward difference method. In Figure \ref{fig:Diffusion} we see that the softer repulsion of the Hertzian model leads to a gradual reduction of the diffusion coefficient as the configurations escape the minima, but we find this drop in $D$ to happen more rapidly in the Lennard-Jones model. Surprisingly, we find that the breakdown of our prediction appears to occur universally when $D$ drops below $\sim0.003$. This cutoff appears to be set by both the number of realizations and the temperature used for our simulation.

\begin{figure}[H]
    \centering
    \includegraphics[width=\textwidth]{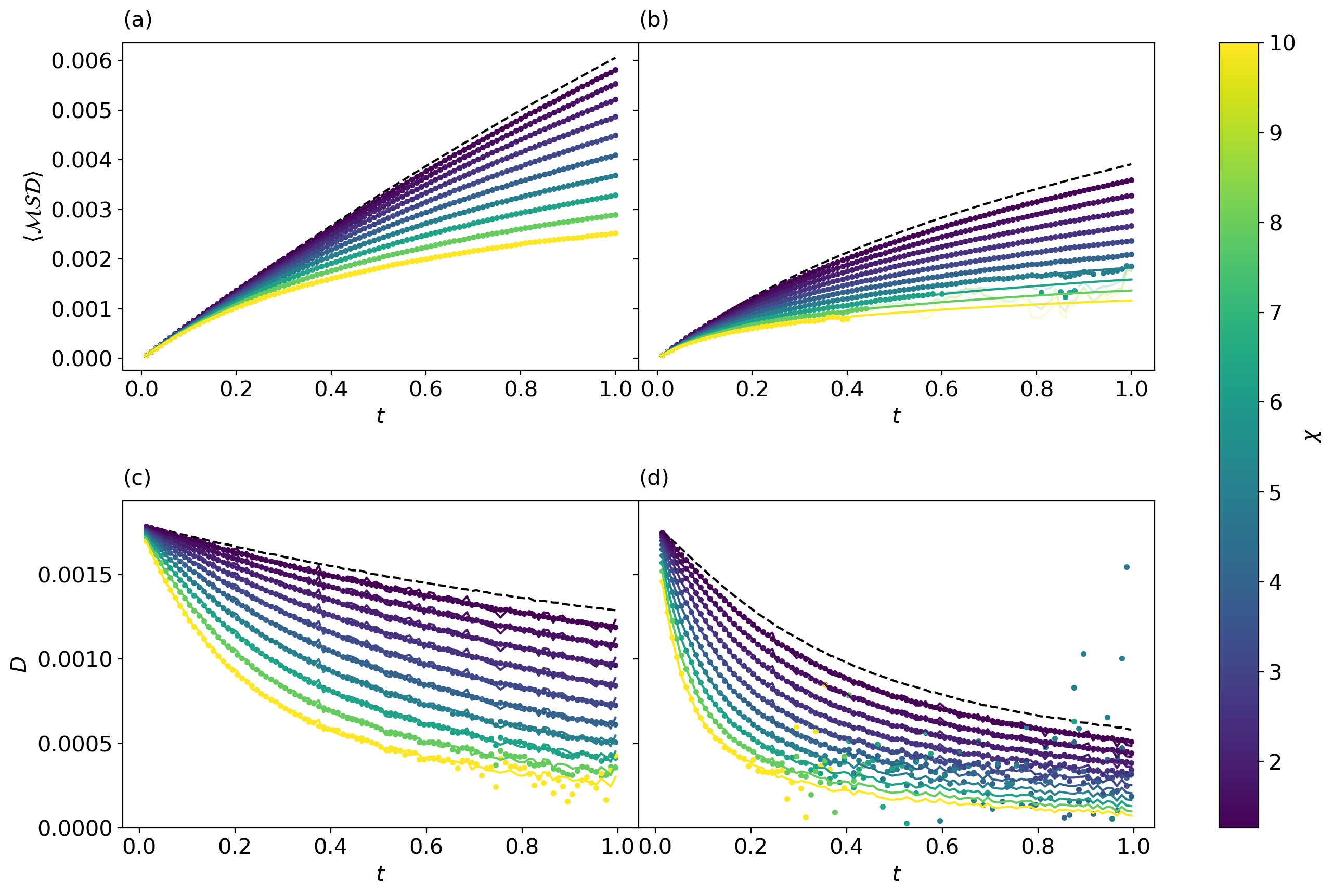}
    \caption{$\mathcal{MSD}$ for the Hertzian (a) and Lennard-Jones (b) models, at $k_B T = 10^{-2}$. Corresponding instantaneous diffusion coefficient in the Hertzian (c) and Lennard-Jones (d) models. Y-scales in both sets of plots are shared. In all cases, prediction points are plotted as dots, and validation as solid lines.}
    \label{fig:LowT}
\end{figure}

Lastly, we examine how these behaviors change through lowering $\tilde{\epsilon}$ and $k_B T$ simultaneously by an order of magnitude. By reducing the temperature of the system to $k_B T = 10^{-2}$, we observe improved accuracy in our results given the same length of time in our simulations at $k_B T = 10^{-1}$. In Figure \ref{fig:LowT}(a,b), our predictions track almost perfectly at all observed $\chi$ for both the Hertzian and Lennard-Jones models. Though this is mostly unsurprising, as we expect this equal reduction in the temperature and energy scale to yield a similar effective temperature of simulations and push our observations out farther in time. In Figure \ref{fig:LowT}(c,d), we find that the threshold $D$ in which we observed the breakdown in our predictions is further reduced by an order of magnitude, matching well with our reduction in temperature. This is a curious result, as it seems that this method of modeling possesses an inherent limitation on the slow dynamical processes we can resolve  that is dependent on our choices of temperature and realizations employed.

\begin{figure}[H]
    \centering
    \includegraphics[width=0.9\textwidth]{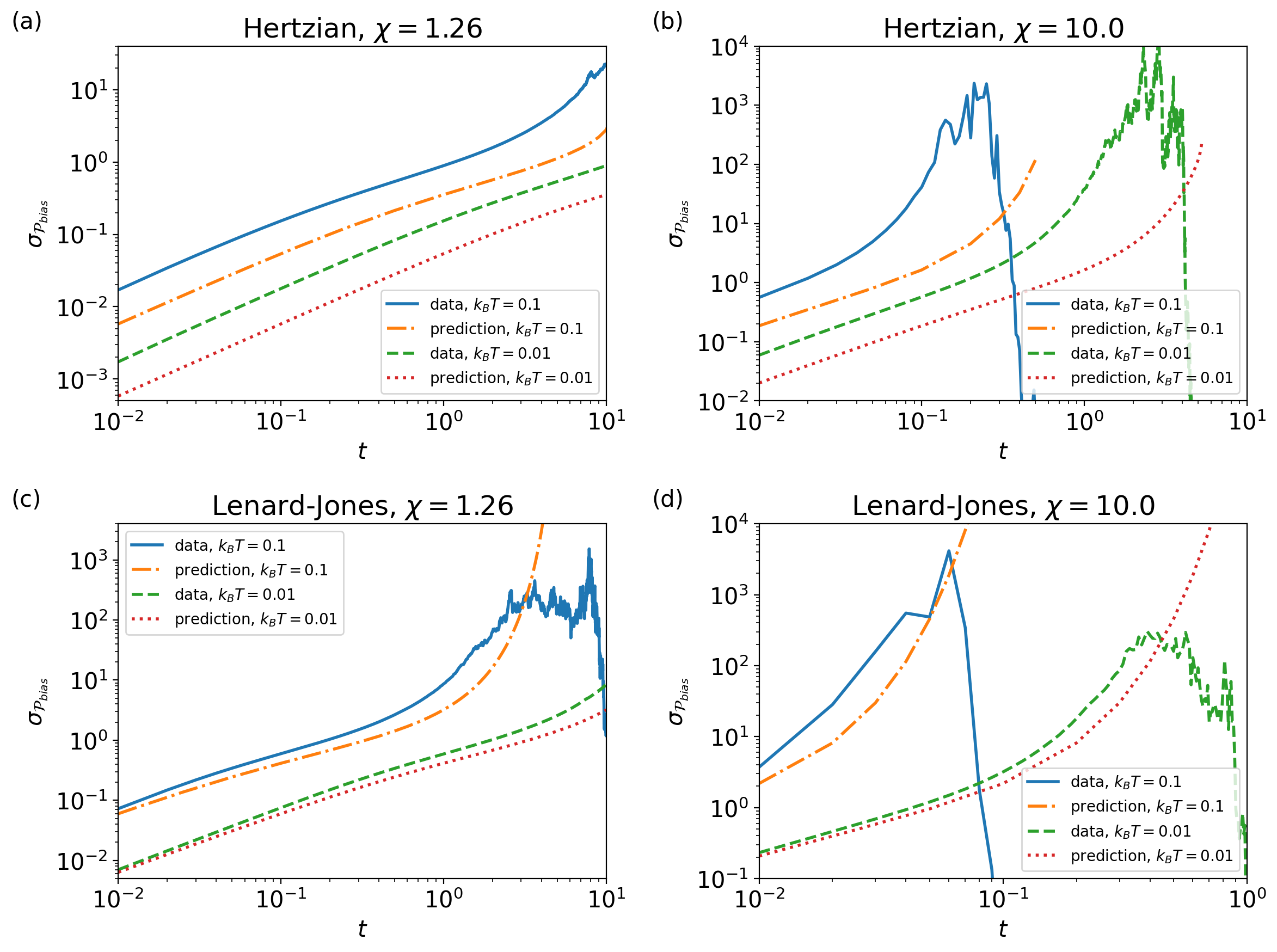}
    \caption{Validation of the uncertainty estimates for the Hertzian potential with (a) $\chi = 1.26$, (b) $\chi = 10$, and for the Lennard-Jones potential with (c) $\chi = 1.26$, (d) $\chi = 10$. The time discretization for these estimates is set as $n_T=100$ and the reference trajectory is chosen as the initial configuration.}
    \label{fig:Hertzian_LJ_UQ}
\end{figure}

\subsection{Uncertainty quantification}

In the following, we compare the UQ estimates for the standard deviation of $\mathcal{P}_{bias}$ with the data, for the two extreme biases considered, i.e., $\chi=1.26$ and $\chi=10.0$, of the Hertzian and the Lennard-Jones systems. These are shown in Figure \ref{fig:Hertzian_LJ_UQ} for both temperatures studied, i.e., $k_BT=0.1$ and $k_BT=0.01$. Here, we use $n_T=100$ for the time discretization and choose the initial (equilibrium) configuration as the reference trajectory. We recall that such estimates were built upon a quadratic approximation of the bias potential, while the biases for the systems considered are themselves Hertzian and Lennard-Jones. These potentials deviate from simple quadratic forms, particularly the Hertzian, which does not have a unique minimum but rather an extended zero plateau beyond a certain threshold. Moreover, the phenomenon here studied is that of diffusion, whereby particle trajectories will visit all parts of the potential, as opposed to be constrained to the neighborhood of a given position. Yet, despite all these rather unfavorable circumstances, the UQ estimates provide remarkably good predictions of the Lennard-Jones system, and mostly results within the same order of magnitude for the Hertzian system. To be more specific, we provide in Table~\ref{Tab:Hertz_LJ_UQ} below the ratio between the data and the predictions for $\sigma_{\mathcal{P}_{bias}}$ at two significant times of the predictions for all cases shown in Figure \ref{fig:Hertzian_LJ_UQ}. These times are chosen to be the lowest time depicted ($t=10^{-2}$) and the time prior to the plateau and subsequent decay of the standard deviation of $\mathcal{P}_{bias}$ observed in the data (this time is case dependent, and hence specified in the table). This behavior is a strong signature of the reduced number of relevant trajectories participating in the ensemble averages, which in turn implies that the estimate of $\sigma_{\mathcal{P}_{bias}}$ obtained from the data ceases to be accurate. Comparisons between the data and the predictions beyond this point are therefore not meaningful.

\begin{table}
\begin{tabular}{ |c|c|c|c|c|c|  }
\cline{3-6}
\multicolumn{2}{c|}{\multirow{3}{*}{}}  & \multicolumn{4}{c|}{$\sigma_{\mathcal{P}_{bias}}(\text{data})/\sigma_{\mathcal{P}_{bias}}(\text{predictions}) $} \\
\cline{3-6} 
\multicolumn{2}{c|}{}  & \multicolumn{2}{c|}{$\chi=1.26$} & \multicolumn{2}{c|}{$\chi=10.0$}\\
\cline{3-6}
\multicolumn{2}{c|}{}  & low time & large time & low time & large time \\
\hline
\multirow{2}{*}{Lennard-Jones} & $k_BT=0.01$ & 1.10 ($t=0.01$) & 2.60 ($t=10$) & 1.12 ($t=$0.01) & 2.84 ($t=0.3$) \\
\cline{2-6}
& $k_BT=0.1$ & 1.34 ($t=0.01$) & 2.69 ($t=1$) & 1.70 ($t=0.01$) & 5.29 ($t=0.03$) \\
\hline
\multirow{2}{*}{Hertzian} & $k_BT=0.01$ & 2.97 ($t=0.01$) & 2.52 ($t=10$) & 2.96 ($t=0.01$) & 23.5 ($t=1$) \\
\cline{2-6}
& $k_BT=0.1$ & 2.95 ($t=0.01$) & 8.10 ($t=10$) & 3.02 ($t=0.01$) & 25.2 ($t=0.1$)\\
\hline
\end{tabular}
\caption{Ratio of the standard deviation of $\mathcal{P}_{bias}$ between the data and UQ prediction for the Lenard-Jones and Hertzian potentials.}
\label{Tab:Hertz_LJ_UQ}
\end{table}

Finally, we make some remarks regarding the computation of the UQ estimates. These calculations were found to be extremely fast and only requiring about a minute for each case in Matlab on a laptop. This is for the time discretization used, $n_T=100$,  which was sufficient to guarantee convergence in all cases, and for $\sim$100 time points. Yet, it is important to remark that, as the difference between the target and simulated potentials increases, or as time increases, the spectrum of the eigenvalues of matrix $\mathbf{A}_{sq}$ become increasingly large, which can lead to a poorly conditioned matrix and associated numerical issues. Indeed, a direct evaluation of Eq.~\eqref{Eq:VarPb_} using Matlab, will not produce results after a given time, as could be implied from Figure \ref{fig:Hertzian_LJ_UQ}. Although these numerical issues could be potentially resolved by recourse to more sophisticated strategies, these were not pursued here, as the time range of the predictions already surpassed the point of failure of the path reweighting strategy.

\section{Conclusions} \label{Sec:Conclusions}

In this paper, we provide a statistical mechanics approach with quantified uncertainty to extrapolate material behavior to distinct loading conditions or material systems. The approach is based on the reweighting of the probability density for trajectories, building up on the ideas of \cite{chen2007exact}, and enables the calculation of ensemble averages of arbitrary observables of system/process $S$ from simulations (or, potentially, experimental data) of system/process $\tilde{S}$. The formalism is \emph{a priori} exact and possess many attractive features, such as acceleration of the dynamics under a suitable choice of bias potential, enabling trivial time parallelization, or the full exploration of a family of potentials at virtually zero added computational cost. Yet, it suffers from sampling issues as the ``distance'' between the predicted and simulated system increase, which become more apparent for large times or large particulate systems. In other words, for a fixed number of realizations of system $\tilde{S}$, the uncertainty in the predictions for $S$ become increasingly large with the bias potential for a fixed time, or with time, for a fixed bias potential. Remarkably though, the uncertainty of such predictions can be estimated \emph{a priori} without requiring any simulations of $S$. Specifically, analytical formulas for estimating the uncertainty were here derived based on a quadratic approximation of the bias potential (difference between the potential of systems $\tilde{S}$ and $S$). These estimates proved to be remarkably accurate for systems with a strong quadratic bias (and markedly more accurate than classical formulas for the propagation of uncertainty), and deliver estimates for the errors within the good order of magnitude for realistic potentials.

The above path reweighting strategy and uncertainty quantification estimates have been applied to two illustrative examples. The first example is a one-dimensional mass-spring chain, often used as prototype for polymer chains or biological macromolecules. This simple example is used to showcase the versatility of the approach, both, in the type of observables that can be predicted (microscopic or macroscopic, and instantaneous or path dependent), and the types of inference that can be made from one system to another (e.g., from one potential to another, from equilibrium to non-equilibrium, or the extreme case of predicting the non-equilibrium behavior of an interacting particle system from independent Brownian particles). The second example, is a two-dimensional glass-former system with Hertzian or Lennard-Jones potential, where 
the emergence of particle caging is predicted from the liquid phase as the strength of the potential is increased. These two rather distinct examples (one elastic and the other diffusive) illustrate the possibility of extrapolating under far-from-equilibrium conditions with a high degree of accuracy for small systems and short times.

\section{Acknowledgements}
The authors gratefully acknowledge support from ACS, USA PRF 61793-ND10 Grant (S.~H.) and from NSF CAREER Award, CMMI-2047506 (C.~R). S.~F. acknowledges support from the UK EPSRC, grant no. EP/R005974/1, and helpful discussions with Profs. T.~Ala-Nissil\"a and A.~Archer. I.~G. and P.~E.~A. acknowledge support from NSF DMR (Penn MRSEC) 1720530.

\appendix

\section{Path integral representation of Langevin dynamics} \label{Sec:ProofPath}
We here consider the main equation discussed in the narrative, mainly,
\begin{equation} \label{Eq:Langevin1D}
\eta \dot{x} = -\frac{\partial V}{\partial x} +\sqrt{\sigma}\, \dot{\xi},
\end{equation}
where, for convenience, we restrict ourselves to the one dimensional case, and denote $\sigma=2k_B T \eta$. This stochastic differential equation has additive noise, i.e., $\sigma$ is independent of $x$, and hence, its It{\^o} and Stratonovich interpretation coincide.

In this appendix we derive the path integral representation of Eq.~\eqref{Eq:Langevin1D} \citep{chernyak2006path,kieninger2021path}, following both the It{\^o} and Stratonovich interpretation, and demonstrate their equivalence.

Beginning with the It{\^o} interpretation, we discretize Eq.~\eqref{Eq:Langevin1D} with a constant time step $\Delta t$, according to the Euler-Maruyama scheme
\begin{equation}
    \eta (x^{n+1}-x^n) = -V'(x^n)\Delta t + \sqrt{\sigma}(\xi^{n+1}-\xi^n),
\end{equation}
where the superscript $n+1$ is associated to time step $t^{n+1}=t^n+\Delta t$ and $V'=\partial V / \partial x$.
Then, the probability of seeing a trajectory $\{x\}_0^{n_T}\coloneqq\{x^0,x^1, ...., x^{n_T}\}$ given the initial conditions, is given by the product of the transition probabilities

\begin{equation}
\begin{split}
      &\mathcal{P}\left(\{x\}_0^{n_T} | x^0 \right)= \prod_{n=0}^{n_T-1} Q_n(n+1,n)\\
      &Q_n(n+1,n) = \frac{1}{Z^n} \exp\left[ -\frac{1}{2\sigma} \left(\eta\frac{x^{n+1}-x^n}{\Delta t}+V'(x^n) \right)^2\Delta t\right],
\end{split}
\end{equation}
where the factors $Z^n$ must ensure that the probability distribution is normalized to 1, i.e.,
\begin{equation}
    Z^n = \int_{-\infty}^{\infty}\exp\left[ -\frac{1}{2\sigma} \left(\eta\frac{x^{n+1}-x^n}{\Delta t}+V'(x^n) \right)^2\Delta t\right]\, dx^{n+1}.
\end{equation}
These are simple Gaussian integrals, which can be readily computed as
\begin{equation}
    Z^n = \frac{\sqrt{2\pi \sigma \Delta t}}{\eta}.
\end{equation}
Hence, the path probability distribution reads

\begin{equation} \label{Eq:PathIto}
    \mathcal{P}\left(\{x\}_0^{n_T} | x^0 \right)= \left(\prod_{n=0}^{n_T-1} \frac{\eta}{\sqrt{2\pi \sigma \Delta t}}\right) \exp\left[ -\frac{1}{2\sigma} \sum_{n=0}^{n_T-1} \left(\eta\frac{x^{n+1}-x^n}{\Delta t}+V'(x^n) \right)^2\Delta t\right],
\end{equation}
or equivalently,
\begin{equation} \label{Eq:PathItoCompact}
    \mathcal{P}\left(\{x\}_0^{n_T} | x^0 \right)= \left(\prod_{n=0}^{n_T-1} \frac{\eta}{\sqrt{2\pi \sigma \Delta t}}\right) e^{-\beta \mathcal{I}}, \quad \text{with} \quad \mathcal{I} = \frac{1}{4\eta}\sum_{n=0}^{n_T-1} \left(\eta\frac{x^{n+1}-x^n}{\Delta t}+V'(x^n) \right)^2\Delta t.
\end{equation}

If, in contrast, we had started with a Stratonovich interpretation of Eq.~\eqref{Eq:Langevin1D}, then, its discretized version would read
\begin{equation}
    \eta (x^{n+1}-x^n) = -V'(x^{n+1/2})\Delta t + \sqrt{\sigma}(\xi^{n+1}-\xi^n),
\end{equation}
where $x^{n+1/2}$ is defined as $\frac{x^{n+1}+x^n}{2}$. The probability of observing a trajectory would then be expressed as
\begin{equation}
\begin{split}
      &\mathcal{P}\left(\{x\}_0^{n_T} | x^0 \right)= \prod_{n=0}^{n_T-1} Q_n(n+1,n), \quad \text{with}\\
      &Q_n(n+1,n) = \frac{1}{Z^n} \exp\left[ -\frac{1}{2\sigma} \left(\eta\frac{x^{n+1}-x^n}{\Delta t}+V'(x^{n+1/2}) \right)^2\Delta t\right].
\end{split}
\end{equation}
Similarly, the normalization factors $Z^n$ must satisfy
\begin{equation}
    Z^n = \int_{-\infty}^{\infty}\exp\left[ -\frac{1}{2\sigma} \left(\eta\frac{x^{n+1}-x^n}{\Delta t}+V'(x^{n+1/2}) \right)^2\Delta t\right]\, dx^{n+1}.
\end{equation}
We approximate this integral by doing a Taylor expansion of $V'(x^{n+1/2})$ around $x^n$, as 
\begin{equation}
    V'(x^{n+1/2}) \simeq V'(x^n)+ V''(x^n) \left(x^{n+1/2}-x^n\right)  \simeq V'(x^n)+ V''(x^n) \frac{x^{n+1}-x^n}{2} .
\end{equation}
Then, a simple Gaussian integration delivers
\begin{equation}
    Z^n \simeq \frac{\sqrt{2\pi\sigma\Delta t}}{\eta\left(1+ V''(x^n)\frac{\Delta t}{2\eta}\right)}.
\end{equation}
The term in parenthesis may be further approximated by an exponential, and similarly, $V''$ may be evaluated at $x^{n+1/2}$ to first order, i.e.
\begin{equation}
    Z^n\simeq\frac{\sqrt{2\pi\sigma\Delta t}}{\eta} \exp \left[-\frac{1}{2\eta}V''(x^{n+1/2}) \Delta t \right].
\end{equation}

The path probability distribution (to first order in $\Delta t$ in the exponential) may then be expressed as

\begin{equation} \label{Eq:PathStrato}
    \mathcal{P}\left(\{x\}_0^{n_T} | x^0 \right)= \left(\prod_{n=0}^{n_T-1} \frac{\eta}{\sqrt{2\pi \sigma \Delta t}}\right) \exp\left[ -\frac{1}{2\sigma} \sum_{n=0}^{n_T-1} \left(\eta\frac{x^{n+1}-x^n}{\Delta t}+V'(x^{n+1/2}) \right)^2\Delta t+\frac{1}{2\eta}V''(x^{n+1/2}) \Delta t\right].
\end{equation}

Although this expression is, in appearance, distinct to Eq.~\eqref{Eq:PathIto}, their exponents are actually identical to first order in $\Delta t$. Indeed, after expanding the squares in \eqref{Eq:PathStrato}, the cross term is the only one that requires special consideration. In particular, from It\^o's formula, it follows that
\begin{equation}
\begin{split}
    2\eta \frac{x^{n+1}-x^n}{\Delta t}V'(x^{n+1/2}) & \simeq 2\eta \frac{x^{n+1}-x^n}{\Delta t} V'(x^n)+ 2 \eta \frac{x^{n+1}-x^n}{\Delta t} V''(x^n)\frac{x^{n+1}-x^n}{2} +\mathcal{O}(\sqrt{\Delta t}) \\
    & \simeq 2\eta \frac{x^{n+1}-x^n}{\Delta t} V'(x^n)+ \frac{\sigma}{\eta} V''(x^n)+\mathcal{O}(\sqrt{\Delta t}).
    \end{split}
\end{equation}
The second term involving $V''(x^n)$ then cancels with the last term in the exponent of Eq.~\eqref{Eq:PathStrato}, recovering Eq.~\eqref{Eq:PathIto}. In views of its simplicity, Eq.~\eqref{Eq:PathIto} is chosen for implementation purposes, as well as in the derivations in the following appendices.


\section{Path integral transformation under change of potential} \label{Sec:ChenHoringProof}

Following the discrete representation of the path integrals in It\^o form used in \ref{Sec:ProofPath}, we here prove Eq.~\eqref{Eq:Bias}, for $N=1$ in one dimension, without lost of generality. Towards that goal, we define $V_{bias}=\tilde{V}-V$ and replace $V$ by $\tilde{V}-V_{bias}$ in the rate functional $\mathcal{I}_{bias}$ of Eq.~\eqref{Eq:PathItoCompact}, and expand the squares as
\begin{equation}
\begin{split}
    \mathcal{I} &= \frac{1}{4\eta}\sum_{n=0}^{n_T-1} \left(\eta\frac{x^{n+1}-x^n}{\Delta t}+\tilde{V}'(x^n)-V'_{bias}(x^n) \right)^2\Delta t \\
    &=\frac{1}{4\eta}\sum_{n=0}^{n_T-1} \left[\left(\eta\frac{x^{n+1}-x^n}{\Delta t}+\tilde{V}'(x^n) \right)^2 + \left(V'_{bias}(x^n)\right)^2 -2 V'_{bias}(x^n)\left(\eta\frac{x^{n+1}-x^n}{\Delta t}+\tilde{V}'(x^n)\right)\right]\Delta t.
 \end{split}
\end{equation}
Here, $\tilde{\mathcal{I}}$ may be readily identified in the first term, and hence
\begin{equation}
\mathcal{I}_{bias} =   \mathcal{I} -\mathcal{\tilde{I}}  = \frac{1}{4\eta} \sum_{n=0}^{n_T-1}  V'_{bias}(x^n) \left[ V'_{bias}(x^n)\Delta t - 2\sqrt{2k_B T\eta} \left(\xi^{n+1}-\xi^n \right)\right] , 
\end{equation}
where we have used the Langevin equation associated to $\tilde{V}$.

\section{Nonlinear uncertainty quantification estimate} \label{Seq:UQ}

In this appendix we provide the detailed calculations that lead to the variance estimate for $\mathcal{P}_{bias}$, given by  Eq.~\eqref{Eq:VarPb_} in the narrative.

We begin by deriving an approximation for $\mathcal{P}_{bias}$ resulting from the expansions given in Eqs.~\eqref{Eq:dVExpansion} and \eqref{Eq:noise_approx}. Directly inserting such expansions into the integrand of $\mathcal{I}_{bias}$, this may be approximated as
\begin{equation}
\begin{split}
    & \nabla V_{bias}(\mathbf{x}^n,t^n) \cdot
    \left(\nabla V_{bias}(\mathbf{x}^n,t^n) \Delta t 
    - 2\sqrt{\sigma} \Delta \boldsymbol\xi^n \right) \\
    & \simeq \left \| \left. \nabla V_{bias} \right|_{\mathbf{x}_r^n (t)} 
    + \left. \nabla \nabla V_{bias}\right|_{\mathbf{x}_r^n (t)}  \delta \mathbf{x}^n \right\| ^2 \Delta t \\
    & \quad 
    - 2 \left( \left. \nabla V_{bias} \right|_{\mathbf{x}_r^n (t)} 
    +  \left. \nabla \nabla V_{bias}\right|_{\mathbf{x}_r^n (t)}  \delta \mathbf{x}^n  \right) \cdot 
    \left[ \eta \left( \Delta \mathbf{x}^{n}_r +  \delta \mathbf{x}^{n+1} 
    - \delta \mathbf{x}^{n} \right)
    + \left. \nabla \tilde{V} \right|_{\mathbf{x}_r^n} \Delta t
    + \left. \nabla \nabla \tilde{V} \right|_{\mathbf{x}_r^n}
    \delta \mathbf{x}^{n}  \Delta t
    \right] 
\end{split}
\end{equation}
Expanding the products, and noting that $V_{bias}-\tilde{V}=-V$,
\begin{equation}
\begin{split}
& \nabla V_{bias}(\mathbf{x}^n,t^n) \cdot
    \left(\nabla V_{bias}(\mathbf{x}^n,t^n) \Delta t 
    - 2\sqrt{\sigma} \Delta \boldsymbol\xi^n \right) \\
    & \simeq \left \| \left. \nabla V_{bias} \right|_{\mathbf{x}_r^n (t)} 
    \right\| ^2 \Delta t
    - 2 \left. \nabla V_{bias} \right|_{\mathbf{x}_r^n (t)}  \cdot
    \left[ \eta \Delta \mathbf{x}^{n}_r
    + \left. \nabla \tilde{V} \right|_{\mathbf{x}_r^n} \Delta t
    \right] \\
    & \quad 
    + 2 \left. \nabla V_{bias} \right|_{\mathbf{x}_r^n (t)} 
    \cdot \left(\left. \nabla \nabla V_{bias}\right|_{\mathbf{x}_r^n (t)} \delta \mathbf{x}^n \Delta t \right)\\
    & \quad 
    - 2 \left. \nabla V_{bias} \right|_{\mathbf{x}_r^n (t)} \cdot
    \left[ \eta \left( \delta \mathbf{x}^{n+1} 
    - \delta \mathbf{x}^{n} \right)
    + \left. \nabla \nabla \tilde{V} \right|_{\mathbf{x}_r^n}
     \delta \mathbf{x}^{n} \Delta t
    \right] 
    - 2 \left(\left. \nabla \nabla V_{bias}\right|_{\mathbf{x}_r^n (t)}  
     \delta \mathbf{x}^n \right) \cdot \left[ \eta \Delta \mathbf{x}^{n}_r
    + \left. \nabla \tilde{V} \right|_{\mathbf{x}_r^n} \Delta t
    \right]     \\
    & \quad 
    - 2 \left( \left. \nabla \nabla V_{bias}\right|_{\mathbf{x}_r^n (t)} \delta \mathbf{x}^n  \right) \cdot
    \left[ \eta \left( \delta \mathbf{x}^{n+1} 
    - \delta \mathbf{x}^{n} \right)
    + \left. \nabla \nabla \tilde{V} \right|_{\mathbf{x}_r^n}
     \delta \mathbf{x}^{n}  \Delta t
    \right] \\
    & \quad 
     + \left \| \left. \nabla \nabla V_{bias}\right|_{\mathbf{x}_r^n (t)}  \delta \mathbf{x}^n \right\| ^2 \Delta t \\
    & = \left \| \left. \nabla V_{bias} \right|_{\mathbf{x}_r^n (t)} 
    \right\| ^2 \Delta t
    - 2 \left. \nabla V_{bias} \right|_{\mathbf{x}_r^n (t)}  \cdot
    \left[ \eta \Delta \mathbf{x}^{n}_r
    + \left. \nabla \tilde{V} \right|_{\mathbf{x}_r^n} \Delta t
    \right] \\
    & \quad 
    - 2 \left. \nabla V_{bias} \right|_{\mathbf{x}_r^n (t)} \cdot
    \left[ \eta \left( \delta \mathbf{x}^{n+1} 
    - \delta \mathbf{x}^{n} \right)
    + \left. \nabla \nabla V \right|_{\mathbf{x}_r^n}
     \delta \mathbf{x}^{n} \Delta t
    \right] 
    - 2 \left(\left. \nabla \nabla V_{bias}\right|_{\mathbf{x}_r^n (t)}  
     \delta \mathbf{x}^n \right) \cdot \left[ \eta \Delta \mathbf{x}^{n}_r
    + \left. \nabla \tilde{V} \right|_{\mathbf{x}_r^n} \Delta t
    \right]     \\
    & \quad 
    - 2 \left( \left. \nabla \nabla V_{bias}\right|_{\mathbf{x}_r^n (t)}  \delta \mathbf{x}^n \right) \cdot
    \left[ \eta \left( \delta \mathbf{x}^{n+1} 
    - \delta \mathbf{x}^{n} \right)
    + \left. \nabla \nabla \tilde{V} \right|_{\mathbf{x}_r^n}
     \delta \mathbf{x}^{n}  \Delta t
    \right] \\
    & \quad 
     + \left \| \left.\nabla \nabla V_{bias}\right|_{\mathbf{x}_r^n (t)} \delta \mathbf{x}^n \right\| ^2 \Delta t 
\end{split}
\end{equation}

Next, we define $\delta \mathbf{y}^n = \left( \eta / \sqrt{\sigma \Delta t} \right) \delta \mathbf{x}^n$ and recall that form the definition of $\sigma$, $\frac{\beta}{4\eta}=\frac{1}{2\sigma}$. Then, $\mathcal{P}_{bias}$ may be approximately written as
\begin{equation}
\begin{split}
    \mathcal{P}_{bias} & = e^{-\beta \mathcal{I}_{bias}} = \exp \left[ - \frac{1}{2\sigma} \sum_{n=0}^{n_T - 1} \nabla V_{bias}(\mathbf{x}^n,t^n) \cdot
    \left(\nabla V_{bias}(\mathbf{x}^n,t^n) \Delta t 
    - 2\sqrt{\sigma} \Delta \boldsymbol\xi^n \right) \right] \\
    & \simeq \exp \left[ - \sum_{n=0}^{n_T - 1} \left\{ \frac{1}{2\sigma} \left[ \left \| \left. \nabla V_{bias} \right|_{\mathbf{x}_r^n}
    \right\| ^2
    - 2 \left. \nabla V_{bias} \right|_{\mathbf{x}_r^n }  \cdot
    \left( \eta \frac{\Delta \mathbf{x}^{n}_r}{\Delta t}
    + \left. \nabla \tilde{V} \right|_{\mathbf{x}_r^n}
    \right) \right] \Delta t \right. \right.\\
    & \quad 
    - \sqrt{\frac{\Delta t}{\sigma}} \left[ \left. \nabla V_{bias} \right|_{\mathbf{x}_r^n} \cdot
    \left( \delta \mathbf{y}^{n+1}
    - \delta \mathbf{y}^{n} 
    + \left. \nabla \nabla V \right|_{\mathbf{x}_r^n}
     \delta \mathbf{y}^{n} \frac{\Delta t}{\eta}
    \right)
    + \left(\left. \nabla \nabla V_{bias}\right|_{\mathbf{x}_r^n}  
     \delta \mathbf{y}^n \right) \cdot \left( \Delta \mathbf{x}^{n}_r
    + \left. \nabla \tilde{V} \right|_{\mathbf{x}_r^n} \frac{\Delta t}{\eta}
    \right) \right]
     \\
    & \quad 
    - \frac{1}{\eta} \left( 
    \left. \nabla \nabla V_{bias}\right|_{\mathbf{x}_r^n}  \delta \mathbf{y}^n \right) \cdot
    \left[ \delta \mathbf{y}^{n+1} 
    - \delta \mathbf{y}^{n}
    + \frac{1}{\eta} \left. \nabla \nabla \tilde{V} \right|_{\mathbf{x}_r^n}
    \delta \mathbf{y}^{n}  \Delta t
    \right] \Delta t \\
    & \quad  \left. \left. 
     + \frac{1}{2 \eta^2} \left \|  \left. \nabla \nabla V_{bias}\right|_{\mathbf{x}_r^n} \delta \mathbf{y}^n \right \|^2 \Delta t^2
       \right\} \right].
\end{split}
\end{equation}
    
We now assume that $\mathbf{x}_r(t^0) = \mathbf{0}$ such that $\delta \mathbf{y}^0 = \eta/\sqrt{(\sigma \Delta t)} \delta \mathbf{x}^0 = \mathbf{0}$, and define $\Delta \left( \left. \nabla V_{bias} \right|_{\mathbf{x}_r^{n-1}} \right)\coloneqq\left. \nabla V_{bias} \right|_{\mathbf{x}_r^{n}} -\left. \nabla V_{bias} \right|_{\mathbf{x}_r^{n-1}}$, to rewrite the exponent in $\mathcal{P}_{bias}$ in a quadratic form, i.e.,  
\begin{equation}
\begin{split}
 \mathcal{P}_{bias}   & \simeq  \exp \left\{ -\frac{1}{2\sigma} \sum_{n=0}^{n_T - 1} \left[ \left \| \left. \nabla V_{bias} \right|_{\mathbf{x}_r^n}
    \right\| ^2
    - 2 \left. \nabla V_{bias} \right|_{\mathbf{x}_r^n}  \cdot
    \left( \eta \frac{\Delta \mathbf{x}^{n}_r}{\Delta t}
    + \left. \nabla \tilde{V} \right|_{\mathbf{x}_r^n}
    \right) \right] \Delta t \right. \\
    & \quad 
    + \sqrt{\frac{\Delta t}{\sigma}}  \sum_{n=1}^{n_T - 1} \left[ 
    - \Delta \left( \left. \nabla V_{bias} \right|_{\mathbf{x}_r^{n-1}} \right)
    +  \left. \nabla \nabla V \right|_{\mathbf{x}_r^n} \left. \nabla V_{bias} \right|_{\mathbf{x}_r^n}  \frac{\Delta t}{\eta}
    +  \left. \nabla \nabla V_{bias}\right|_{\mathbf{x}_r^n}  \left(  \Delta \mathbf{x}^{n}_r
    + \left. \nabla \tilde{V} \right|_{\mathbf{x}_r^n} \frac{\Delta t}{\eta}
    \right) 
   \right]  \cdot \delta \mathbf{y}^n \\
    & \quad 
    + \sqrt{\frac{\Delta t}{\sigma}} 
    \left. \nabla V_{bias} \right|_{\mathbf{x}_r^{n_T-1}} \cdot
    \delta \mathbf{y}^{n_T}
     \\
    & \quad 
    + \frac{1}{2 } \sum_{n=1}^{n_T - 1}  \frac{1}{\eta}
    \left( \left. \nabla \nabla V_{bias}\right|_{\mathbf{x}_r^n}  \delta \mathbf{y}^n \right) \cdot
    \left[ \left( - 2 \mathbf{I} \Delta t
    + \frac{2}{\eta} \left. \nabla \nabla \tilde{V} \right|_{\mathbf{x}_r^n} \Delta t^2
    - \frac{1}{\eta} \left. \nabla \nabla V_{bias}\right|_{\mathbf{x}_r^n} \Delta t^2
    \right)  \delta \mathbf{y}^{n} \right] \\
    & \quad
    \left. + \frac{1}{2} \sum_{n=1}^{n_T - 1}  \frac{2}{\eta} 
    \left( \left. \nabla \nabla V_{bias}\right|_{\mathbf{x}_r^n} \delta \mathbf{y}^n  \right) \cdot 
    \delta \mathbf{y}^{n+1}  \Delta t \right\}.
\end{split}
\end{equation}

We recall that the path probability distribution for $\mathbf{x}(t)$ in system $\tilde{S}$ is
\begin{equation}
    \tilde{\mathcal{P}} \left(\mathbf{x}|\mathbf{x}^0\right)
    = \left(\frac{\eta}{\sqrt{2\pi \sigma \Delta t}}\right)^{Nd n_T} \exp \left(-\frac{1}{2 \sigma \Delta t} 
    \sum_{n=0}^{n_T-1} 
    \left\| \eta \Delta \mathbf{x}^n + \nabla \tilde{V}(\mathbf{x}^n, t^n) \Delta t \right\|^2 \right).
\end{equation}
Expanding $\nabla \tilde{V}$ with respect to the reference path, and using the change of variables previously introduced, $\delta \mathbf{y}^n = \left( \eta / \sqrt{\sigma \Delta t} \right) \delta \mathbf{x}^n$, the path probability distribution for $\delta \mathbf{y}(t)$ reads
\begin{equation}
\label{Eq:PathP1y}
    \tilde{\mathcal{P}}\left(\delta \mathbf{y}(t)|\delta \mathbf{y}^0 \right) 
     \simeq \left(\frac{1}{\sqrt{2\pi}}\right)^{\hspace{-0.1cm}Nd n_T} \hspace{-0.4cm}
    \exp \hspace{-0.1cm}\left(\hspace{-0.1cm}-\frac{1}{2} 
    \sum_{n=0}^{n_T-1} 
    \left\| \frac{\eta}{\sqrt{\sigma \Delta t}} \Delta \mathbf{x}^{n}_r
    + \sqrt{\frac{\Delta t}{\sigma}} \left. \nabla \tilde{V} \right|_{\mathbf{x}_r^n} \hspace{-0.2cm}
    + \delta \mathbf{y}^{n+1} - \delta \mathbf{y}^{n}
    + \frac{\Delta t}{\eta} \left. \nabla \nabla \tilde{V} \right|_{\mathbf{x}_r^n} \hspace{-0.2cm}  \delta \mathbf{y}^{n} \right\|^2 \right)\hspace{-0.1cm}. 
\end{equation}
We remark that the prefactor has been modified as well to ensure that the probability distribution is normalized to one. Expanding the squares in the exponential, one obtains
\begin{equation}
\begin{split}
 \tilde{\mathcal{P}}\left(\delta \mathbf{y}(t)|\delta \mathbf{y}^0 \right)   
    & \simeq \left( \frac{1}{\sqrt{2\pi}}\right)^{Nd n_T} 
    \exp \left\{ -\frac{1}{2 \sigma } 
    \sum_{n=0}^{n_T-1} 
    \left\| \eta \frac{\Delta \mathbf{x}^{n}_r}{ \Delta t} 
    + \left. \nabla \tilde{V} \right|_{\mathbf{x}_r^n}
    \right\|^2 \Delta t \right. \\
    & \quad
    - \sum_{n=0}^{n_T-1} 
    \sqrt{\frac{\Delta t}{\sigma}}
    \left( \eta \frac{\Delta \mathbf{x}^{n}_r}{ \Delta t} 
    + \left. \nabla \tilde{V} \right|_{\mathbf{x}_r^n} \right)
    \cdot \left( \delta \mathbf{y}^{n+1} - \delta \mathbf{y}^{n}
    + \frac{\Delta t}{\eta} \left. \nabla \nabla \tilde{V} \right|_{\mathbf{x}_r^n}  \delta \mathbf{y}^{n} \right) \\
    & \quad
    -\frac{1}{2} 
    \sum_{n=0}^{n_T-1} 
    \left[ \left\|\delta \mathbf{y}^{n+1} \right\|^2 
    - 2 \delta \mathbf{y}^{n+1} \cdot \left(
    \left( \mathbf{I} 
    - \frac{\Delta t}{\eta} 
    \left. \nabla \nabla \tilde{V} \right|_{\mathbf{x}_r^n} \right) 
     \delta \mathbf{y}^{n}  \right. \right) \\
    & \quad 
     \left. \left.
    + \delta \mathbf{y}^{n} \cdot 
    \left( \left( \mathbf{I} 
    - 2 \frac{\Delta t}{\eta} 
    \left. \nabla \nabla \tilde{V} \right|_{\mathbf{x}_r^n}
    + \frac{\Delta t^2}{\eta^2} 
    \left. \nabla \nabla \tilde{V} \right|_{\mathbf{x}_r^n}  
    \left. \nabla \nabla \tilde{V} \right|_{\mathbf{x}_r^n} \right) 
     \delta \mathbf{y}^{n}\right) \right] \right\} \\
\end{split}
\end{equation}

Similarly to what we did for $\mathcal{P}_{bias}$, we assume $\delta \mathbf{y}^0=\mathbf{0}$, and manipulate the exponent to convert it into a quadratic form in $\delta \mathbf{y}^n$, i.e., 
\begin{equation}
\begin{split}
\tilde{\mathcal{P}}\left(\delta \mathbf{y}(t)|\delta \mathbf{y}^0 \right)    
    & \simeq \left( \frac{1}{\sqrt{2\pi}}\right)^{Nd n_T} 
    \exp \left\{ -\frac{1}{2 \sigma } 
    \sum_{n=0}^{n_T-1} 
    \left\| \eta \frac{\Delta \mathbf{x}^{n}_r}{ \Delta t} 
    + \left. \nabla \tilde{V} \right|_{\mathbf{x}_r^n}
    \right\|^2 \Delta t \right. \\
    & \quad
    - \sum_{n=1}^{n_T-1} 
    \sqrt{\frac{\Delta t}{\sigma}}
    \left( \eta \frac{\Delta \mathbf{x}^{n}_r}{ \Delta t} 
    + \left. \nabla \tilde{V} \right|_{\mathbf{x}_r^n} \right)
    \cdot \left(\left( - \mathbf{I}
    + \frac{\Delta t}{\eta} \left. \nabla \nabla \tilde{V} \right|_{\mathbf{x}_r^n} \right)
     \delta \mathbf{y}^{n} \right) \\
    & \quad
    - \sum_{n=1}^{n_T} 
    \sqrt{\frac{\Delta t}{\sigma}}
    \left( \eta \frac{\Delta \mathbf{x}^{n-1}_r}{ \Delta t} 
    + \left. \nabla \tilde{V} \right|_{\mathbf{x}_r^{n-1}} \right)
    \cdot \delta \mathbf{y}^{n} \\
    & \quad
    -\frac{1}{2} 
    \sum_{n=1}^{n_T-1} 
    \left[  - 2 \delta \mathbf{y}^{n+1} \cdot  \left(
    \left( \mathbf{I} 
    - \frac{\Delta t}{\eta} 
    \left. \nabla \nabla \tilde{V} \right|_{\mathbf{x}_r^n} \right) 
     \delta \mathbf{y}^{n}  \right. \right) \\
    & \quad 
      \left.
    + \delta \mathbf{y}^{n} \cdot \left(
    \left( \mathbf{I} 
    - 2 \frac{\Delta t}{\eta} 
    \left. \nabla \nabla \tilde{V} \right|_{\mathbf{x}_r^n}
    + \frac{\Delta t^2}{\eta^2} 
    \left. \nabla \nabla \tilde{V} \right|_{\mathbf{x}_r^n} 
    \left. \nabla \nabla \tilde{V} \right|_{\mathbf{x}_r^n} \right) 
     \delta \mathbf{y}^{n} \right) \right]  \\
    & \quad \left.
    -\frac{1}{2} 
    \sum_{n=1}^{n_T} 
    \left\| \delta \mathbf{y}^{n} \right\|^2 \right\}\\
    & = \left( \frac{1}{\sqrt{2\pi}}\right)^{Nd n_T} 
    \exp \left\{ -\frac{1}{2 \sigma } 
    \sum_{n=0}^{n_T-1} 
    \left\| \eta \frac{\Delta \mathbf{x}^{n}_r}{ \Delta t} 
    + \left. \nabla \tilde{V} \right|_{\mathbf{x}_r^n}
    \right\|^2 \Delta t \right. \\
    & \quad
    - \sum_{n=1}^{n_T-1} 
    \sqrt{\frac{\Delta t}{\sigma}}
    \left[ 
     \left( 
    \frac{\Delta t}{\eta} \left. \nabla \nabla \tilde{V} \right|_{\mathbf{x}_r^n} \right) \left( \eta \frac{\Delta \mathbf{x}^{n}_r}{ \Delta t} 
    + \left. \nabla \tilde{V} \right|_{\mathbf{x}_r^n} \right)
    - \Delta \left( \eta \frac{\Delta \mathbf{x}^{n-1}_r}{ \Delta t} 
    + \left. \nabla \tilde{V} \right|_{\mathbf{x}_r^{n-1}} \right) \right]
    \cdot \delta \mathbf{y}^{n}  \\
    & \quad
    - \sqrt{\frac{\Delta t}{\sigma}}
    \left( \eta \frac{\Delta \mathbf{x}^{n_T-1}_r}{ \Delta t} 
    + \left. \nabla \tilde{V} \right|_{\mathbf{x}_r^{n_T-1}} \right)
    \cdot \delta \mathbf{y}^{n_T} \\
    & \quad
    -\frac{1}{2} 
    \sum_{n=1}^{n_T-1} 
    \left[  - 2 \delta \mathbf{y}^{n+1} \cdot \left( 
    \left( \mathbf{I} 
    - \frac{\Delta t}{\eta} 
    \left. \nabla \nabla \tilde{V} \right|_{\mathbf{x}_r^n} \right) 
     \delta \mathbf{y}^{n}  \right. \right) \\
    & \quad 
     \left.
    + \delta \mathbf{y}^{n} \cdot \left(
    \left( 2 \mathbf{I} 
    - 2 \frac{\Delta t}{\eta} 
    \left. \nabla \nabla \tilde{V} \right|_{\mathbf{x}_r^n}
    + \frac{\Delta t^2}{\eta^2} 
    \left. \nabla \nabla \tilde{V} \right|_{\mathbf{x}_r^n}  
    \left. \nabla \nabla \tilde{V} \right|_{\mathbf{x}_r^n} \right) 
     \delta \mathbf{y}^{n} \right)\right] \\
    & \quad \left. 
    - \frac{1}{2} 
    \left\| \delta \mathbf{y}^{n_T} \right\|^2  \right\}.
\end{split}
\end{equation}

Therefore, the average of $\mathcal{P}_{bias}$ can be approximated as,
\begin{equation}
\begin{split}
    \left< \mathcal{P}_{bias} \right>_{\tilde{S}} 
    & = \int\cdots \int 
     \mathcal{P}_{bias}  
    \tilde{\mathcal{P}}\left(\delta \mathbf{y}(t)|\delta \mathbf{y}^0 \right)
    d\delta y^1 \cdots d\delta y^{n_T}  \\
    & \simeq  \int\cdots \int \left( \frac{1}{\sqrt{2\pi}} \right)^{Nd n_T}
    \exp \left\{ 
    -\frac{1}{2\sigma} \sum_{n=0}^{n_T - 1} \left[ 
     \left \| \left. \nabla V_{bias} \right|_{\mathbf{x}_r^n}
    \right\| ^2
    - 2 \left. \nabla V_{bias} \right|_{\mathbf{x}_r^n}  \cdot
    \left( \eta \frac{\Delta \mathbf{x}^{n}_r}{\Delta t}
    + \left. \nabla \tilde{V} \right|_{\mathbf{x}_r^n}
    \right)  \right. \right. \\
    & \quad 
    + \left.
    \left\| \eta \frac{\Delta \mathbf{x}^{n}_r}{ \Delta t} 
    + \left. \nabla \tilde{V} \right|_{\mathbf{x}_r^n}
    \right\|^2 \right] \Delta t \\
    & \quad 
    + \sqrt{\frac{\Delta t}{\sigma}}  \sum_{n=1}^{n_T - 1} \left[ 
     - \Delta \left( \left. \nabla V_{bias} \right|_{\mathbf{x}_r^{n-1}} \right)
    +  \left. \nabla \nabla V \right|_{\mathbf{x}_r^n} \left. \nabla V_{bias} \right|_{\mathbf{x}_r^n} \frac{\Delta t}{\eta}
    +\left. \nabla \nabla V_{bias}\right|_{\mathbf{x}_r^n} \left( \Delta \mathbf{x}^{n}_r
    + \left. \nabla \tilde{V} \right|_{\mathbf{x}_r^n} \frac{\Delta t}{\eta}
    \right)
    \right. \\
    & \quad
    \left. - 
      \left( 
    \frac{\Delta t}{\eta} \left. \nabla \nabla \tilde{V} \right|_{\mathbf{x}_r^n} \right) \left( \eta \frac{\Delta \mathbf{x}^{n}_r}{ \Delta t} 
    + \left. \nabla \tilde{V} \right|_{\mathbf{x}_r^n} \right)
    + \Delta \left( \eta \frac{\Delta \mathbf{x}^{n-1}_r}{ \Delta t} 
    + \left. \nabla \tilde{V} \right|_{\mathbf{x}_r^{n-1}} \right) \right]
    \cdot \delta \mathbf{y}^{n}  \\
    & \quad 
    - \sqrt{\frac{\Delta t}{\sigma}}
    \left( \eta \frac{\Delta \mathbf{x}^{n_T-1}_r}{ \Delta t} 
    + \left. \nabla \tilde{V} \right|_{\mathbf{x}_r^{n_T-1}} 
    - \left. \nabla V_{bias} \right|_{\mathbf{x}_r^{n_T-1}} 
    \right)
    \cdot \delta \mathbf{y}^{n_T} \\
    & \quad 
    + \frac{1}{2} \sum_{n=1}^{n_T - 1} \left(\delta\mathbf{y}^n\right)^\text{T}  
    \left[  \frac{1}{\eta} 
    \left. \nabla \nabla V_{bias}\right|_{\mathbf{x}_r^n } 
    \left( - 2 \mathbf{I} \Delta t
    + \frac{2}{\eta} \left. \nabla \nabla \tilde{V} \right|_{\mathbf{x}_r^n} \Delta t^2
    - \frac{1}{\eta} \left. \nabla \nabla V_{bias}\right|_{\mathbf{x}_r^n} \Delta t^2
    \right)  \right.  \\
    & \quad 
     \left.
    - \left( 2 \mathbf{I} 
    - 2 \frac{\Delta t}{\eta} 
    \left. \nabla \nabla \tilde{V} \right|_{\mathbf{x}_r^n}
    + \frac{\Delta t^2}{\eta^2} 
    \left. \nabla \nabla \tilde{V} \right|_{\mathbf{x}_r^n} 
    \left. \nabla \nabla \tilde{V} \right|_{\mathbf{x}_r^n} \right) 
    \right] \delta \mathbf{y}^{n}  \\
    & \quad
    -\frac{1}{2} 
    \sum_{n=1}^{n_T-1} 
     \left( \delta \mathbf{y}^{n+1} \right)^\text{T}  
    \left( - 2 \mathbf{I} 
    + \frac{2}{\eta} 
    \left. \nabla \nabla \tilde{V} \right|_{\mathbf{x}_r^n} \Delta t
    - \frac{2}{\eta} 
    \left. \nabla \nabla V_{bias}\right|_{\mathbf{x}_r^n } \Delta t 
    \right)
    \delta \mathbf{y}^{n} \\
    & \quad \left. 
    - \frac{1}{2} 
    \left\| \delta \mathbf{y}^{n_T} \right\|^2 
    \right\}
    d\delta y^1 \cdots d\delta y^{n_T}. \\
\end{split}
\end{equation}
    
Recalling that $\tilde{V}-V_{bias}=V$, the resulting expression may be simplified to    
\begin{equation}
\begin{split}
    \left< \mathcal{P}_{bias} \right>_{\tilde{S}} 
    & \simeq \int\cdots \int \left( \frac{1}{\sqrt{2\pi}} \right)^{Nd n_T}
    \exp \left\{ 
    -\frac{1}{2\sigma} \sum_{n=0}^{n_T - 1}  \left\| \eta \frac{\Delta \mathbf{x}^{n}_r}{ \Delta t} 
    + \left. \nabla V \right|_{\mathbf{x}_r^n}
    \right\|^2  \Delta t \right.\\
    & \quad 
    + \sqrt{\frac{\Delta t}{\sigma}}  \sum_{n=1}^{n_T - 1} \left[ 
    -   \left( 
    \frac{\Delta t}{\eta} \left. \nabla \nabla V \right|_{\mathbf{x}_r^n} \right) \left( \eta \frac{\Delta \mathbf{x}^{n}_r}{ \Delta t} 
    + \left. \nabla V \right|_{\mathbf{x}_r^n} \right)
    + \Delta \left( \eta \frac{\Delta \mathbf{x}^{n-1}_r}{ \Delta t} 
    + \left. \nabla V \right|_{\mathbf{x}_r^{n-1}} \right) \right]
    \cdot \delta \mathbf{y}^{n}  \\
    & \quad 
    - \sqrt{\frac{\Delta t}{\sigma}}
    \left( \eta \frac{\Delta \mathbf{x}^{n_T-1}_r}{ \Delta t} 
    + \left. \nabla V \right|_{\mathbf{x}_r^{n_T-1}} 
    \right)
    \cdot \delta \mathbf{y}^{n_T} \\
    & \quad 
    - \frac{1}{2} \sum_{n=1}^{n_T - 1} \left( \delta \mathbf{y}^n \right)^\text{T}  
     \left( 2\mathbf{I} 
    - 2 \frac{\Delta t}{\eta} 
    \left. \nabla \nabla V \right|_{\mathbf{x}_r^n}
    + \frac{\Delta t^2}{\eta^2} 
    \left. \nabla \nabla V \right|_{\mathbf{x}_r^n} 
    \left. \nabla \nabla V \right|_{\mathbf{x}_r^n} \right) 
     \delta \mathbf{y}^{n}  \\
    & \quad
    + \frac{1}{2} 
    \sum_{n=1}^{n_T-1} 
     \left( \delta \mathbf{y}^{n+1} \right)^\text{T} 
    2 \left( \mathbf{I} 
    - \frac{1}{\eta} 
    \left. \nabla \nabla V \right|_{\mathbf{x}_r^n} \Delta t
    \right)
    \delta \mathbf{y}^{n} \\
    & \quad \left. 
    - \frac{1}{2} 
    \left\| \delta \mathbf{y}^{n_T} \right\|^2 
    \right\}
    d\delta y^1 \cdots d\delta y^{n_T} \\
    & = \int \left( \frac{1}{\sqrt{2\pi}} \right)^{Nd n_T}
    e^{ - \frac{1}{2}\mathbf{y}^\text{T} \mathbf{A y} + \mathbf{b} \cdot \mathbf{y} + c  }
    d\mathbf{y} \\
    & =  \frac{1}{\sqrt{\det(\mathbf{A})}} 
    e^{ \frac{1}{2} \mathbf{b}^\text{T} \mathbf{A}^{-1} \mathbf{b} + c  },
\end{split}
\end{equation}
where $c$ is a constant defined as,
\begin{equation}
     c = -\frac{1}{2\sigma} \sum_{n=0}^{n_T - 1}
    \left\| \eta \frac{\Delta \mathbf{x}^{n}_r}{ \Delta t} 
    + \left. \nabla V \right|_{\mathbf{x}_r^n}
    \right\|^2 \Delta t .
\end{equation}
Both vectors $\mathbf{y}$ and $\mathbf{b}$ consist of $n_T$ small $Nd$-dimensional vectors,
\begin{equation}
\begin{split}
    \mathbf{y}  =  
    \begin{pmatrix}
    \mathbf{y}^{1} \\
    \mathbf{y}^{2} \\
    \vdots  \\
    \mathbf{y}^{n_T}
    \end{pmatrix}
    \quad \text{and} \quad
    \mathbf{b}  =  
    \begin{pmatrix}
    \mathbf{b}^{1} \\
    \mathbf{b}^{2} \\
    \vdots  \\
    \mathbf{b}^{n_T}
    \end{pmatrix},
\end{split}
\end{equation}
where, 
\begin{equation}
\begin{split}
    \mathbf{b}^n = 
     \left\{ \begin{aligned}
    & - \sqrt{\frac{\Delta t}{\sigma}}  \left[ 
    \frac{\Delta t}{\eta} \left. \nabla \nabla V \right|_{\mathbf{x}_r^n} \left( \eta \frac{\Delta \mathbf{x}^{n}_r}{ \Delta t} 
    + \left. \nabla V \right|_{\mathbf{x}_r^n} \right)
    - \Delta \left( \eta \frac{\Delta \mathbf{x}^{n-1}_r}{ \Delta t} 
    + \left. \nabla V \right|_{\mathbf{x}_r^{n-1}} \right) \right] ,
    \quad && n < n_T \\
    & - \sqrt{\frac{\Delta t}{\sigma}}
    \left( \eta \frac{\Delta \mathbf{x}^{n_T-1}_r}{ \Delta t} 
    + \left. \nabla V \right|_{\mathbf{x}_r^{n_T-1}} 
    \right) , 
    \quad && n = n_T
    \end{aligned} \right.
\end{split}
\end{equation}
The matrix $\mathbf{A}$ is written as, 
\begin{equation}
\begin{split}
    \mathbf{A} &  =  
    \begin{pmatrix}
    \mathbf{A}^{11} & \mathbf{A}^{12} & \cdots & \mathbf{A}^{1 n_T} \\
    \mathbf{A}^{21} & \mathbf{A}^{22} & \cdots & \mathbf{A}^{2 n_T} \\
    \vdots & \vdots & \ddots & \vdots \\
    \mathbf{A}^{n_T 1} & \mathbf{A}^{n_T 2} & \cdots & \mathbf{A}^{n_T n_T}
    \end{pmatrix}
\end{split}
\end{equation}
with each $Nd \times Nd$ matrix block defined as,
\begin{equation}
\begin{split}
    \mathbf{A}^{pq} & =  
    \left\{ \begin{aligned}
    & \mathbf{I}+\mathbf{\Gamma}^n \mathbf{\Gamma}^n, 
    \quad \quad 
    && (p,q) = (n, n) \text{ with } 
    n = 1, \cdots, n_T-1\\
    & - \mathbf{\Gamma}^n,
    && (p,q) = (n+1, n) \text{ or } (n, n+1) \text{ with } 
    n = 1, \cdots, n_T-1\\
    & \mathbf{I},
    && (p,q) = (n_T, n_T) \\
    & \mathbf{0},
    && \text{otherwise} \\
    \end{aligned} \right. 
\end{split}
\end{equation}
with
\begin{equation} \label{Eq:AppGamma}
    \mathbf{\Gamma}^n
    = \mathbf{I} 
    - \frac{1}{\eta} 
    \left. \nabla \nabla V \right|_{\mathbf{x}_r^n} \Delta t.
\end{equation}
The matrices $\boldsymbol \Gamma^n$ are symmetric, and, hence, so is the matrix $\mathbf{A}$. Furthermore, $\det \left(\mathbf{A}\right) =1$, as is shown bellow following an iterative procedure, by which we add row $n$, multiplied by $\boldsymbol \Gamma^{n-1}$, to row $n-1$ from $n=n_T$ to $n=2$. That is,
\begin{equation}
\begin{split}
    \det(\mathbf{A}) 
    & = \begin{vmatrix}
    \mathbf{I}+\mathbf{\Gamma}^1 \mathbf{\Gamma}^1 
    & -\mathbf{\Gamma}^1
    & \mathbf{0}
    & \cdots 
    & \mathbf{0}
    & \mathbf{0}
    & \mathbf{0} \\
    -\mathbf{\Gamma}^1 
    & \mathbf{I}+\mathbf{\Gamma}^2 \mathbf{\Gamma}^2
    & -\mathbf{\Gamma}^2
    & \cdots 
    & \mathbf{0}
    & \mathbf{0}
    & \mathbf{0} \\
    -\mathbf{0} 
    & -\mathbf{\Gamma}^2
    & \mathbf{I}+\mathbf{\Gamma}^3 \mathbf{\Gamma}^3
    & \cdots 
    & \mathbf{0}
    & \mathbf{0}
    & \mathbf{0} \\
    \vdots & \vdots & \vdots 
    &\ddots & \vdots & \vdots & \vdots \\
    \mathbf{0} 
    & \mathbf{0}
    & \mathbf{0}
    & \cdots
    & \mathbf{I}+\mathbf{\Gamma}^{n_T-2} \mathbf{\Gamma}^{n_T-2}
    & -\mathbf{\Gamma}^{n_T-2}
    & \mathbf{0} \\
    \mathbf{0} 
    & \mathbf{0}
    & \mathbf{0}
    & \cdots
    & -\mathbf{\Gamma}^{n_T-2}
    & \mathbf{I}+\mathbf{\Gamma}^{n_T-1} \mathbf{\Gamma}^{n_T-1}
    & -\mathbf{\Gamma}^{n_T-1} \\
    \mathbf{0} 
    & \mathbf{0}
    & \mathbf{0}
    & \cdots
    & \mathbf{0}
    & -\mathbf{\Gamma}^{n_T-1}
    & \mathbf{I}
    \end{vmatrix} \\
    & = \begin{vmatrix}
    \mathbf{I}+\mathbf{\Gamma}^1 \mathbf{\Gamma}^1 
    & -\mathbf{\Gamma}^1
    & \mathbf{0}
    & \cdots 
    & \mathbf{0}
    & \mathbf{0}
    & \mathbf{0} \\
    -\mathbf{\Gamma}^1 
    & \mathbf{I}+\mathbf{\Gamma}^2 \mathbf{\Gamma}^2
    & -\mathbf{\Gamma}^2
    & \cdots 
    & \mathbf{0}
    & \mathbf{0}
    & \mathbf{0} \\
    -\mathbf{0} 
    & -\mathbf{\Gamma}^2
    & \mathbf{I}+\mathbf{\Gamma}^3 \mathbf{\Gamma}^3
    & \cdots 
    & \mathbf{0}
    & \mathbf{0}
    & \mathbf{0} \\
    \vdots & \vdots & \vdots 
    &\ddots & \vdots & \vdots & \vdots \\
    \mathbf{0} 
    & \mathbf{0}
    & \mathbf{0}
    & \cdots
    & \mathbf{I}+\mathbf{\Gamma}^{n_T-2} \mathbf{\Gamma}^{n_T-2}
    & -\mathbf{\Gamma}^{n_T-2}
    & \mathbf{0} \\
    \mathbf{0} 
    & \mathbf{0}
    & \mathbf{0}
    & \cdots
    & -\mathbf{\Gamma}^{n_T-2}
    & \mathbf{I}
    & \mathbf{0}\\
    \mathbf{0} 
    & \mathbf{0}
    & \mathbf{0}
    & \cdots
    & \mathbf{0}
    & -\mathbf{\Gamma}^{n_T-1}
    & \mathbf{I}
    \end{vmatrix} \\
    & = \begin{vmatrix}
    \mathbf{I}+\mathbf{\Gamma}^1 \mathbf{\Gamma}^1 
    & -\mathbf{\Gamma}^1
    & \mathbf{0}
    & \cdots 
    & \mathbf{0}
    & \mathbf{0}
    & \mathbf{0} \\
    -\mathbf{\Gamma}^1 
    & \mathbf{I}+\mathbf{\Gamma}^2 \mathbf{\Gamma}^2
    & -\mathbf{\Gamma}^2
    & \cdots 
    & \mathbf{0}
    & \mathbf{0}
    & \mathbf{0} \\
    -\mathbf{0} 
    & -\mathbf{\Gamma}^2
    & \mathbf{I}+\mathbf{\Gamma}^3 \mathbf{\Gamma}^3
    & \cdots 
    & \mathbf{0}
    & \mathbf{0}
    & \mathbf{0} \\
    \vdots & \vdots & \vdots 
    &\ddots & \vdots & \vdots & \vdots \\
    \mathbf{0} 
    & \mathbf{0}
    & \mathbf{0}
    & \cdots
    & \mathbf{I}
    & \mathbf{0}
    & \mathbf{0} \\
    \mathbf{0} 
    & \mathbf{0}
    & \mathbf{0}
    & \cdots
    & -\mathbf{\Gamma}^{n_T-2}
    & \mathbf{I}
    & \mathbf{0}\\
    \mathbf{0} 
    & \mathbf{0}
    & \mathbf{0}
    & \cdots
    & \mathbf{0}
    & -\mathbf{\Gamma}^{n_T-1}
    & \mathbf{I}
    \end{vmatrix} \\
    & = \cdots \\
    & = \begin{vmatrix}
    \mathbf{I}
    & \mathbf{0}
    & \mathbf{0}
    & \cdots 
    & \mathbf{0}
    & \mathbf{0}
    & \mathbf{0} \\
    -\mathbf{\Gamma}^1 
    & \mathbf{I}
    & \mathbf{0}
    & \cdots 
    & \mathbf{0}
    & \mathbf{0}
    & \mathbf{0} \\
    -\mathbf{0} 
    & -\mathbf{\Gamma}^2
    & \mathbf{I}
    & \cdots 
    & \mathbf{0}
    & \mathbf{0}
    & \mathbf{0} \\
    \vdots & \vdots & \vdots 
    &\ddots & \vdots & \vdots & \vdots \\
    \mathbf{0} 
    & \mathbf{0}
    & \mathbf{0}
    & \cdots
    & \mathbf{I}
    & \mathbf{0}
    & \mathbf{0} \\
    \mathbf{0} 
    & \mathbf{0}
    & \mathbf{0}
    & \cdots
    & -\mathbf{\Gamma}^{n_T-2}
    & \mathbf{I}
    & \mathbf{0}\\
    \mathbf{0} 
    & \mathbf{0}
    & \mathbf{0}
    & \cdots
    & \mathbf{0}
    & -\mathbf{\Gamma}^{n_T-1}
    & \mathbf{I}
    \end{vmatrix} \\
    & = 1.
\end{split}
\end{equation}

Next, we show that $\mathbf{b}^\text{T} \mathbf{A}^{-1} \mathbf{b} + 2c = 0$, which leads to $\left< \mathcal{P}_{bias} \right> = 1$. To do so,  we introduce a vector $\mathbf{u}$ with dimension $N d n_T$ such that $\mathbf{Au}=\mathbf{b}$. Then, using as previously the notation $\Delta x^n=x^{n+1}-x^n$ and defining a backward finite difference $\Delta_b x^n=x^{n}-x^{n-1}$, and hence $\Delta_b \Delta x^{n}=\Delta_b x^{n+1} - \Delta_b x^n=x^{n+1}-2x^n+x^{n-1}$, the $n^{th}$ timestep of $\mathbf{Au}=\mathbf{b}$ reads
\begin{equation}
\begin{split}
    (\mathbf{Au})^n &= -\boldsymbol \Gamma^{n-1} \mathbf{u}^{n-1} + \left( \mathbf{I}+\boldsymbol \Gamma^n \boldsymbol \Gamma^n \right)\mathbf{u}^n-\boldsymbol \Gamma^n \mathbf{u}^{n+1}\\
    &= - \Delta_b \Delta \mathbf{u}^{n} 
    + \frac{\Delta t}{\eta} \left[ - \Delta_b \left( \left. \nabla \nabla V\right|_{\mathbf{x}_r^{n}} \mathbf{u}^{n} \right) 
    + \left. \nabla \nabla V\right|_{\mathbf{x}_r^{n}} \Delta \mathbf{u}^n \right] 
    + \frac{\Delta t^2}{\eta^2} \left. \nabla \nabla V\right|_{\mathbf{x}_r^{n}}  \left. \nabla \nabla V\right|_{\mathbf{x}_r^{n}} \mathbf{u}^n
    \\
    & = \mathbf{b}^n 
    = - \sqrt{\frac{\Delta t}{\sigma}}  \left[ 
    \frac{\Delta t}{\eta} \left. \nabla \nabla V \right|_{\mathbf{x}_r^n}
    \left( \eta \frac{\Delta \mathbf{x}^{n}_r}{ \Delta t} 
    + \left. \nabla V \right|_{\mathbf{x}_r^n} \right)
    - \Delta_b \left( \eta \frac{\Delta \mathbf{x}^{n}_r}{ \Delta t} 
    + \left. \nabla V \right|_{\mathbf{x}_r^{n}} \right) \right],
    \quad \text{for } n<n_T \\
    (\mathbf{Au})^{n_T} 
    & = - \boldsymbol \Gamma^{n_T-1} \mathbf{u}^{n_T-1}+\mathbf{u}^{n_T}\\
    & =  \Delta \mathbf{u}^{n_T-1} 
    + \frac{\Delta t}{\eta} \left. \nabla \nabla V\right|_{\mathbf{x}_r^{n_T-1}} \mathbf{u}^{n_T-1} \\
    & = \mathbf{b}^{n_T}
    = - \sqrt{\frac{\Delta t}{\sigma}}
    \left( \eta \frac{\Delta \mathbf{x}^{n_T-1}_r}{ \Delta t} 
    + \left. \nabla V \right|_{\mathbf{x}_r^{n_T-1}}
    \right), 
    \quad \text{for } n = n_T.
\end{split}
\end{equation}
Here, it is assumed that $\mathbf{u}^0=\mathbf{0}$ so that the expressions remain valid for $n=1$. Equivalently, dividing by $\Delta t^2$ and $\Delta t $, respectively, these equations may be written as
\begin{equation}
\label{Eq:2nd_ODE_disc}
\begin{split}
    \frac{\Delta_b}{\Delta t} \frac{\Delta \mathbf{u}^{n}}{\Delta t} 
    - & \frac{1}{\eta} 
    \left[ - \frac{\Delta_b}{\Delta t} \left( \left. \nabla \nabla V\right|_{\mathbf{x}_r^{n}} \mathbf{u}^{n} \right) 
    + \left. \nabla \nabla V\right|_{\mathbf{x}_r^{n}} 
    \frac{\Delta \mathbf{u}^n}{\Delta t} \right]
    - \frac{1}{\eta^2} \left. \nabla \nabla V\right|_{\mathbf{x}_r^{n}}  \left. \nabla \nabla V\right|_{\mathbf{x}_r^{n}} \mathbf{u}^n \\
    & = \frac{\Delta_b}{\Delta t} 
    \left( \frac{\Delta \mathbf{u}^{n}}{\Delta t} 
    + \frac{1}{\eta} \left. \nabla \nabla V\right|_{\mathbf{x}_r^{n}} \mathbf{u}^{n} \right)
    - \frac{1}{\eta} \left. \nabla \nabla V\right|_{\mathbf{x}_r^{n}} 
    \left( \frac{\Delta \mathbf{u}^{n}}{\Delta t} 
    + \frac{1}{\eta} \left. \nabla \nabla V\right|_{\mathbf{x}_r^{n}} \mathbf{u}^{n} \right) \\
    & = \frac{1}{\sqrt{\sigma \Delta t}}
    \left[ \frac{1}{\eta} \left. \nabla \nabla V \right|_{\mathbf{x}_r^n} 
    \left( \eta \frac{\Delta \mathbf{x}^{n}_r}{\Delta t} 
    + \left. \nabla V \right|_{\mathbf{x}_r^n} \right)
    -  \frac{\Delta_b}{\Delta t} \left( \eta \frac{\Delta \mathbf{x}^{n}_r}{ \Delta t} 
    + \left. \nabla V \right|_{\mathbf{x}_r^{n}} \right) \right],
    \quad \text{for } n<n_T, \\
    \frac{\Delta \mathbf{u}^{n_T-1}}{\Delta t} 
    & + \frac{1}{\eta} \left. \nabla \nabla V\right|_{\mathbf{x}_r^{n_T-1}} \mathbf{u}^{n_T-1}
    = - \frac{1}{\sqrt{\sigma \Delta t}}
    \left( \eta \frac{\Delta \mathbf{x}^{n_T-1}_r}{ \Delta t} 
    + \left. \nabla V \right|_{\mathbf{x}_r^{n_T-1}} 
    \right), 
    \quad \text{for } n = n_T.
\end{split}
\end{equation}
Introducing the following two operators,
\begin{equation}
    L^+_n = \frac{\Delta }{\Delta t} + \frac{1}{\eta} \left. \nabla \nabla V \right|_{\mathbf{x}_r^n} 
    \quad \text{and} \quad
    L^-_n = \frac{\Delta_b}{\Delta t} - \frac{1}{\eta} \left. \nabla \nabla V \right|_{\mathbf{x}_r^n}  , 
\end{equation}
the above equations can be simplified as,
\begin{equation}
    \left\{ \begin{aligned}
    & L^-_n L^+_n \mathbf{u}^n
    = - \frac{1}{\sqrt{\sigma \Delta t}} L^-_n\left( \eta \frac{\Delta \mathbf{x}_r^n}{\Delta t} 
    + \left. \nabla V \right|_{\mathbf{x}_r^n} \right),
    \quad \text{for } n < n_T    
     \\
    &  L^+_{n_T-1} \mathbf{u}^{n_T-1}
    = - \frac{1}{\sqrt{\sigma \Delta t}} \left( \eta \frac{\Delta \mathbf{x}_r^{n_T-1} }{\Delta t}
    + \left. \nabla V \right|_{\mathbf{x}_r^{n_T-1}} 
    \right)
    \end{aligned} \right. 
\end{equation}
By inspection, this second order equations can be simplified to a first order equation as,
\begin{equation}
\label{Eq:1ODE_disc}
    L^+_n \mathbf{u}^n
    = - \frac{1}{\sqrt{\sigma \Delta t}} 
    \left( \eta \frac{\Delta \mathbf{x}_r^n }{\Delta t}
    + \left. \nabla V \right|_{\mathbf{x}_r^n} \right),
    \quad \text{for } n < n_T .   
\end{equation}
Moreover,
\begin{equation}
\begin{split}
    \mathbf{b}^\text{T} \mathbf{A}^{-1}& \mathbf{b} 
     = \sum_{n=1}^{n_T} \mathbf{b}^n \cdot \mathbf{u}^n \\
    & = \sqrt{\frac{\Delta t}{\sigma}}  \sum_{n=1}^{n_T-1}  \mathbf{u}^n \cdot L^-_n 
    \left( \eta \frac{\Delta \mathbf{x}_r^n }{\Delta t}
    + \left. \nabla V \right|_{\mathbf{x}_r^n} \right) \Delta t
    - \sqrt{\frac{\Delta t}{\sigma}}  \mathbf{u}^{n_T} \cdot
    \left( \eta \frac{\Delta \mathbf{x}_r^{n_T-1}}{\Delta t} 
    + \left. \nabla V \right|_{\mathbf{x}_r^{n_T-1}} \right) \\
    & = \sqrt{\frac{\Delta t}{\sigma}} \left[ \sum_{n=1}^{n_T-1}  \mathbf{u}^n \cdot \left( \frac{\Delta_b}{\Delta t} - \frac{1}{\eta} \left. \nabla \nabla V \right|_{\mathbf{x}_r^n}  \right) 
    \left( \eta \frac{\Delta \mathbf{x}_r^n }{\Delta t}
    + \left. \nabla V \right|_{\mathbf{x}_r^n} \right) \Delta t 
    - \mathbf{u}^{n_T} \cdot
    \left( \eta \frac{\Delta \mathbf{x}_r^{n_T-1}}{\Delta t} 
    + \left. \nabla V \right|_{\mathbf{x}_r^{n_T-1}} \right) \right] \\
    & = - \sqrt{\frac{\Delta t}{\sigma}} \sum_{n=1}^{n_T-1}  \left( \eta \frac{\Delta \mathbf{x}_r^n }{\Delta t}
    + \left. \nabla V \right|_{\mathbf{x}_r^n} \right) \cdot \left( \frac{\Delta}{\Delta t} + \frac{1}{\eta} \left. \nabla \nabla V \right|_{\mathbf{x}_r^n}  \right) \mathbf{u}^n \Delta t     \\
    & = -  \sqrt{\frac{\Delta t}{\sigma}} \sum_{n=0}^{n_T-1}  
    \left( \eta \frac{\Delta \mathbf{x}_r^n}{\Delta t} 
    + \left. \nabla V \right|_{\mathbf{x}_r^n} \right)
    \cdot L^+_n \mathbf{u}^n \Delta t \\
    & = \frac{1}{\sigma} \sum_{n=0}^{n_T-1} 
    \left\| \eta \frac{\Delta \mathbf{x}_r^n }{\Delta t}
    + \left. \nabla V \right|_{\mathbf{x}_r^n} \right\|^2 \Delta t.
\end{split}
\end{equation}
The previous to last equality represents the discrete analogue of integration by parts, while for the last equality we have made use of Eq.~\eqref{Eq:1ODE_disc}. Therefore $\mathbf{b}^\text{T} \mathbf{A}^{-1} \mathbf{b} + 2c = 0$, which leads to $\left< \mathcal{P}_{bias} \right>_{\tilde{S}} = 1$, as previously anticipated.

Next, we follow a similar procedure to find the ensemble average of $\mathcal{P}_{bias}^2$ as
\begin{equation}
\begin{split}
    \left< \mathcal{P}_{bias}^2 \right>_{\tilde{S}} 
    & = \int\cdots \int 
     \mathcal{P}_{bias}^2 
    \tilde{\mathcal{P}} \left(\delta \mathbf{y}(t)|\delta \mathbf{y}^0 \right)
    d\delta y^1 \cdots d\delta y^{n_T}  \\
    & \simeq  \int\cdots \int \left( \frac{1}{\sqrt{2\pi}} \right)^{Nd n_T}
    \exp \left\{ 
    -\frac{1}{2\sigma} \sum_{n=0}^{n_T - 1} \left[ 
    2 \left \| \left. \nabla V_{bias} \right|_{\mathbf{x}_r^n}
    \right\| ^2
    - 4 \left. \nabla V_{bias} \right|_{\mathbf{x}_r^n}  \cdot
    \left( \eta \frac{\Delta \mathbf{x}^{n}_r}{\Delta t}
    + \left. \nabla \tilde{V} \right|_{\mathbf{x}_r^n}
    \right)  \right. \right. \\
    & \quad 
    + \left.
    \left\| \eta \frac{\Delta \mathbf{x}^{n}_r}{ \Delta t} 
    + \left. \nabla \tilde{V} \right|_{\mathbf{x}_r^n}
    \right\|^2 \right] \Delta t \\
    & \quad 
    + \sqrt{\frac{\Delta t}{\sigma}}  \sum_{n=1}^{n_T - 1} \left[ 
    - 2 \Delta \left( \left. \nabla V_{bias} \right|_{\mathbf{x}_r^{n-1}} \right)
    + 2  \left. \nabla \nabla V \right|_{\mathbf{x}_r^n} \left. \nabla V_{bias} \right|_{\mathbf{x}_r^n} \frac{\Delta t}{\eta}
    + 2 \left. \nabla \nabla V_{bias}\right|_{\mathbf{x}_r^n} \left( \Delta \mathbf{x}^{n}_r
    + \left. \nabla \tilde{V} \right|_{\mathbf{x}_r^n} \frac{\Delta t}{\eta}
    \right)
    \right. \\
    & \quad
    \left. - 
     \left( 
    \frac{\Delta t}{\eta} \left. \nabla \nabla \tilde{V} \right|_{\mathbf{x}_r^n} \right)\left( \eta \frac{\Delta \mathbf{x}^{n}_r}{ \Delta t} 
    + \left. \nabla \tilde{V} \right|_{\mathbf{x}_r^n} \right)
    + \Delta \left( \eta \frac{\Delta \mathbf{x}^{n-1}_r}{ \Delta t} 
    + \left. \nabla \tilde{V} \right|_{\mathbf{x}_r^{n-1}} \right) \right]
    \cdot \delta \mathbf{y}^{n}  \\
    & \quad 
    - \sqrt{\frac{\Delta t}{\sigma}}
    \left( \eta \frac{\Delta \mathbf{x}^{n_T-1}_r}{ \Delta t} 
    + \left. \nabla \tilde{V} \right|_{\mathbf{x}_r^{n_T-1}} 
     - 2 \left. \nabla V_{bias} \right|_{\mathbf{x}_r^{n_T-1}} 
    \right)
    \cdot \delta \mathbf{y}^{n_T} \\
    & \quad 
    + \frac{1}{2} \sum_{n=1}^{n_T - 1} \left(\delta \mathbf{y}^n \right)^\text{T}  
    \left[  \frac{1}{\eta} 
    \left. \nabla \nabla V_{bias}\right|_{\mathbf{x}_r^n }  
    \left( - 4 \mathbf{I} \Delta t
    + \frac{4}{\eta} \left. \nabla \nabla \tilde{V} \right|_{\mathbf{x}_r^n} \Delta t^2
    - \frac{2}{\eta} \left. \nabla \nabla V_{bias}\right|_{\mathbf{x}_r^n} \Delta t^2
    \right)  \right.  \\
    & \quad 
     \left.
    - \left( 2 \mathbf{I} 
    - 2 \frac{\Delta t}{\eta} 
    \left. \nabla \nabla \tilde{V} \right|_{\mathbf{x}_r^n}
    + \frac{\Delta t^2}{\eta^2} 
    \left. \nabla \nabla \tilde{V} \right|_{\mathbf{x}_r^n}  
    \left. \nabla \nabla \tilde{V} \right|_{\mathbf{x}_r^n} \right) 
    \right] \delta \mathbf{y}^{n}  \\
    & \quad
    -\frac{1}{2} 
    \sum_{n=1}^{n_T-1} 
    \left( \delta \mathbf{y}^{n+1} \right)^\text{T} 
    \left( - 2 \mathbf{I} 
    + \frac{2}{\eta} 
    \left. \nabla \nabla \tilde{V} \right|_{\mathbf{x}_r^n} \Delta t
    - \frac{4}{\eta} 
    \left. \nabla \nabla V_{bias}\right|_{\mathbf{x}_r^n } \Delta t 
    \right)
    \delta \mathbf{y}^{n} \\
    & \quad \left. 
    - \frac{1}{2} 
    \left\| \mathbf{y}^{n_T} \right\|^2 
    \right\}
    d\delta y^1 \cdots d\delta y^{n_T}.
\end{split}
\end{equation}

Recalling that $\tilde{V}-V_{bias}=V$, and denoting by $ \tilde{\boldsymbol \Gamma}= \mathbf{I} 
    - \frac{1}{\eta} 
    \left. \nabla \nabla \tilde{V} \right|_{\mathbf{x}_r^n} \Delta t$ in analogy to \eqref{Eq:AppGamma}, the resulting expression may be simplified to  
\begin{equation}
\begin{split}
    \left< \mathcal{P}_{bias}^2 \right>_{\tilde{S}} 
    & \simeq \int\cdots \int \left( \frac{1}{\sqrt{2\pi}} \right)^{Nd n_T}
    \exp \left\{ 
    -\frac{1}{2\sigma} \sum_{n=0}^{n_T - 1}  \left[ 2 \left\| \eta \frac{\Delta \mathbf{x}^{n}_r}{ \Delta t} 
    + \left. \nabla V \right|_{\mathbf{x}_r^n}
    \right\|^2 
    - 
    \left\| \eta \frac{\Delta \mathbf{x}^{n}_r}{ \Delta t} 
    + \left. \nabla \tilde{V} \right|_{\mathbf{x}_r^n}
    \right\|^2  
    \right] \Delta t \right. \\
    & \quad 
    + \sqrt{\frac{\Delta t}{\sigma}}  \sum_{n=1}^{n_T - 1} \left[ 
    -  2  \left( 
    \frac{\Delta t}{\eta} \left. \nabla \nabla V \right|_{\mathbf{x}_r^n} \right) 
    \left( \eta \frac{\Delta \mathbf{x}^{n}_r}{ \Delta t} 
    + \left. \nabla V \right|_{\mathbf{x}_r^n} \right)
    + 2 \Delta \left( \eta \frac{\Delta \mathbf{x}^{n-1}_r}{ \Delta t} 
    + \left. \nabla V \right|_{\mathbf{x}_r^{n-1}} \right)\right. \\
    & \quad
    \left. + 
      \left( 
    \frac{\Delta t}{\eta} \left. \nabla \nabla \tilde{V} \right|_{\mathbf{x}_r^n} \right) \left( \eta \frac{\Delta \mathbf{x}^{n}_r}{ \Delta t} 
    + \left. \nabla \tilde{V} \right|_{\mathbf{x}_r^n} \right)
    - \Delta \left( \eta \frac{\Delta \mathbf{x}^{n-1}_r}{ \Delta t} 
    + \left. \nabla \tilde{V} \right|_{\mathbf{x}_r^{n-1}} \right) \right]
    \cdot \delta \mathbf{y}^{n}  \\
    & \quad 
    - \sqrt{\frac{\Delta t}{\sigma}}
    \left( 2 \eta \frac{\Delta \mathbf{x}^{n_T-1}_r}{ \Delta t} 
    + 2 \left. \nabla V \right|_{\mathbf{x}_r^{n_T-1}} 
    - \eta \frac{\Delta \mathbf{x}^{n_T-1}_r}{ \Delta t} 
    - \left. \nabla \tilde{V} \right|_{\mathbf{x}_r^{n_T-1}} \right)
    \cdot \delta \mathbf{y}^{n_T} \\
    & \quad 
    - \frac{1}{2} \sum_{n=1}^{n_T - 1} \left( \delta \mathbf{y}^n \right)^\text{T}  
     \left( 2 \mathbf{I} + 2 \mathbf{\Gamma}^n \mathbf{\Gamma}^n 
    - \mathbf{I} - \mathbf{\tilde{\Gamma}}^n \mathbf{\tilde{\Gamma}}^n \right) 
     \delta \mathbf{y}^{n}  \\
    & \quad
    + \frac{1}{2} 
    \sum_{n=1}^{n_T-1} 
     \left( \delta \mathbf{y}^{n+1}\right)^\text{T} 
    2 \left( 2 \mathbf{\Gamma}^n
    - \mathbf{\tilde{\Gamma}}^n \right)
    \delta \mathbf{y}^{n} \\
    & \quad \left. 
    - \frac{1}{2} 
    \left\| \delta \mathbf{y}^{n_T} \right\|^2 
    \right\}
    d \delta y^1 \cdots d \delta y^{n_T} \\
    & = \int \left( \frac{1}{\sqrt{2\pi}} \right)^{Nd n_T}
    e^{ - \frac{1}{2}\mathbf{\delta y}^\text{T} \mathbf{A}_{sq} \delta \mathbf{y} + \mathbf{b}_{sq} \cdot \delta \mathbf{y} + c_{sq}  }
    d\delta \mathbf{y} \\
    & =  \frac{1}{\sqrt{\det (\mathbf{A}_{sq})}} 
    e^{ \frac{1}{2} \mathbf{b}^\text{T}_{sq} \mathbf{A}_{sq}^{-1} \mathbf{b}_{sq} + c_{sq} },
\end{split}
\end{equation}
where the new coefficients $\mathbf{A}_{sq}$, $\mathbf{b}_{sq}$ and $c_{sq}$ can be expressed as,
\begin{equation}
\begin{split}
    &\mathbf{A}_{sq} = 2\mathbf{A} - \tilde{\mathbf{A}},  \\
    &\mathbf{b}_{sq} = 2\mathbf{b} -\tilde{\mathbf{b}},\\ 
    &c_{sq} = 2c - \tilde{c}.
\end{split}
\end{equation}
Here, $\tilde{\mathbf{A}}$, $\tilde{\mathbf{b}}$ and $\tilde{c}$ are the defined as the analogues of $\mathbf{A}$, $\mathbf{b}$, and $c$, respectively, with the potential $V$ replaced by $\tilde{V}$.

Therefore, the variance of the bias probability is,
\begin{equation}
\label{Eq:VarPb}
\begin{split}
    \sigma^2_{\mathcal{P}_{bias}}
    & = \left< \mathcal{P}_{bias}^2 \right>_{\tilde{S}} 
    -  \left< \mathcal{P}_{bias} \right>_{\tilde{S}}^2 \\
    &  = \frac{1}{\sqrt{\det (\mathbf{A}_{sq})}} 
    e^{ \frac{1}{2} \mathbf{b}_{sq}^\text{T} \mathbf{A}_{sq}^{-1} \mathbf{b}_{sq} + c_{sq} }
    - 1.
\end{split}
\end{equation}

\section{From the nonlinear to the linear uncertainty quantification estimate} \label{Sec:Connection}

This appendix provides a detailed calculation of Eqs.~\eqref{Eq:<Pbias>2} and \eqref{Eq:<Pbias2>} provided in Section \ref{Sec:Nonlinar2Linear}. Following the approximations given by Eq.~\eqref{Eq:Ibas} for the the integrand of $\mathcal{I}_{bias}$ and the ones described right after, $\mathcal{P}_{bias}$ may be approximated by
\begin{equation}
\begin{split}
    \mathcal{P}_{bias} & = e^{-\beta \mathcal{I}_{bias}} = \exp \left[ - \frac{1}{2\sigma} \sum_{n=0}^{n_T - 1} \nabla V_{bias}(\mathbf{x}^n,t^n) \cdot
    \left(\nabla V_{bias}(\mathbf{x}^n,t^n) \Delta t 
    - 2\sqrt{\sigma} \Delta \boldsymbol\xi^n \right) \right] \\
    & \simeq \exp \left[ - \sum_{n=0}^{n_T - 1}  \frac{1}{2\sigma} \left \| \left. \nabla V_{bias}(\mathbf{x}^n, t^n) \right|_{\Delta \boldsymbol\xi = \mathbf{0}}
    \right\| ^2 \Delta t \right. \\
    & \quad 
     + \frac{1}{\sqrt{\sigma}} \sum_{n=0}^{n_T - 1} 
    \left. \left( \left.\nabla V_{bias}(\mathbf{x}^n, t^n)\right|_{\Delta \boldsymbol\xi = \mathbf{0}} \cdot \Delta \boldsymbol\xi^n 
    - \frac{1}{\eta}
    \left. \nabla \nabla V_{bias}(\mathbf{x}^n, t^n)\right|_{\Delta \boldsymbol\xi = \mathbf{0}} \left. \nabla V_{bias}(\mathbf{x}^n, t^n)\right|_{\Delta \boldsymbol\xi = \mathbf{0}} \cdot \sum_{m=0}^{n-1}  \Delta \boldsymbol\xi^m \Delta t \right) \right]. \\
\end{split}
\end{equation}
Next, the last two sums over $n$ and $m$ can be interchanged ($\sum_{n=0}^{n_T-1}\sum_{m=0}^{n-1}=\sum_{m=0}^{n_T-2}\sum_{n=m+1}^{n_T-1}$) and the labels $m$ and $n$ can be swapped, leading to
\begin{equation}
\begin{split}
    \mathcal{P}_{bias} & = e^{-\beta \mathcal{I}_{bias}} = \exp \left[ - \frac{1}{2\sigma} \sum_{n=0}^{n_T - 1} \nabla V_{bias}(\mathbf{x}^n,t^n) \cdot
    \left(\nabla V_{bias}(\mathbf{x}^n,t^n) \Delta t 
    - 2\sqrt{\sigma} \Delta \boldsymbol\xi^n \right) \right] \\
    & \simeq \exp \left[ - \sum_{n=0}^{n_T - 1}  \frac{1}{2\sigma} \left \| \left. \nabla V_{bias}(\mathbf{x}^n, t^n) \right|_{\Delta \boldsymbol\xi = \mathbf{0}}
    \right\| ^2 \Delta t \right. \\
    & \quad 
    \left. + \frac{1}{\sqrt{\sigma}} \sum_{n=0}^{n_T - 1} 
    \left. \left( \nabla V_{bias}(\mathbf{x}^n, t^n)
    - \frac{1}{\eta} \sum_{m=n+1}^{n_T-1} 
    \nabla \nabla V_{bias}(\mathbf{x}^m, t^m) \nabla V_{bias}(\mathbf{x}^m, t^m) \Delta t \right)
    \right|_{\Delta \boldsymbol\xi = \mathbf{0}}
    \cdot \Delta \boldsymbol\xi^n \right]. \\
\end{split}
\end{equation}
Recalling now that the path probability density for the noise $\Delta \boldsymbol \xi$ can be formally written as
\begin{equation}
    \tilde{\mathcal{P}} \left( \Delta \boldsymbol \xi \right)
    = \left(\frac{\eta}{\sqrt{2\pi \Delta t}}\right)^{Nd n_T} 
    \exp \left(-\frac{1}{2 \Delta t} 
    \sum_{n=0}^{n_T-1} 
    \left\| \Delta \boldsymbol\xi^n \right\|^2 \right),
\end{equation}
the average of $\mathcal{P}_{bias}$ can be found by
\begin{equation}
\begin{split}
    \left< \mathcal{P}_{bias} \right>_{\tilde{S}} 
    & = \int\cdots \int \mathcal{P}_{bias}
    \tilde{\mathcal{P}}\left( \Delta \boldsymbol \xi \right)
    d\Delta \boldsymbol\xi^0 \cdots d\Delta \boldsymbol\xi^{n_T-1}  \\
    & \simeq \int\cdots \int \left( \frac{1}{\sqrt{2\pi \Delta t}} \right)^{Nd n_T}
    \exp \left[ - \sum_{n=0}^{n_T - 1}  \frac{1}{2\sigma} \left \| \left. \nabla V_{bias}(\mathbf{x}^n, t^n) \right|_{\Delta \boldsymbol\xi = \mathbf{0}}
    \right\| ^2 \Delta t \right. \\
    & \quad 
     + \frac{1}{\sqrt{\sigma}} \sum_{n=0}^{n_T - 1} 
    \left. \left( \nabla V_{bias}(\mathbf{x}^n, t^n)
    - \frac{1}{\eta} \sum_{m=n+1}^{n_T-1} 
    \nabla \nabla V_{bias}(\mathbf{x}^m, t^m)\nabla V_{bias}(\mathbf{x}^m, t^m) \Delta t \right)
    \right|_{\Delta \boldsymbol\xi = \mathbf{0}}
    \cdot \Delta \boldsymbol\xi^n \\
    & \quad 
    \left. -\frac{1}{2 \Delta t} 
    \sum_{n=0}^{n_T-1} 
    \left\| \Delta \boldsymbol\xi^n \right\|^2 \right]
    d\Delta \boldsymbol\xi^0 \cdots d\Delta \boldsymbol\xi^{n_T-1} \\
    & = \prod_{n=0}^{n_T-1}
    \int \left( \frac{1}{\sqrt{2\pi \Delta t}} \right)^{Nd}
    \exp \left[ - \frac{1}{2\sigma} \left \| \left. \nabla V_{bias}(\mathbf{x}^n, t^n) \right|_{\Delta \boldsymbol\xi = \mathbf{0}}
    \right\| ^2 \Delta t \right. \\
    & \quad 
     + \frac{1}{\sqrt{\sigma}}
    \left. \left( \nabla V_{bias}(\mathbf{x}^n, t^n)
    - \frac{1}{\eta} \sum_{m=n+1}^{n_T-1} 
    \nabla \nabla V_{bias}(\mathbf{x}^m, t^m) \nabla V_{bias}(\mathbf{x}^m, t^m)\Delta t \right)
    \right|_{\Delta \boldsymbol\xi = \mathbf{0}}
    \cdot \Delta \boldsymbol\xi^n \\
    & \quad 
    \left. -\frac{1}{2 \Delta t} 
    \left\| \Delta \boldsymbol\xi^n \right\|^2 \right]
    d\Delta \boldsymbol\xi^n \\
    & = \left. \exp \left[ - \frac{1}{2\sigma}  \sum_{n=0}^{n_T-1} \left \|  \nabla V_{bias}(\mathbf{x}^n, t^n) 
    \right\| ^2 \Delta t \right]
    \right|_{\Delta \boldsymbol\xi = \mathbf{0}} \\
    & \quad 
    \exp \left[ 
     \frac{\Delta t}{2 \sigma} \sum_{n=0}^{n_T-1} 
    \left. \left( \nabla V_{bias}(\mathbf{x}^n, t^n)
    - \frac{1}{\eta} \sum_{m=n+1}^{n_T-1} 
    \nabla \nabla V_{bias}(\mathbf{x}^m, t^m) \nabla V_{bias}(\mathbf{x}^m, t^m) \Delta t \right)^2  \right|_{\Delta \boldsymbol\xi = \mathbf{0}}\right].
\end{split}
\end{equation}
Similarly, the average of $\mathcal{P}_{bias}^2$ is given by
\begin{equation}
\begin{split}
    \left< \mathcal{P}_{bias}^2 \right>_{\tilde{S}} 
    & = \int\cdots \int \mathcal{P}_{bias}^2
    \tilde{\mathcal{P}}\left( \Delta \boldsymbol \xi \right)
    d\Delta \boldsymbol\xi^0 \cdots d\Delta \boldsymbol\xi^{n_T-1}  \\
    & \simeq \int\cdots \int \left( \frac{1}{\sqrt{2\pi \Delta t}} \right)^{Nd n_T} \exp \left[ - \sum_{n=0}^{n_T - 1}  \frac{1}{\sigma} \left \| \left. \nabla V_{bias}(\mathbf{x}^n, t^n) \right|_{\Delta \boldsymbol\xi = \mathbf{0}}
    \right\| ^2 \Delta t \right. \\
    & \quad 
     + \frac{2}{\sqrt{\sigma}} \sum_{n=0}^{n_T - 1} 
    \left. \left( \nabla V_{bias}(\mathbf{x}^n, t^n)
    - \frac{1}{\eta} \sum_{m=n+1}^{n_T-1} 
    \nabla \nabla V_{bias}(\mathbf{x}^m, t^m) \nabla V_{bias}(\mathbf{x}^m, t^m) \Delta t \right)
    \right|_{\Delta \boldsymbol\xi = \mathbf{0}}
    \cdot \Delta \boldsymbol\xi^n  \\
    & \quad 
    \left. -\frac{1}{2 \Delta t} 
    \left\| \Delta \boldsymbol\xi^n \right\|^2 \right]
    d\Delta \boldsymbol\xi^0 \cdots d\Delta \boldsymbol\xi^{n_T-1} \\
    & = \left. \exp \left[ - \frac{1}{\sigma}  \sum_{n=0}^{n_T-1} \left \|  \nabla V_{bias}(\mathbf{x}^n, t^n) 
    \right\| ^2 \Delta t \right]
    \right|_{\Delta \boldsymbol\xi = \mathbf{0}} \\
    & \quad 
    \exp \left[ 
     \frac{2\Delta t}{\sigma} \sum_{n=0}^{n_T-1} 
    \left. \left( \nabla V_{bias}(\mathbf{x}^n, t^n)
    - \frac{1}{\eta} \sum_{m=n+1}^{n_T-1} 
    \nabla \nabla V_{bias}(\mathbf{x}^m, t^m) \nabla V_{bias}(\mathbf{x}^m, t^m) \Delta t \right)^2
    \right|_{\Delta \boldsymbol\xi = \mathbf{0}}  \right],
\end{split}
\end{equation}
recovering the sought-after expressions.


\begin{thebibliography}{10}

\bibitem{chen2007exact}
L.~Y. Chen and N.~J.~M. Horing.
\newblock An exact formulation of hyperdynamics simulations.
\newblock {\em The Journal of Chemical Physics}, 126(22):224103, 2007.

\bibitem{nummela2007exact}
Jeremiah Nummela and Ioan Andricioaei.
\newblock Exact low-force kinetics from high-force single-molecule unfolding
  events.
\newblock {\em Biophysical Journal}, 93(10):3373--3381, 2007.

\bibitem{kieninger2021path}
Stefanie Kieninger and Bettina~G. Keller.
\newblock Path probability ratios for langevin dynamics—exact and
  approximate.
\newblock {\em The Journal of Chemical Physics}, 154(9):094102, 2021.

\bibitem{yan2016time}
Xin Yan and Pradeep Sharma.
\newblock Time-scaling in atomistics and the rate-dependent mechanical behavior
  of nanostructures.
\newblock {\em Nano Letters}, 16(6):3487--3492, 2016.

\bibitem{collin2005verification}
Delphine Collin, Felix Ritort, Christopher Jarzynski, Steven~B. Smith, Ignacio
  Tinoco, and Carlos Bustamante.
\newblock Verification of the crooks fluctuation theorem and recovery of {RNA}
  folding free energies.
\newblock {\em Nature}, 437(7056):231--234, 2005.

\bibitem{hashin1963variational}
Zvi Hashin and Shmuel Shtrikman.
\newblock A variational approach to the theory of the elastic behaviour of
  multiphase materials.
\newblock {\em Journal of the Mechanics and Physics of Solids}, 11(2):127--140,
  1963.

\bibitem{zwanzig1954}
Robert~W. Zwanzig.
\newblock High-temperature equation of state by a perturbation method. {I}.
  {N}onpolar gases.
\newblock {\em The Journal of Chemical Physics}, 22(8):1420--1426, 1954.

\bibitem{rickman2002free}
J.~M. Rickman and R.~LeSar.
\newblock Free-energy calculations in materials research.
\newblock {\em Annual Review of Materials Research}, 32(1):195--217, 2002.

\bibitem{torrie1977nonphysical}
Glenn~M. Torrie and John~P. Valleau.
\newblock Nonphysical sampling distributions in monte carlo free-energy
  estimation: Umbrella sampling.
\newblock {\em Journal of Computational Physics}, 23(2):187--199, 1977.

\bibitem{laio2002escaping}
Alessandro Laio and Michele Parrinello.
\newblock Escaping free-energy minima.
\newblock {\em Proceedings of the National Academy of Sciences},
  99(20):12562--12566, 2002.

\bibitem{laio2008metadynamics}
Alessandro Laio and Francesco~L. Gervasio.
\newblock Metadynamics: a method to simulate rare events and reconstruct the
  free energy in biophysics, chemistry and material science.
\newblock {\em Reports on Progress in Physics}, 71(12):126601, 2008.

\bibitem{jarzynski1997equilibrium}
Christopher Jarzynski.
\newblock Equilibrium free-energy differences from nonequilibrium measurements:
  A master-equation approach.
\newblock {\em Physical Review E}, 56(5):5018, 1997.

\bibitem{crooks1999entropy}
Gavin~E. Crooks.
\newblock Entropy production fluctuation theorem and the nonequilibrium work
  relation for free energy differences.
\newblock {\em Physical Review E}, 60(3):2721, 1999.

\bibitem{voter}
Arthur~F. Voter.
\newblock Hyperdynamics: Accelerated molecular dynamics of infrequent events.
\newblock {\em Physical Review Letters}, 78(20):3908, 1997.

\bibitem{kim2013local}
Soo~Young Kim, Danny Perez, and Arthur~F. Voter.
\newblock Local hyperdynamics.
\newblock {\em The Journal of Chemical Physics}, 139(14):144110, 2013.

\bibitem{bolhuis2002transition}
Peter~G. Bolhuis, David Chandler, Christoph Dellago, and Phillip~L. Geissler.
\newblock Transition path sampling: Throwing ropes over rough mountain passes,
  in the dark.
\newblock {\em Annual Review of Physical Chemistry}, 53(1):291--318, 2002.

\bibitem{allen2009forward}
Rosalind~J. Allen, Chantal Valeriani, and Pieter~Rein Ten~Wolde.
\newblock Forward flux sampling for rare event simulations.
\newblock {\em Journal of Physics: Condensed Matter}, 21(46):463102, 2009.

\bibitem{faradjian2004computing}
Anton~K. Faradjian and Ron Elber.
\newblock Computing time scales from reaction coordinates by milestoning.
\newblock {\em The Journal of Chemical Physics}, 120(23):10880--10889, 2004.

\bibitem{voter1998parallel}
Arthur~F. Voter.
\newblock Parallel replica method for dynamics of infrequent events.
\newblock {\em Physical Review B}, 57(22):R13985, 1998.

\bibitem{perez2016long}
Danny Perez, Ekin~D. Cubuk, Amos Waterland, Efthimios Kaxiras, and Arthur~F.
  Voter.
\newblock Long-time dynamics through parallel trajectory splicing.
\newblock {\em Journal of Chemical Theory and Computation}, 12(1):18--28, 2016.

\bibitem{perez2009accelerated}
Danny Perez, Blas~P. Uberuaga, Yunsic Shim, Jacques~G. Amar, and Arthur~F.
  Voter.
\newblock Accelerated molecular dynamics methods: introduction and recent
  developments.
\newblock {\em Annual Reports in Computational Chemistry}, 5:79--98, 2009.

\bibitem{voter2002extending}
Arthur~F. Voter, Francesco Montalenti, and Timothy~C. Germann.
\newblock Extending the time scale in atomistic simulation of materials.
\newblock {\em Annual Review of Materials Research}, 32(1):321--346, 2002.

\bibitem{zuckerman1999dynamic}
Daniel~M. Zuckerman and Thomas~B. Woolf.
\newblock Dynamic reaction paths and rates through importance-sampled
  stochastic dynamics.
\newblock {\em The Journal of Chemical Physics}, 111(21):9475--9484, 1999.

\bibitem{shin2010polymer}
Jaeoh Shin, Timo Ikonen, Mahendra~D. Khandkar, Tapio Ala-Nissila, and Wokyung
  Sung.
\newblock Polymer escape from a metastable kramers potential: Path integral
  hyperdynamics study.
\newblock {\em The Journal of Chemical Physics}, 133(18):184902, 2010.

\bibitem{ikonen2011diffusion}
T.~Ikonen, M.~D. Khandkar, L.~Y. Chen, S.~C. Ying, and T.~Ala-Nissila.
\newblock Diffusion in periodic potentials with path integral hyperdynamics.
\newblock {\em Physical Review E}, 84(2):026703, 2011.

\bibitem{raj2011phase}
Ritwik Raj and Prashant~K. Purohit.
\newblock Phase boundaries as agents of structural change in macromolecules.
\newblock {\em Journal of the Mechanics and Physics of Solids},
  59(10):2044--2069, 2011.

\bibitem{markutsya2014characterization}
Sergiy Markutsya, Rodney~O. Fox, and Shankar Subramaniam.
\newblock Characterization of sheared colloidal aggregation using langevin
  dynamics simulation.
\newblock {\em Physical Review E}, 89(6):062312, 2014.

\bibitem{hijazi2018fast}
Mahdi Hijazi, David~M. Wilkins, and Michele Ceriotti.
\newblock Fast-forward langevin dynamics with momentum flips.
\newblock {\em The Journal of Chemical Physics}, 148(18):184109, 2018.

\bibitem{chaichian2018path}
Masud Chaichian and Andrei Demichev.
\newblock {\em Path Integrals in Physics: Volume II Quantum Field Theory,
  Statistical Physics and other Modern Applications}.
\newblock CRC Press, 2018.

\bibitem{kloeden1992stochastic}
Peter~E. Kloeden and Eckhard Platen.
\newblock {\em Numerical Solution of Stochastic Differential Equations}.
\newblock Springer, 1999.

\bibitem{taylor1997introduction}
John Taylor.
\newblock {\em Introduction to error analysis, the study of uncertainties in
  physical measurements}.
\newblock University Science Books, 1997.

\bibitem{doi1988theory}
Masao Doi and Samuel~Frederick Edwards.
\newblock {\em The theory of polymer dynamics}, volume~73.
\newblock Oxford University Press, 1988.

\bibitem{huang2021harnessing}
Shenglin Huang, Chuanpeng Sun, Prashant~K Purohit, and Celia Reina.
\newblock Harnessing fluctuation theorems to discover free energy and
  dissipation potentials from non-equilibrium data.
\newblock {\em Journal of the Mechanics and Physics of Solids}, page 104323,
  2021.

\bibitem{Stillinger2013}
Frank~H. Stillinger and Pablo~G. Debenedetti.
\newblock Glass transition thermodynamics and kinetics.
\newblock {\em Annual Review of Condensed Matter Physics}, 4(1):263--285, 2013.

\bibitem{Char2017}
Patrick Charbonneau, Jorge Kurchan, Giorgio Parisi, Pierfrancesco Urbani, and
  Francesco Zamponi.
\newblock Glass and jamming transitions: From exact results to
  finite-dimensional descriptions.
\newblock {\em Annual Review of Condensed Matter Physics}, 8(1):265--288, 2017.

\bibitem{Bonn2017}
Daniel Bonn, Morton~M. Denn, Ludovic Berthier, Thibaut Divoux, and S\'ebastien
  Manneville.
\newblock Yield stress materials in soft condensed matter.
\newblock {\em Reviews of Modern Physics}, 89:035005, 2017.

\bibitem{Nic2018}
Alexandre Nicolas, Ezequiel~E. Ferrero, Kirsten Martens, and Jean-Louis Barrat.
\newblock Deformation and flow of amorphous solids: Insights from elastoplastic
  models.
\newblock {\em Reviews of Modern Physics}, 90:045006, 2018.

\bibitem{berthier2011theoretical}
Ludovic Berthier and Giulio Biroli.
\newblock Theoretical perspective on the glass transition and amorphous
  materials.
\newblock {\em Reviews of Modern Physics}, 83(2):587, 2011.

\bibitem{Bruening2008}
Ralf Brüning, Denis~A. St-Onge, Steve Patterson, and Walter Kob.
\newblock Glass transitions in one-, two-, three-, and four-dimensional binary
  lennard-jones systems.
\newblock {\em Journal of Physics: Condensed Matter}, 21(3):035117, 2008.

\bibitem{Flenner2015}
Elijah Flenner and Grzegorz Szamel.
\newblock Fundamental differences between glassy dynamics in two and three
  dimensions.
\newblock {\em Nature Communications}, 6(1):7392, 2015.

\bibitem{Tsalikis2008}
Dimitrios~G. Tsalikis, Nikolaos Lempesis, Georgios~C. Boulougouris, and
  Doros~N. Theodorou.
\newblock On the role of inherent structures in glass-forming materials: I.
  {T}he vitrification process.
\newblock {\em The Journal of Physical Chemistry B}, 112(34):10619--10627,
  2008.

\bibitem{chernyak2006path}
Vladimir~Y. Chernyak, Michael Chertkov, and Christopher Jarzynski.
\newblock Path-integral analysis of fluctuation theorems for general langevin
  processes.
\newblock {\em Journal of Statistical Mechanics: Theory and Experiment},
  2006(08):P08001, 2006.

\end{thebibliography}

\end{document}